\documentclass[twocolumn]{aastex701} 

\usepackage{hyperref}
\usepackage{enumitem}
\usepackage{amssymb}
\usepackage{lipsum}
\usepackage{lineno}

\newcommand{\spherex}{SPHEREx}

\newdimen\saa  \newdimen\sbb
\def\arcsec{\ifmmode {^{\scriptstyle\prime\prime}}
          \else $^{\scriptstyle\prime\prime}$\fi}
\def\arcmin{\ifmmode {^{\scriptstyle\prime}}
          \else $^{\scriptstyle\prime}$\fi}               
\def\parcs{\saa=.07em \sbb=.03em
     \ifmmode \hbox{\rlap{.}}^{\scriptstyle\prime\kern -\sbb\prime}\hbox{\kern -\saa}
     \else \rlap{.}$^{\scriptstyle\prime\kern -\sbb\prime}$\kern -\saa\fi}
\def\deg{\ifmmode^\circ\else$^\circ$\fi}

\submitjournal{ApJS}
\shorttitle{The SPHEREx Instrument}
\shortauthors{Korngut et al.}

\begin{document}
\makebox[17cm][r]{\textcopyright 2025 all rights reserved.}

\title{The SPHEREx Instrument: Calibration, testing and performance measurements of the NIR spectroscopic surveyor from the laboratory to in-orbit commissioning.  }

\author[0009-0003-8869-3651]{Phil~M.~Korngut}%
\affiliation{Department of Physics, California Institute of Technology, 1200 E. California Boulevard, Pasadena, CA 91125, USA}%
\affiliation{Contact author, pkorngut@caltech.edu}%
\affiliation{Jet Propulsion Laboratory, California Institute of Technology, 4800 Oak Grove Drive, Pasadena, CA 91109, USA}%
\email{placeholder@caltech.edu}%

\author[0000-0002-5710-5212]{James~J.~Bock}%
\affiliation{Department of Physics, California Institute of Technology, 1200 E. California Boulevard, Pasadena, CA 91125, USA}%
\affiliation{Jet Propulsion Laboratory, California Institute of Technology, 4800 Oak Grove Drive, Pasadena, CA 91109, USA}%
\email{placeholder@jpl.nasa.gov}

\author[0000-0003-4255-3650]{Samuel~Condon}%
\affiliation{Department of Physics, California Institute of Technology, 1200 E. California Boulevard, Pasadena, CA 91125, USA}%
\email{placeholder@jpl.nasa.gov}

\author[0009-0002-0098-6183]{C.~Darren~Dowell}%
\affiliation{Jet Propulsion Laboratory, California Institute of Technology, 4800 Oak Grove Drive, Pasadena, CA 91109, USA}%
\affiliation{Department of Physics, California Institute of Technology, 1200 E. California Boulevard, Pasadena, CA 91125, USA}%
\email{placeholder@caltech.edu}%

\author[0000-0001-9925-0146]{Candice~M.~Fazar}%
\affiliation{School of Physics and Astronomy, Rochester Institute of Technology, 1 Lomb Memorial Dr., Rochester, NY 14623, USA}%
\email{placeholder@caltech.edu}%

\author[0000-0001-5812-1903]{Howard~Hui}%
\affiliation{Department of Physics, California Institute of Technology, 1200 E. California Boulevard, Pasadena, CA 91125, USA}%
\affiliation{Jet Propulsion Laboratory, California Institute of Technology, 4800 Oak Grove Drive, Pasadena, CA 91109, USA}%
\email{placeholder@caltech.edu}%

\author{Bradley D. Moore}
\affiliation{Jet Propulsion Laboratory, California Institute of Technology, 4800 Oak Grove Drive, Pasadena, CA 91109, USA}%
\email{placeholder@jpl.nasa.gov}

\author{Bret J. Naylor}
\affiliation{Jet Propulsion Laboratory, California Institute of Technology, 4800 Oak Grove Drive, Pasadena, CA 91109, USA}%
\email{placeholder@jpl.nasa.gov}

\author[0000-0001-9368-3186]{Chi~H.~Nguyen}%
\affiliation{Department of Physics, California Institute of Technology, 1200 E. California Boulevard, Pasadena, CA 91125, USA}%
\email{placeholder@caltech.edu}%

\author[0009-0001-9993-4393]{Stephen~Padin}%
\affiliation{Department of Physics, California Institute of Technology, 1200 E. California Boulevard, Pasadena, CA 91125, USA}%
\email{placeholder@caltech.edu}

\author{James Wincentsen}
\affiliation{Jet Propulsion Laboratory, California Institute of Technology, 4800 Oak Grove Drive, Pasadena, CA 91109, USA}%
\email{placeholder@jpl.nasa.gov}

\author{Asad M. Aboobaker}
\affiliation{Jet Propulsion Laboratory, California Institute of Technology, 4800 Oak Grove Drive, Pasadena, CA 91109, USA}%
\email{placeholder@jpl.nasa.gov}

\author[0000-0001-9674-1564]{Rachel~Akeson}%
\affiliation{IPAC, California Insitute of Technology, MC 100-22, 1200 E California Blvd Pasadena, CA 91125, USA}%
\email{placeholder@caltech.edu}%

\author{John M. Alred}
\affiliation{Jet Propulsion Laboratory, California Institute of Technology, 4800 Oak Grove Drive, Pasadena, CA 91109, USA}%
\email{placeholder@jpl.nasa.gov}

\author{Farah Alibay}
\affiliation{Jet Propulsion Laboratory, California Institute of Technology, 4800 Oak Grove Drive, Pasadena, CA 91109, USA}%
\email{placeholder@jpl.nasa.gov}

\author[0000-0002-3993-0745]{Matthew~L.~N.~Ashby}%
\affiliation{Center for Astrophysics $|$ Harvard \& Smithsonian, Optical and Infrared Astronomy Division, Cambridge, MA 01238, USA}%
\email{placeholder@caltech.edu}%

\author[0000-0002-2618-1124]{Yoonsoo~P.~Bach}%
\affiliation{Korea Astronomy and Space Science Institute (KASI), 776 Daedeok-daero, Yuseong-gu, Daejeon 34055, Republic of Korea}%
\email{placeholder@caltech.edu}%

\author{Joseph~Bichel}%
\affiliation{Department of Physics, California Institute of Technology, 1200 E. California Boulevard, Pasadena, CA 91125, USA}%
\email{placeholder@caltech.edu}%

\author{Douglas~Bolton}
\affiliation{Jet Propulsion Laboratory, California Institute of Technology, 4800 Oak Grove Drive, Pasadena, CA 91109, USA}%
\email{placeholder@jpl.nasa.gov}

\author{David F. Braun}
\affiliation{Jet Propulsion Laboratory, California Institute of Technology, 4800 Oak Grove Drive, Pasadena, CA 91109, USA}%
\email{placeholder@jpl.nasa.gov}

\author{Thomas~Brown}
\affiliation{Jet Propulsion Laboratory, California Institute of Technology, 4800 Oak Grove Drive, Pasadena, CA 91109, USA}%
\email{placeholder@jpl.nasa.gov}

\author[0000-0003-4607-9562]{Sean~A.~Bryan}%
\affiliation{School of Earth and Space Exploration, Arizona State University, 781 Terrace Mall, Tempe, AZ 85287 USA}%
\email{placeholder@caltech.edu}%

\author{Jill~Burnham}%
\affiliation{Department of Physics, California Institute of Technology, 1200 E. California Boulevard, Pasadena, CA 91125, USA}%
\email{placeholder@caltech.edu}%

\author{Thomas~A.~Burk}
\affiliation{Jet Propulsion Laboratory, California Institute of Technology, 4800 Oak Grove Drive, Pasadena, CA 91109, USA}%
\email{placeholder@jpl.nasa.gov}

\author{Nicholas~Burke}%
\affiliation{VIAVI solutions inc., 1402 Mariner Way, Santa Rosa, CA 95407 USA}%
\email{placeholder@caltech.edu}%

\author{Ben~Catching}%
\affiliation{VIAVI solutions inc., 1402 Mariner Way, Santa Rosa, CA 95407 USA}%
\email{placeholder@caltech.edu}%

\author[0000-0001-5929-4187]{Tzu-Ching~Chang}%
\affiliation{Jet Propulsion Laboratory, California Institute of Technology, 4800 Oak Grove Drive, Pasadena, CA 91109, USA}%
\affiliation{Department of Physics, California Institute of Technology, 1200 E. California Boulevard, Pasadena, CA 91125, USA}%
\email{placeholder@caltech.edu}%

\author[0009-0000-3415-2203]{Shuang-Shuang~Chen}%
\affiliation{Department of Physics, California Institute of Technology, 1200 E. California Boulevard, Pasadena, CA 91125, USA}%
\email{placeholder@caltech.edu}%

\author[0000-0002-5437-0504]{Yun-Ting~Cheng}%
\affiliation{Department of Physics, California Institute of Technology, 1200 E. California Boulevard, Pasadena, CA 91125, USA}%
\affiliation{Jet Propulsion Laboratory, California Institute of Technology, 4800 Oak Grove Drive, Pasadena, CA 91109, USA}%
\email{ycheng3@caltech.edu}%

\author[0000-0001-6320-261X]{Yi-Kuan~Chiang}%
\affiliation{Academia Sinica Institute of Astronomy and Astrophysics (ASIAA), No. 1, Section 4, Roosevelt Road, Taipei 10617, Taiwan}%
\email{placeholder@caltech.edu}%

\author{Yong~Chong}
\affiliation{Jet Propulsion Laboratory, California Institute of Technology, 4800 Oak Grove Drive, Pasadena, CA 91109, USA}%
\email{placeholder@jpl.nasa.gov}

\author[0000-0002-3892-0190]{Asantha~Cooray}%
\affiliation{University of California Irvine, Irvine, CA 92697, USA }%
\email{placeholder@caltech.edu }%

\author{Walter~R.~Cook}
\affiliation{Department of Physics, California Institute of Technology, 1200 E. California Boulevard, Pasadena, CA 91125, USA}%
\email{placeholder@caltech.edu}

\author{Velibor~Cormarkovic}
\affiliation{Jet Propulsion Laboratory, California Institute of Technology, 4800 Oak Grove Drive, Pasadena, CA 91109, USA}%
\email{placeholder@caltech.edu}

\author[0000-0002-4650-8518]{Brendan~P.~Crill}%
\affiliation{Jet Propulsion Laboratory, California Institute of Technology, 4800 Oak Grove Drive, Pasadena, CA 91109, USA}%
\affiliation{Department of Physics, California Institute of Technology, 1200 E. California Boulevard, Pasadena, CA 91125, USA}%
\email{placeholder@caltech.edu}%

\author[0000-0002-7471-719X]{Ari~J.~Cukierman}%
\affiliation{Department of Physics, California Institute of Technology, 1200 E. California Boulevard, Pasadena, CA 91125, USA}%
\email{placeholder@caltech.edu}%

\author{Andrew~Davis}%
\affiliation{Department of Physics, California Institute of Technology, 1200 E. California Boulevard, Pasadena, CA 91125, USA}%
\email{placeholder@caltech.edu}%

\author{Dan~Darga}
\affiliation{Jet Propulsion Laboratory, California Institute of Technology, 4800 Oak Grove Drive, Pasadena, CA 91109, USA}%
\email{placeholder@caltech.edu}

\author{Thomas~Disarro}
\affiliation{Jet Propulsion Laboratory, California Institute of Technology, 4800 Oak Grove Drive, Pasadena, CA 91109, USA}%
\email{placeholder@caltech.edu}

\author[0000-0001-7432-2932]{Olivier~Dor\'{e}}%
\affiliation{Jet Propulsion Laboratory, California Institute of Technology, 4800 Oak Grove Drive, Pasadena, CA 91109, USA}%
\affiliation{Department of Physics, California Institute of Technology, 1200 E. California Boulevard, Pasadena, CA 91125, USA}%
\email{placeholder@caltech.edu}%

\author{Beth E. Fabinsky}
\affiliation{Jet Propulsion Laboratory, California Institute of Technology, 4800 Oak Grove Drive, Pasadena, CA 91109, USA}%
\email{placeholder@caltech.edu}

\author[0000-0002-9382-9832]{Andreas~L.~Faisst}%
\affiliation{IPAC, California Insitute of Technology, MC 100-22, 1200 E California Blvd Pasadena, CA 91125, USA}%
\email{placeholder@caltech.edu}%

\author{James~L.~Fanson}
\affiliation{Jet Propulsion Laboratory, California Institute of Technology, 4800 Oak Grove Drive, Pasadena, CA 91109, USA}%
\email{placeholder@caltech.edu}

\author{Allen~H.~Farrington}
\affiliation{Jet Propulsion Laboratory, California Institute of Technology, 4800 Oak Grove Drive, Pasadena, CA 91109, USA}%
\email{placeholder@caltech.edu}

\author[0000-0002-0665-5759]{Tamim~Fatahi}%
\affiliation{IPAC, California Insitute of Technology, MC 100-22, 1200 E California Blvd Pasadena, CA 91125, USA}%
\email{placeholder@caltech.edu}%

\author[0000-0002-9330-8738]{Richard~M.~Feder}%
\affiliation{University of California at Berkeley, Berkeley, CA 94720, USA}%
\email{placeholder@caltech.edu}%

\author{Eric~H.~Frater}
\affiliation{BAE Systems, Inc., Space and Mission Systems, 1600 Commerce St, Boulder, CO 80301, USA}
\email{placeholder@caltech.edu}

\author[0009-0003-5316-5562]{Tatiana~Goldina}%
\affiliation{IPAC, California Insitute of Technology, MC 100-22, 1200 E California Blvd Pasadena, CA 91125, USA}%
\email{placeholder@caltech.edu}%

\author[0000-0002-8990-2101]{Varoujan~Gorjian}
\affiliation{Jet Propulsion Laboratory, California Institute of Technology, 4800 Oak Grove Drive, Pasadena, CA 91109, USA}%
\email{placeholder@jpl.nasa.gov}

\author{William~G.~Hart}
\affiliation{Jet Propulsion Laboratory, California Institute of Technology, 4800 Oak Grove Drive, Pasadena, CA 91109, USA}%
\email{placeholder@jpl.nasa.gov}

\author{Warren~Hendricks}%
\affiliation{VIAVI solutions inc., 1402 Mariner Way, Santa Rosa, CA 95407 USA}%
\email{placeholder@caltech.edu}%

\author[0000-0002-5599-4650]{Joseph~L.~Hora}%
\affiliation{Center for Astrophysics $|$ Harvard \& Smithsonian, Optical and Infrared Astronomy Division, Cambridge, MA 01238, USA}%
\email{placeholder@caltech.edu}%

\author{Viktor~Hristov}%
\affiliation{Department of Physics, California Institute of Technology, 1200 E. California Boulevard, Pasadena, CA 91125, USA}%
\email{placeholder@caltech.edu}%

\author[0009-0009-1219-5128]{Zhaoyu~Huai}%
\affiliation{Department of Physics, California Institute of Technology, 1200 E. California Boulevard, Pasadena, CA 91125, USA}%
\email{placeholder@caltech.edu}%

\author{Charles~A.~Hulse}%
\affiliation{VIAVI solutions inc., 1402 Mariner Way, Santa Rosa, CA 95407 USA}%
\email{placeholder@caltech.edu}%

\author[0000-0003-3574-1784]{Young-Soo~Jo}%
\affiliation{Korea Astronomy and Space Science Institute (KASI), 776 Daedeok-daero, Yuseong-gu, Daejeon 34055, Republic of Korea}%
\email{placeholder@caltech.edu}%

\author[0000-0002-2770-808X]{Woong-Seob~Jeong}%
\affiliation{Korea Astronomy and Space Science Institute (KASI), 776 Daedeok-daero, Yuseong-gu, Daejeon 34055, Republic of Korea}%
\email{placeholder@caltech.edu}%

\author{Makenzie~L.~Kamei}
\affiliation{Jet Propulsion Laboratory, California Institute of Technology, 4800 Oak Grove Drive, Pasadena, CA 91109, USA}%
\email{placeholder@caltech.edu}

\author[0000-0002-3470-2954]{Jae~Hwan~Kang}%
\affiliation{Department of Physics, California Institute of Technology, 1200 E. California Boulevard, Pasadena, CA 91125, USA}%
\email{placeholder@caltech.edu}%

\author{Branislav~Kecman}
\affiliation{Department of Physics, California Institute of Technology, 1200 E. California Boulevard, Pasadena, CA 91125, USA}%
\email{kecman@srl.caltech.edu}

\author{Will~Marchant} 
\affiliation{Space Sciences Laboratory, University of California Berkeley, 7 Gauss Way, Berkeley, CA 94720}%
\email{marchant@ssl.berkeley.edu}

\author{Giacomo~Mariani} 
\affiliation{Jet Propulsion Laboratory, California Institute of Technology, 4800 Oak Grove Drive, Pasadena, CA 91109, USA}%
\email{giacomo.mariani@jpl.nasa.gov}

\author[0000-0001-5382-6138]{Daniel~C.~Masters}%
\affiliation{IPAC, California Insitute of Technology, MC 100-22, 1200 E California Blvd Pasadena, CA 91125, USA}%
\email{dmasters@ipac.caltech.edu}%

\author[0000-0002-6025-0680]{Gary~J.~Melnick}%
\affiliation{Center for Astrophysics $|$ Harvard \& Smithsonian, Optical and Infrared Astronomy Division, Cambridge, MA 01238, USA}%
\email{gmelnick@cfa.harvard.edu}%

\author{Hiromasa~Miyasaka}
\affiliation{Department of Physics, California Institute of Technology, 1200 E. California Boulevard, Pasadena, CA 91125, USA}%
\email{placeholder@jpl.nasa.gov}

\author[0009-0002-0149-9328]{Giulia~Murgia}%
\affiliation{Department of Physics, California Institute of Technology, 1200 E. California Boulevard, Pasadena, CA 91125, USA}%
\email{placeholder@caltech.edu}%

\author[0000-0002-5713-3803]{Christina~Nelson}%
\affiliation{IPAC, California Insitute of Technology, MC 100-22, 1200 E California Blvd Pasadena, CA 91125, USA}%
\email{placeholder@caltech.edu}%

\author[0000-0001-9368-3186]{Hien~T.~Nguyen}%
\affiliation{Jet Propulsion Laboratory, California Institute of Technology, 4800 Oak Grove Drive, Pasadena, CA 91109, USA}%
\affiliation{University of Science, Viet Nam National University Ho Chi Minh City, 227 Nguyen Van Cu, Ho Chi Minh City, Vietnam 700000}%
\email{placeholder@jpl.nasa.gov}

\author{Christopher~Owen}
\affiliation{Jet Propulsion Laboratory, California Institute of Technology, 4800 Oak Grove Drive, Pasadena, CA 91109, USA}%
\email{christopher.j.owen-122899@jpl.nasa.gov}

\author[0000-0002-5158-243X]{Roberta~Paladini}%
\affiliation{IPAC, California Insitute of Technology, MC 100-22, 1200 E California Blvd Pasadena, CA 91125, USA}%
\email{placeholder@caltech.edu}%

\author{Sung-Joon~Park}
\affiliation{Korea Astronomy and Space Science Institute (KASI), 776 Daedeok-daero, Yuseong-gu, Daejeon 34055, Republic of Korea}
\email{placeholder@caltech.edu}

\author{Harshad~Patil}%
\affiliation{VIAVI solutions inc., 1402 Mariner Way, Santa Rosa, CA 95407 USA}%
\email{placeholder@caltech.edu}%

\author{Konstantin~Penanen}
\affiliation{Jet Propulsion Laboratory, California Institute of Technology, 4800 Oak Grove Drive, Pasadena, CA 91109, USA}%
\email{placeholder@caltech.edu}

\author{Chris~Piazzo}%
\affiliation{VIAVI solutions inc., 1402 Mariner Way, Santa Rosa, CA 95407 USA}%
\email{placeholder@caltech.edu}%

\author[0000-0001-9937-8270]{Jeonghyun~Pyo}%
\affiliation{Korea Astronomy and Space Science Institute (KASI), 776 Daedeok-daero, Yuseong-gu, Daejeon 34055, Republic of Korea}%
\email{placeholder@caltech.edu}%

\author{Amelia~Quon}
\affiliation{Jet Propulsion Laboratory, California Institute of Technology, 4800 Oak Grove Drive, Pasadena, CA 91109, USA}%
\email{placeholder@caltech.edu}

\author{Keshav~Ramanathan}
\affiliation{Jet Propulsion Laboratory, California Institute of Technology, 4800 Oak Grove Drive, Pasadena, CA 91109, USA}%
\email{placeholder@caltech.edu}

\author[0000-0003-4408-0463]{Zafar~Rustamkulov}
\affiliation{IPAC, California Insitute of Technology, MC 100-22, 1200 E California Blvd Pasadena, CA 91125, USA}%
\email{placeholder@caltech.edu}

\author{Daniel~J.~Reiley}
\affiliation{Department of Physics, California Institute of Technology, 1200 E. California Boulevard, Pasadena, CA 91125, USA}%
\email{placeholder@caltech.edu }

\author{Eric~B.~Rice}
\affiliation{Jet Propulsion Laboratory, California Institute of Technology, 4800 Oak Grove Drive, Pasadena, CA 91109, USA}%
\email{placeholder@caltech.edu}

\author{Flora~Ridenhour}
\affiliation{Jet Propulsion Laboratory, California Institute of Technology, 4800 Oak Grove Drive, Pasadena, CA 91109, USA}%
\email{placeholder@caltech.edu}

\author{Amber~Roberts}%
\affiliation{VIAVI solutions inc., 1402 Mariner Way, Santa Rosa, CA 95407 USA}%
\email{placeholder@caltech.edu}%

\author{Jennifer~M.~Rocca}
\affiliation{Jet Propulsion Laboratory, California Institute of Technology, 4800 Oak Grove Drive, Pasadena, CA 91109, USA}%
\email{placeholder@jpl.nasa.gov}

\author{Alessandro~Signorini}%
\affiliation{Department of Physics, California Institute of Technology, 1200 E. California Boulevard, Pasadena, CA 91125, USA}%
\email{placeholder@caltech.edu}%

\author{Sara~Susca}
\affiliation{Jet Propulsion Laboratory, California Institute of Technology, 4800 Oak Grove Drive, Pasadena, CA 91109, USA}%
\email{placeholder@caltech.edu}

\author[0000-0003-1841-2241]{Volker~Tolls}%
\affiliation{Center for Astrophysics $|$ Harvard \& Smithsonian, Optical and Infrared Astronomy Division, Cambridge, MA 01238, USA}%
\email{placeholder@caltech.edu}%

\author[0009-0004-2580-3624]{Phani~Velicheti}%
\affiliation{IPAC, California Insitute of Technology, MC 100-22, 1200 E California Blvd Pasadena, CA 91125, USA}%
\email{placeholder@caltech.edu}%

\author[0009-0005-3796-2312]{Pao-Yu~Wang}%
\affiliation{School of Earth and Space Exploration, Arizona State University, 781 Terrace Mall, Tempe, AZ 85287 USA}%
\email{placeholder@caltech.edu}%

\author[0000-0003-4990-189X]{Michael~W.~Werner}%
\affiliation{Jet Propulsion Laboratory, California Institute of Technology, 4800 Oak Grove Drive, Pasadena, CA 91109, USA}%
\email{placeholder@jpl.nasa.gov}

\author{Casey~White}
\affiliation{VIAVI solutions inc., 1402 Mariner Way, Santa Rosa, CA 95407 USA}%
\email{placeholder@caltech.edu}

\author{Ross~Williamson}
\affiliation{Jet Propulsion Laboratory, California Institute of Technology, 4800 Oak Grove Drive, Pasadena, CA 91109, USA}%
\email{placeholder@caltech.edu}

\author[0000-0003-3078-2763]{Yujin~Yang}%
\affiliation{Korea Astronomy and Space Science Institute (KASI), 776 Daedeok-daero, Yuseong-gu, Daejeon 34055, Republic of Korea}%
\email{placeholder@caltech.edu}%

\author[0000-0001-8253-1451]{Michael~Zemcov}%
\affiliation{School of Physics and Astronomy, Rochester Institute of Technology, 1 Lomb Memorial Dr., Rochester, NY 14623, USA}%
\affiliation{Jet Propulsion Laboratory, California Institute of Technology, 4800 Oak Grove Drive, Pasadena, CA 91109, USA}%
\email{placeholder@caltech.edu}

\begin{abstract}
The SPHEREx near-infrared space telescope is an all-sky spectroscopic survey mission launched on March 12th, 2025 UTC.  In addition to providing the community with a spectral database applicable to a wide range of investigations, it is optimized to address three core science goals: to survey the large scale structure of the Universe for signatures of non-Gaussianity during inflation; to conduct intensity mapping studies of the extragalactic background light for probing the history of galaxy evolution; and to survey the plane of the Milky Way for the prevalence and distribution of water and other biogenic ices.  Each of these science goals imposes unique requirements on the performance of the instrument.  We detail the design and testing strategies and report the performance results for the full instrument test campaign, ranging from component-level screening to in-orbit tests during the commissioning phase.  The instrument, currently operating in full science survey mode, meets all of its driving requirements including optical performance, point source sensitivity, thermal stability and correlated noise minimization.  
\end{abstract}

\keywords{cosmology: Diffuse radiation –- Near-infrared astronomy -- Large-scale structure of the universe -- Galaxy evolution -- Interstellar ices -- Infrared instrumentation}

\section{Introduction} 
\label{sec:intro}
\linenumbers
The SpectroPhotometer for the History of the Universe, Epoch of Reionization and ices Explorer, (SPHEREx) was selected for implementation by NASA in February of 2019, as the next medium-class explorer mission in astrophysics. 
At its core, SPHEREx is an all-sky spectroscopic survey, designed to collect a low-resolution spectrum that spans the near-infrared (NIR), towards every 6\farcs15 pixel on the celestial sphere.
The mission and spectrometer design were optimized to simultaneously address three specific astrophysics goals, as well as produce a legacy survey of spectroscopic data for the broader community to use for decades to come.  The ultimate design is the result of a series of performance trades, with certain characteristics geared towards each individual goal, brought together for execution with a single observing mode, in an all-sky spectroscopic survey. 

The fabrication and integration of the SPHEREx instrument was carried out in a staged approach, in a campaign that spanned from mid 2019 through early 2025. Tests, alignments and calibrations were inserted at all levels of instrument fidelity, from screening of individual components to maneuvers with an operational orbital observatory.  In all cases, the test design flowed directly from branches of a requirements tree intended to track and control the interaction of key performance parameters that enable the challenging science goals of the mission.

This paper is one in a series that report on the design and implementation of the SPHEREx instrument.  \cite{jamiemission} serves as the highest level paper, describing the mission as a whole. It motivates the design choices taken to address the scientific goals and includes overviews of the instrument along with other facets such as operation, survey strategy and spacecraft bus capabilities.  Additionally, a set of highly detailed papers with narrow scope describe individual measurements and characteristics.  These include studies of noise (\cite{chinoise}, \cite{grig}), telescope alignment (\cite{frater23}), spectral calibration (\cite{howard}), focus testing (\cite{condon24}), image persistence (\cite{candicepersistence}) and stray light (\cite{darren}).

In this work, we report on the instrument test campaign as a complete picture, detailing the key measurements at each level of mission integration, from component (Section~\ref{sec:component}), to subsystem (Section~\ref{sec:subsystem}), instrument and observatory (Section~\ref{sec:inst}).  At each level, we detail the measurements and how they relate to each other and directly address the mission's scientific objectives. In Section~\ref{S:instover}, we provide an overview of the instrument design to provide context for the campaign. The tests which are featured in their own papers are summarized here (e.g. Sections~ \ref{ssec:tel}, \ref{ssec:focus}, \ref{ssec:sppeccal}), and we focus on how their results feed into the instrument performance understanding and influence downstream measurements and decisions.  Sections~\ref{ssec:lvf}, ~\ref{ssec:fpas}, and \ref{ssec:ice}  include more detailed hardware descriptions along with the test strategies for elements which were not featured in their own dedicated publications.  In Sections \ref{sec:ioc} and \ref{sec:conc}, we report on the in-orbit commissioning of the observatory, and the demonstrated performance of the instrument in the science survey. 

\section{Instrument Overview}
\label{S:instover}

\begin{figure*}
    \centering
    \includegraphics[width=1.00\linewidth]{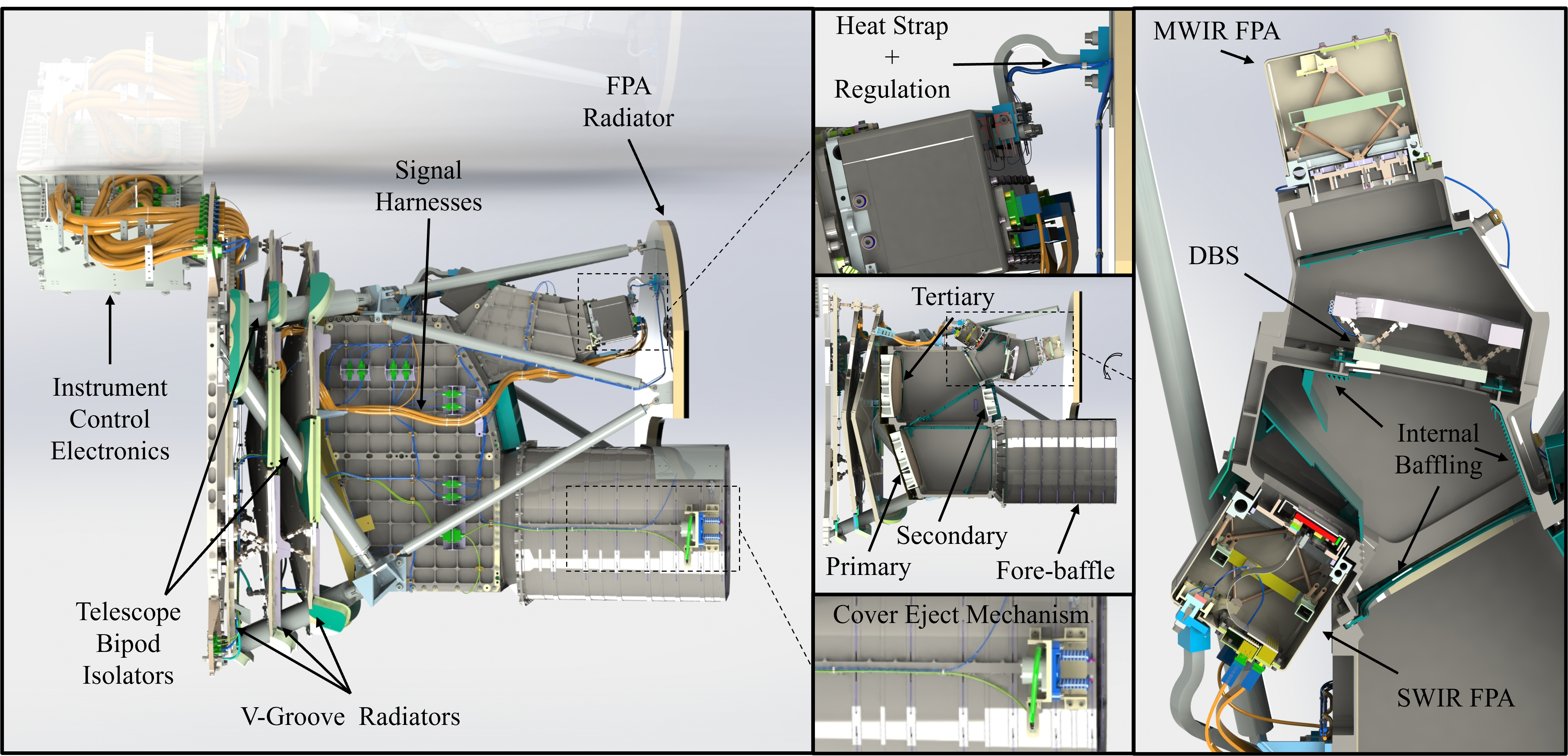}
    \caption{\textit{Left:} Rendering of the SPHEREx Instrument with key components and design elements called out. \textit{Center Top}: A zoomed-in view of the MWIR FPA with its thermal heat sinks and active thermal stabilization components highlighted. \textit{Center Middle}: Central cross section view of the cold instrument with optical elements labeled.  \textit{Center Bottom}: A zoomed in view of the dust cover eject mechanism with harnessing and spring separation device. \textit{Right}: Cross section view of the DBS ``doghouse" and FPAs including the internal optical baffling.}
    \label{fig:inst}
\end{figure*}

Figure~\ref{fig:inst} shows a detailed rendering of the SPHEREx instrument. The telescope is an f/3 off-axis free-form reflective triplet telescope with an $11.3^{\circ} \times 3.5^{\circ}$ FOV, and a circular virtual entrance pupil 0.2~m in diameter.  The primary mirror is physically 0.4~m in its longest dimension to accommodate movement of the pupil's footprint across the FOV. 
The field of view (FOV) is sampled by two co-aligned focal plane assemblies (FPAs) that view the sky simultaneously through a dichroic beamsplitter (DBS).  The DBS reflects wavelengths shorter than $2.42~\mu$m and transmits at longer wavelengths. 
The off-axis nature of the telescope was desirable for SPHEREx as it provides a wide FOV in a compact package with an un-obscured aperture.  This design eliminates azimuthally asymmetric diffraction spikes and allows for a housing that can be filled with internal vanes to block unwanted off-axis signals. A vaned fore-baffle extends over the primary to provide additional stray light control. To limit the amount of particulate contamination reaching optical surfaces during testing, and exposure to the launch vehicle fairing environment, the telescope's baffle was equipped with an ejectable dust cover.  The cover remained fixed to the baffle through launch and separation, and was ejected by command once in-orbit.   The telescope and dust cover assembly were both fabricated by BAE Systems in Boulder Colorado. 

The spectroscopic power of SPHEREx comes from the use of linear variable filters (LVFs).  These custom interference filters are sapphire substrates with coatings that behave like narrow bandpass filters whose central wavelength varies with spatial position along the substrate. Therefore, the position within the focal plane that a source lands on during an exposure determines the wavelength at which it is measured.  To produce a complete spectrum of the sky, each position on the celestial sphere must land on each spectral channel of the focal plane.  Spectral channels are approximately rectangular $3.5^{\circ} \times 0.2^\circ$ in dimension.

The LVFs are integrated into the FPAs, with three distinct bands in each, that we denote as bands 1 through 6 in increasing wavelength order.  The full spectral range spans from 0.75~$\mu m$ to 5.0~$\mu m$.  Each LVF is placed directly above a Teledyne HAWAII-2-RG (H2RG) detector array (\cite{h2rg}).  Bands 1, 2 and 3 reside in the short-wave infrared (SWIR) FPA with a spectral resolving power of 41, and bands 4, 5 and 6 in the mid-wave infrared (MWIR) FPA with resolving powers of 35, 110 and 130 respectively.  The H2RGs are read out in 32 electrical channels that are 64 pixels wide each. The 2048$\times$2048 pixel arrays contain 4 rows and columns on the border that are non-photosensitive, used for reference subtraction.

To eliminate significant self-emission from our instrument and control dark current, the entire instrument must be cooled to cryogenic temperatures.  During operation we required that the telescope housing, all optics and the SWIR FPA be lower than $80$~K with the MWIR FPA lower than $55$~K. This was accomplished entirely by using a passive cooling system consisting of three V-groove radiators for primary cooling, with an additional radiator coupled to the MWIR FPA (\cite{bradthesis}).  
Because SPHEREx operates in low Earth orbit (LEO), a series of three conical photon shields were used in concert with the V-groove radiators to block thermal emission from the Earth and Sun such that the heat can be radiated out to cold space.  See \cite{jamiemission} for a depiction of the full spacecraft including the photon shields.

The Instrument Control Electronics (ICE) provide the readout, on-board data processing, thermal control, and spacecraft communications for the instrument.  It is comprised of three types of circuit boards connected via a backplane mechanically housed in a box mounted to the side of the spacecraft bus. A series of six readout boards (ROBs) clock, bias and sample the detector array outputs, one for each band. Each ROB contains four Video 8 chips that provide amplification and integration designed specifically for SPHEREx. The central electronics board (CEB) communicates with the spacecraft and performs all housekeeping, data compression and thermal regulation functions.  A low voltage power supply board (LVPS) conditions and distributes power throughout the ICE.  

 \section{Astrophysical goals and instrument performance metrics}
 \label{S:sci}
In addition to delivering an all-sky spectroscopic survey, SPHEREx was designed to specifically address three core scientific themes.  For a detailed description of them, see \cite{jamiemission}. For context, we summarize them below.

\begin{enumerate}
    \item \textbf{Measurement of $f_{NL}$:} To probe the fields which drove inflation in the early Universe by searching for primordial non-Gaussianity, as parameterized by $f_{NL}\,$. We are required to measure or provide upper limits on this parameter with an accuracy of $\sigma_{f_{NL}} < 0.5$ after a two-year survey. The signal of interest manifests itself as a correlation between galaxy positions distributed over large cosmological distances. Therefore to accomplish this goal, we must measure the three dimensional positions of hundreds of millions of galaxies with precisely controlled systematic errors on large angular scales.  
    
     \item \textbf{The Extragalactic Background Light, and constraints on galaxy evolution:} 
     To probe the history of galaxy formation by disentangling the component contributions to the The Extragalactic Background Light (EBL).  This light field is the signal generated by the integrated emission across cosmic history.  We accomplish this through spectroscopic intensity mapping, using mosaic maps of two deep fields near the ecliptic poles. We conduct a Fourier analyses on these data, seeking to quantify the clustered galaxy and diffuse signals residing on angular scales between $5\arcmin$ and $ 20\arcmin$. Any spurious sources of noise on these scales must be limited in measurement and precisely quantified. 
    
     \item \textbf{A survey of water and other biogenic ices:}
     To survey the Galactic plane and study the large-scale distribution of biogenic materials throughout the Milky Way.  Evidence supporting the presence of these compounds is found in the absorption line features associated with chemicals understood to be required by life, such as water ice ($3.0~\mu$m), carbon dioxide ($4.3~\mu$m) and carbon monoxide ($4.7~\mu$m). We accomplish this survey by collecting millions of NIR absorption spectra towards background illuminating sources across a range of material in intervening sight lines, with well characterized spectral response functions.
     
\end{enumerate}

Each goal drives SPHEREx to deliver a unique measurement, each susceptible to its own set of systematic and statistical errors. In the following subsections, we walk through the top-level instrument performance metrics, and relate each one's importance to the goals. The test campaign described in greater detail in the body of this paper was ultimately designed to quantify, mitigate and track these metrics to verify that the delivered instrument would meet these goals.  A summary of the performance parameters and the locations in the campaign that they were addressed is given in Table~\ref{tab:testing}.

\subsection{Noise}

The random noise budget, parameterized simply by the standard deviation ($\sigma$) of a signal-less exposure, is essential in assessing the point source sensitivity (PSS).  The PSS determines the number of galaxies whose redshifts can be extracted, which drives our ability to measure $ f_{NL}$. For EBL studies, random noise tracked through $\sigma$ is insufficient, as the spatial correlations in the noise, quantified through its power spectrum, is needed to assess the amplitude on angular scales of interest.  The contributors to the noise are as follows, each needed to be tracked for random and correlated noise characteristics:

\subsubsection{Photon Noise} 
Operating in a LEO environment, the telescope is embedded within the interplanetary dust cloud and each exposure contains a diffuse foreground of Zodiacal light (ZL) which dominates the photocurrent incident on the detectors (\cite{kelsall}).  The Poisson statistics of photon arrival times from ZL is the limiting noise contribution for SPHEREx.  Forecasting this dominant noise source required understanding of the throughput of the system. 

\subsubsection{Detector Read Noise} 
H2RG arrays are comprised of photo-sensitive HgCdTe hybridized to 
read-out integrated circuit (ROIC) boards that read, multiplex, and amplify the cold signal of the detectors.  We require the random read noise associated with this process to be small compared to the irreducible photon noise in each exposure.  Therefore, we tracked the read noise by measuring $\delta Q_{CDS}$, the correlated double sample noise, at various stages.  

\subsubsection{Warm electronics Noise} 
The total noise also contains a small contribution from the warm electronics chain, primarily from the Video 8 chips resident in the ICE. 1/f drifts in detector bias generated in the warm electronics can produce correlated noise in the data. To mitigate these effects, a multi-stage voltage referencing scheme and novel multiplexing readout pattern were implemented, and needed testing at multiple levels to validate. 

\subsection{Optical Properties}
\subsubsection{Core Point Spread Function} 
In the SPHEREx regime, dominated by photon noise from a diffuse foreground, the signal-to-noise ratio (SNR) in photometry is maximized when a source is incident on as few pixels as possible.  Therefore, unlike most imaging telescopes, SPHEREx pixels under-sample the optical point spread function (PSF).  All sources of wavefront error (WFE) that degrade the core PSF must be tracked to prevent impacts to PSS.

\subsubsection{Stray Light Sources}  
The instrument's angular response pattern to sources both on and off of the FOV could potentially be a significant contributor of spurious signal on the angular scales of interest for EBL.  For any telescope, the extended PSF is an unavoidable source of diffuse signal including the irreducible contribution of diffraction. Optical glint paths and scattering from contamination can also enhance off-axis response and must be controlled.
    
Transmissive optics, such as the LVFs and DBS can introduce multiple reflection path signals from bright sources, commonly referred to as ``ghost images."  If the light is not collimated at the point of interaction with transmissive optics, the additional length traversed by rays in ghost images means that they will be out of focus and produce extended signal at the detectors.  Anticipation of these effects and the deployment of mitigating configurations in surface treatment and alignment procedures are essential to limit the contamination of EBL signals by ghost images. The stray light characterization of SPHEREx will be featured in a dedicated study (\cite{darren}).

\subsubsection{Spectroscopic Performance}
The spectral response function of each pixel must be known precisely to interpret measured spectra.  This is especially true for fitting absorption profiles of biogenic ices at the longer wavelengths.  The final response is dominated by the performance of the LVFs, but contains contributions from LVF to H2RG alignment, the DBS, and spectral characteristics of the optical components.  

\subsection{FPA thermal stability}
Thermal drifts in the detectors on orbital to survey timescales could produce gain variations across the sky that must be controlled to mitigate $f_{NL}$ systematics. Thermal fluctuations on exposure timescales could produce spurious clustered structure across the arrays that make systematic errors in EBL studies.  For these reasons, the FPAs were designed with an active thermal stabilization system.

\begin{table*}
    \centering
    \tiny
    \begin{tabular}{|c|c|c|c|c|c|c|}
    \hline
    \textbf{Measurement}& \textbf{Science} & \textbf{Component} & \textbf{Subsystem} & \textbf{Instrument} & \textbf{Observatory} & \textbf{IOC} \\ 
     \textbf{Quantity} & \textbf{Driver} &\textbf{Level} & \textbf{Level} & \textbf{Level} &\textbf{Level} & \textbf{and Survey} \\ \hline \hline
    
    Spectral         & Survey  &LVF screening   &  FPA spectral  &  Spectral response  & & Planetary nebulae       \\ 
    Performance            &  Sampling            & Section~\ref{ssec:lvf}  & response Section~\ref{ssec:fpas} &  in KASI Section~\ref{ssec:sppeccal} & & \cite{howard}       \\ 
                  &  Resolving ice lines       & Figure~\ref{fig:lamprogmeas} &  \cite{howard}           &  \cite{howard} & &        \\ 
                             & &DBS screening       &       & Figure~\ref{fig:lamcompare}  & &        \\ 
                            & &Section~\ref{ssec:dbs}       &       & Figure~\ref{fig:lamresid} & &        \\ \hline
    
    Noise         &  PSS&$\delta Q_{CDS}$    &  Dark FPA testing  &  KASI chamber   & Readout Only& Field differences     \\ 
           & EBL  &Section~\ref{ssec:h2rg}   &  Section~\ref{ssec:fpas}  &  Figure~\ref{fig:phannoise}   & &  \cite{jamiemission}     \\ 
          & systematics  &                        &  \cite{chinoise}  &     &  &   \\ 
            &     &                      &  ICE testing  &     &  &   \\ 
            &      &                     &  \cite{grig}  &     &   &  \\ 
            &       &                    &  Section~\ref{ssec:ice}  &     &   &  \\ \hline
    
    Optical        & PSS &$\eta$ of optical   &    &     &  & Standard stars     \\ 
    Throughput     &  &components  &    &     & & Section~\ref{ssec:practice}      \\ 
       &  &Tables~\ref{tab:lvfscreen},~\ref{tab:detscreen},~\ref{tab:dbs}  &    &     &   &     \\ 
                    &  &Figure~\ref{fig:aeta}  &    &     &   &     \\ \hline
     
    FPA Thermal            &$f_{NL}$   &FPA prototyping  & Dark FPA testing   & KASI chamber    && Orbital measurement       \\ 
    Stability            &EBL  &Section~\ref{sssec:therm} & Section~\ref{sssec:therm}   & Section~\ref{sec:inst}    & &  Section~\ref{ssec:cooldown}     \\ 
               & systematics&Figure~\ref{fig:tempctrl} & Figure~\ref{fig:tempctrl}   & Figure~\ref{fig:tempctrl}    & &  Figure~\ref{fig:tempctrl}     \\ 
                & & &    &     &  & Figure~\ref{fig:6month}     \\ \hline

    Imaging            &  PSS & Mirror WFE  & Telescope alignment   & KASI focus test    & &$N_{eff}$ from stars       \\ 
    Performance                & Confusion&\cite{frater23}   & \cite{frater23}    &   \cite{condon24}  & &Section~\ref{ssec:practice}       \\ 
                 &            & DBS at VIAVI   &   Section~\ref{ssec:tel} & Section~\ref{ssec:focus}    & &Figure~\ref{fig:neff}       \\ 
              &                & Table~\ref{tab:dbs}   &  H2RG Coplanarity  & Figure~\ref{fig:neff}    &  &      \\ 
                               & &HgCdTe flatness & Figure~\ref{fig:fpashim}   &     & &       \\ 
                               & &Table~\ref{tab:detscreen}   &  Section~\ref{sssec:fpamech}  &     & &       \\ \hline
    
    Stray Light      &  Ext PSF &LVF $r_2$ &  H2RG- LVF gap  &     & & Moonshine       \\ 
                    & EBL  &Table~\ref{tab:lvfscreen} & Section~\ref{sssec:fpamech}   &     & & \cite{darren}       \\ 
                &  systematics   & DBS $r_{D1}r_{D2}$ &    &     & & Section~\ref{ssec:moonshine}       \\ 
                     & &Table~\ref{tab:dbs} &    &     & & Figure~\ref{fig:moonmeas}       \\ 
                         & & &    &     & &  Earthshine    \\ \hline

     On-board      & Data Volume &Compression sim & ICE testing   & KASI test    & Functionality& Parameter settings       \\ 
     Processing      & & & Section~\ref{ssec:ice}   &     & & Section~\ref{ssec:surveydiag}       \\ 
                     &  &    & Figure~\ref{fig:compsky}    & &      \\ 
                     & & &    &     & &      \\ \hline

      \end{tabular}
    \caption{Overview of the instrument test campaign and flow of key measurement quantities across configurations.  References to detailed discussions of each topic within this paper and elsewhere are given.  Fidelity of observatory level testing was limited by the higher operating temperature and other logistical considerations.}
    \label{tab:testing}
\end{table*}

\section{Component-Level Testing}
\label{sec:component}
Before the instrument subsystems began construction, a suite of testing was carried out at component-level.  In particular, items that directly touch the signal chain required down-selection and testing at the earliest stages of mission implementation.  In this section, we detail the testing performed on the components.

\subsection{Linear Variable Filters}
\label{ssec:lvf}
The SPHEREx LVFs were developed under a collaborative effort with VIAVI Solutions Inc. (VIAVI), who fabricated and delivered them.  During the fabrication process, the filters were subjected to a suite of measurements to verify compliance with the requirements imposed by the scientific objectives.

\subsubsection{Wavelength Progressions}
\label{sssec:lamprog}

\begin{figure}
    \centering
    \includegraphics[width=1.00\linewidth]{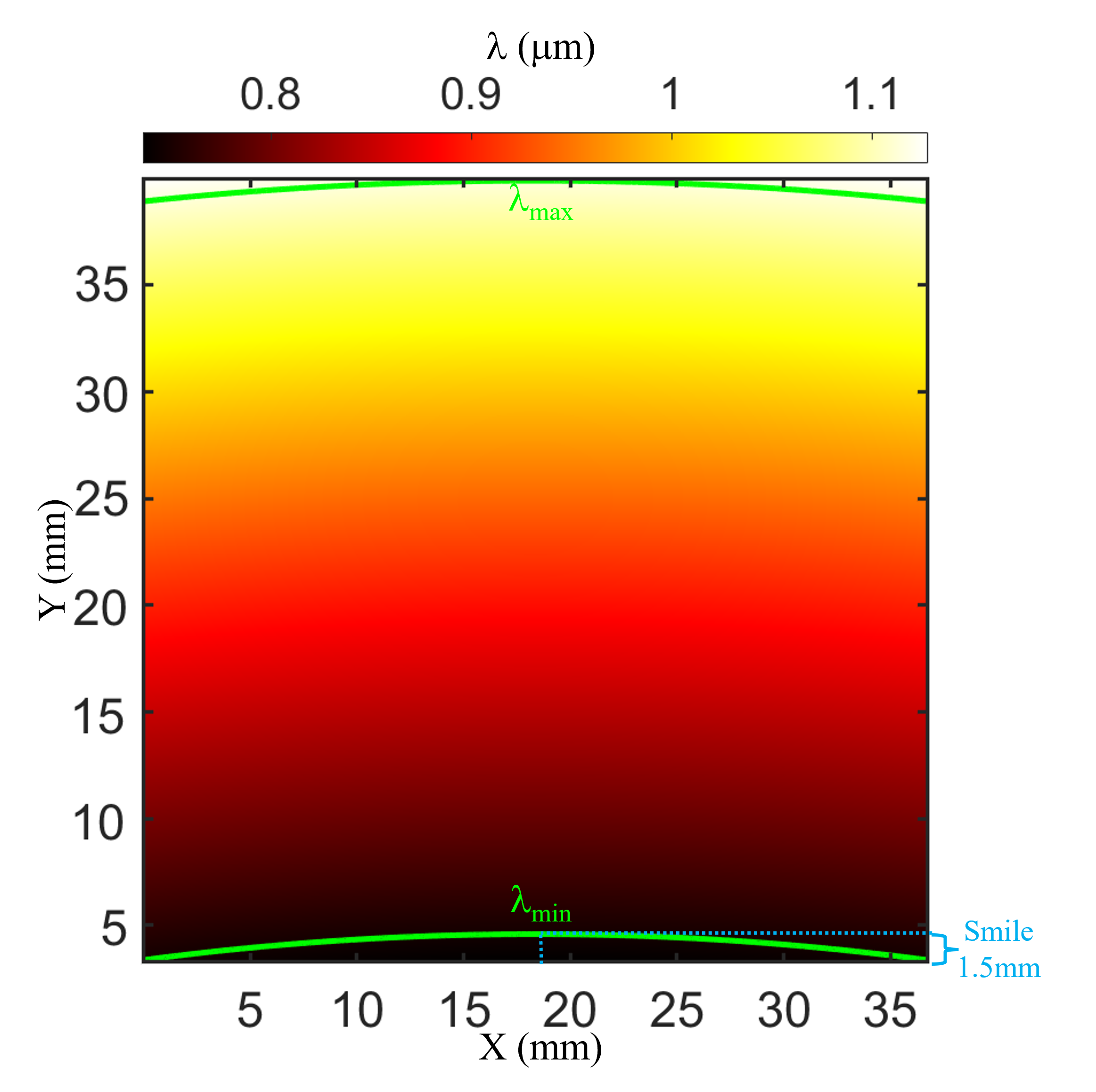}
    \caption{Central wavelength progression for an ideal band 1 LVF.  The wavelength $\lambda_{min}$ corresponds to the iso-wavelength contour that intersects the corner pixel, while the $\lambda_{max}$ contour apex intersects the central X pixel. Smile quantifies the level of curvature in an iso-wavelength contour. }
    \label{fig:lammap}
\end{figure}

LVFs functionally map each location within the telescope's FOV to a wavelength. To accurately execute the spectral sampling of the sky, it was essential to produce LVFs with a well-constrained and characterized wavelength progression.

\begin{figure*}
    \centering
    \includegraphics[width=1.00\linewidth]{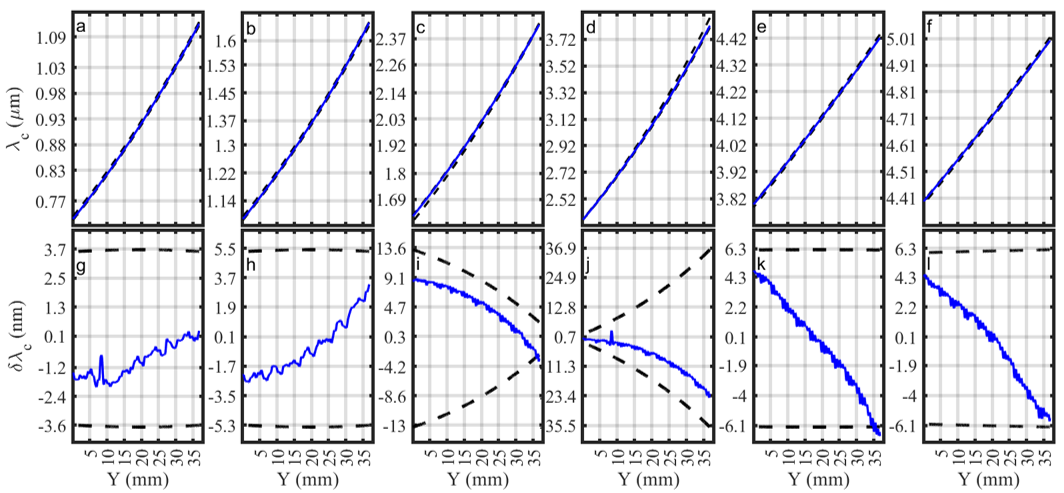}
    \caption{The one-dimensional central wavelength progression for each of the six flight LVFs measured at component-level.  \textit{Panels a-f} show the measured data (blue) and the allowed values based on tolerance specification (black dash) for bands 1 through 6 respectively. \textit{Panels g-l} show the residual of the as-built and the target design wavelength progression (blue) and the allowed tolerance ranges (black dash).}
    \label{fig:lamprogmeas}
\end{figure*}

The term \textit{linear variable filter} is adopted in this paper and throughout the SPHEREx literature, as it is an optics industry standard term.  As the name implies, for many applications, LVFs use a progression in which the central wavelength of a bandpass, $\lambda_c$, increases linearly as a function of distance along one direction on the filter. In the case of SPHEREx, the optimal wavelength progression follows the form 
\begin{equation}
\label{eqn:lamprog}
\lambda_{c,i} (\bar{Y}) = \lambda_{min}\left(\frac{\lambda_{max}}{\lambda_{min}}\right)^{\bar{Y}},
\end{equation}
where $i$ represents each column of detector pixels, and $\lambda_{\rm{min}}$ and $\lambda_{\rm{max}}$ are band-specific boundary wavelengths. $\bar{\rm{Y}}$ represents the Y coordinate along the filter that is normalized to have a value of 0 at $\lambda_{\rm{min}}$ and 1 at  $\lambda_{\rm{max}}$.  Because the LVFs were coated in a machine with a rotational symmetry, the iso-wavelength lines are segments of a circle, whose radius is many times larger than the size of the filter.  We refer to the curvature as ``smile," and quantify its size by the distance from the edge of the filter to the $\lambda_{\rm{min}}$ iso-wavelength contour. Figure~\ref{fig:lammap} gives a graphical depiction of the target design for band~1 with the definition of smile and the boundary wavelengths. The $\lambda_{\rm{min}}$ iso-wavelength contour intersects the corners of the filter, and the  apex of the $\lambda_{\rm{max}}$ contour intersects the edge of the filter at its center X value.  The exponential functional form of Equation~\ref{eqn:lamprog} is preferable because the distance spanned on the filter by one spectral resolution element $\Delta\lambda_{\rm{c}}$ is constant across the wavelength range of a band, as long as the resolving power
\begin{equation}
\label{eqn:R}
\rm{R} = \frac{\lambda_{c}}{\Delta \lambda}
\end{equation}
is held constant for all values of $\lambda_{c}$ in a filter.

Measurements of the wavelength progression were critical to collect at component level to ensure contiguous wavelength coverage across the SPHEREx band.  These were measured in VIAVI facilities using a spectrophotometer with the optic mounted on a translation stage equipped with a precise encoder.  Bandpass measurements were taken as the optic was scanned and the wavelength characteristics were registered with respect to the distance to the edge of the optic.  Figure~\ref{fig:lamprogmeas} shows the results of these measurements compared to the requirements imposed.  The tolerances were specified to encompass the uncertainty incurred in the coating performance as well as mechanical tolerances in trimming the filters.  For all bands besides 3 and 4, the tolerance was distributed evenly between $\lambda_{\rm{max}}$ and $\lambda_{\rm{min}}$.  Bands 3 and 4 were treated differently because they interact with the DBS transition from reflection to transmission. To center the transition point within the overlap region, we placed tighter tolerances on Band 3 $\lambda_{\rm{max}}$ and Band 4 $\lambda_{\rm{min}}$.  Because the total tolerance is split between mechanical cutting and coating progression, Band 3 $\lambda_{\rm{min}}$ and Band 4 $\lambda_{\rm{max}}$ received looser tolerances.
In all cases, the measured wavelength progressions met specification. The targets for $\lambda_{\rm{min}}$ and $\lambda_{\rm{max}}$, given in Table~\ref{tab:lvfscreen}, included a quarter of a channel overlap between bands.  This padding ensured continuous spectral coverage, sufficient to encompass the errors in coated progression as well as alignment between the LVF and H2RG. 

\subsubsection{Optical properties}
\label{sssec:lvfprops}

In addition to the wavelength progression, there were several critical optical transmission measurements performed on the LVF at component level.  These are described below and the specifications are summarized in Table~\ref{tab:lvfscreen}. 
\begin{itemize}
    \item {\textbf{In-band transmission minimum, $T_{min}$:}  
    This ties directly into the optical efficiency of the system, and is a strong contributor to the PSS.}
    \item{\textbf{Resolving power, $R$:} Represented by Equation~\ref{eqn:R}, 
    with $\Delta \lambda = \int_{0.5\mu \rm{m}}^{2.6 \mu \rm{m}}T_{\lambda}d\lambda$ for SWIR and $\Delta \lambda = \int_{0.5\mu \rm{m}}^{5.3 \mu \rm{m}}T_{\lambda}d\lambda$ for MWIR.  In both cases, $T_{\lambda}$ is the peak-normalized transmission as a function of wavelength. The bounds of the integrals are set by the photosensitive regions of the H2RGs.}
    \item {\textbf{Out-of-band blocking, }B\textbf{:} 
    Defined as the mean transmission across the optically sensitive range included in the bounds of the integrals above, excluding the range $-3\Delta \lambda < \lambda_{c} < 3\Delta \lambda$}.
    \item{\textbf{Detector-facing in-band reflection, $r_2$:} The in-band reflection on the surface of the LVF which faces the detectors.  Minimizing this reflection is essential to control the amplitude of ghost images caused by multiple reflections between the detector and filter.  The desire to minimize the radii of these ghosts is what drove the FPA design to place the filters close above the detectors.}
    \item{\textbf{Optics-facing in-band reflection, $r_1$:} The in-band reflection of the top surface of the LVF which faces the up-stream optics. }
\end{itemize}

The optical transmission characteristics were measured at operating temperature using witness samples coated alongside the LVFs that spanned the wavelength range of the LVFs.  In all cases, the delivered filters met these key requirements in both flight and flight spare components.  Achieved $\lambda_{\rm{max}}$ and $\lambda_{\rm{min}}$ are shown in Figure~\ref{fig:lamprogmeas}, measured transmission is reported in Section~\ref{S:opteff} and $R$ and $B$ appear in \cite{howard}.

\begin{table*}
    \centering
    \begin{tabular}{ccccccc}
        \textbf{Property} & \textbf{Band 1} & \textbf{Band 2} & \textbf{Band 3} & \textbf{Band 4} & \textbf{Band 5} & \textbf{Band 6}\\ \hline \hline
        $\lambda_{min}$     &$0.7455 $ & $1.1033 $& $1.6301 $ & $2.4027 $& $3.8113 $& $4.4115$ \\
        ($\mu$m) &$\pm 0.0026 $ & $\pm 0.0038 $& $ \pm 0.0102$ & $\pm 0.0013 $& $ \pm 0.0056 $& $\pm 0.0058$ \\ \hline
        $\lambda_{max}$     & $1.1167 $& $1.6499 $& $2.4346 $ & $3.8473 $ & $4.4300 $ & $5.0096 $\\ 
        ($\mu$m) & $\pm 0.0036 $& $\pm 0.0053 $& $\pm 0.0011 $ & $\pm 0.0268 $ & $\pm 0.0066$ & $\pm 0.0062 $\\ \hline
        $T_{min}$     & $80\,\%$ &$75\,\%$ & $75\,\%$ & $70\,\%$ & $65\,\%$ & $60\,\%$\\ \hline
        $R$     & $41.4 \pm 10\,\%$&$41.4 \pm 10\,\%$ &$41.4 \pm 10\,\%$ & $35 \pm 10\,\%$& $110 \pm 10\,\%$& $130 \pm 7\,\%$\\ \hline
        B     &OD\,$>4$ & OD\,$>4$&OD$>4$ &OD\,$>4$ & OD\,$>4$& OD\,$>4$ \\ \hline
        $r_{2}$    & $<2\,\%$ & $<2\,\%$ & $<2\,\%$ & $<5\,\%$ & $<5\,\%$ &  $<5\,\%$ \\ \hline
        $r_{1}$    & $<10\,\%$&$<10\,\%$ & $<10\,\%$&$<10\,\%$ &$<10\,\%$ & $<10\,\%$ \\ \hline
    \end{tabular}
    \caption{Component-level optical performance requirements for the LVFs.  In each case, both a flight unit and a flight spare were delivered in compliance.}
    \label{tab:lvfscreen}
\end{table*}

\subsection{Detector Arrays}
\label{ssec:h2rg}

\begin{table*}
    \centering
    \begin{tabular}{cccc}
        \textbf{Property} & \textbf{Requirement} & \textbf{Science Driver} & \textbf{Comments}\\ \hline \hline
        $\lambda_{cut}$ & SWIR (2.6 $\pm ~0.1)~\mu\rm{m}$ & SWIR: DBS Coverage & \\ 
                          & MWIR (5.3 $\pm ~0.1)~\mu\rm{m}$  & MWIR: LVF Coverage & \\ \hline
        $\eta_{Det}$ & $>70\,\%,~0.8~\mu \rm{m} < \lambda < 2~\mu \rm{m}$ SWIR & Sensitivity & Measured at 4 discrete $\lambda$\\ 
        $\eta_{Det}$ & $>70\,\%,~0.8~\mu \rm{m} < \lambda < 4.4~\mu \rm{m}$ MWIR & Sensitivity & Measured at 5 discrete $\lambda$\\ \hline
        $\delta Q_{CDS}$ & $\le 18~e^{-}$ SWIR & Sensitivity & \\ 
                         & $\le 15~e^{-}$ MWIR & Sensitivity & \\ \hline
        $i_{DC}$ & $\le 0.05~e^{-}/\rm{s}$  & EBL Fluctuation Power & \\ \hline
        Operability & $> 95\,\%$  & Voxel Completeness &  Measured at 4 discrete $\lambda$\\ \hline
        Flatness of HgCdTe surface & $< 25~\mu \rm{m}$~P-V  & PSF size & \\ \hline

    \end{tabular}
    \caption{Performance parameters that had required acceptance levels in detector component-level screening. In all cases, the 12 flight candidate units delivered complied with these requirements.  $\delta Q_{CDS}$ is the correlated double sample noise.}
    \label{tab:detscreen}
\end{table*}

The testing of flight candidate arrays at Teledyne was consistent with the standard acceptance screening program developed for the H2RG product line.  The set of performance parameters on which specific acceptance criteria were agreed upon for delivery are summarized below and given in Table~\ref{tab:detscreen} along with their requirements.

\begin{itemize}
    \item \textbf{Cutoff wavelength, $\lambda_{cut}$:} The wavelength at which the detector stops being photosensitive on the long-wavelength side.  Compliance here is required to control the out-of-band blocking in tandem with the LVF requirements while maintaining good efficiency in-band. 
     \item \textbf{Quantum efficiency, $\eta_{Det}$ :} The percent of photons converted to electrons in-band.  This is measured at 4 discrete wavelengths in SWIR and 5 in MWIR.  Feeds directly into instrument throughput.
      \item \textbf{Read noise, $\delta Q_{CDS}$ :} The correlated double sample noise of each ROIC+H2RG hybrid.
     \item \textbf{Median dark current, $i_{DC}$ :} The offset in slope measured in the absence of photons. Structure in dark current can provide spurious EBL signal if not removed accurately.  We seek to minimize this.
    \item \textbf{Operability :} The percentage of pixels in the array which meet all other criteria.  Missing pixels lead to holes in the survey completeness, which we seek to minimize.
    \item \textbf{H2RG flatness :} The peak to valley amplitude of deviations from planarity in the HgCdTe substrate.  High deviations limit the ability to accurately focus the instrument.  
    
\end{itemize}

Six flight-candidate units of both SWIR and MWIR arrays were delivered to the project on schedule and all units complied with the requirements in their entirety.

\subsection{Dichroic Beamsplitter}
\label{ssec:dbs}

The DBS enables the two FPAs to sample the sky simultaneously for increased mapping speed.  The transition wavelength between reflection and transmission was selected to be 2.42~$\mu$m.  This wavelength is a sensible location for the division between bands 3 and 4, as it aligns well with the $2.6~\mu m$ cutoff wavelength for the SWIR detectors.  The DBS substrate is made of silicon, which absorbs NIR short of $1.1 \mu m$ (\cite{Green1995}) but transmits well out to $5 \mu m$  (\cite{Palik1985}).  For this reason, the SWIR views the DBS in reflection.
To preserve the wavefront error performance in the MWIR bands, the back surface of the DBS substrate is powered, for refractive corrections.  The DBS is mounted with a tip angle of 22.5$^\circ$ from the chief ray and the incident light cone is f/3.  The back surface of the DBS has an anti-reflection (AR) coating that is designed to minimize the amplitude of a ghost image, present for all bright point sources viewed in DBS transmission as well as Band 3 in reflection.    It results from secondary bounces between the two surfaces of the optic.  For Bands 1 and 2, the DBS ghost is severely attenuated, because the ray path includes travel through the silicon substrate, which absorbs at short wavelengths. Because the DBS receives a converging beam and the optic is tipped, the ghost image is out-of-focus and offset from the primary beam. 

The DBS substrate was procured from Coherent systems, and was coated by VIAVI.  The component-level performance measurements were carried out using spectrophotometry and interferometry at VIAVI.
This campaign was designed to verify the key performance criteria for operation: that the total reflection and transmission in-band have acceptable levels of efficiency loss, and that the WFE be suitable to preserve optical performance. Due to geometrical restrictions in the testing environment, reflection and transmission efficiencies were measured at ambient temperatures.  To obtain the expected performance at operating temperature, the ambient measurements were scaled analytically by a spectral shift factor of 0.992.  This factor encompasses the temperature dependence of the refractive indices of the substrate and coatings.  It was obtained by monitoring the shift of the half power reflection wavelength as a function of temperature on a small witness sample, coated alongside the flight unit. The key performance parameters measured for the DBS are summarized below, and given with their measured values in in Table~\ref{tab:dbs}. 

\begin{itemize}
    \item \textbf{$\Delta WFE$:}  The change in wavefront error incurred on a reflection off the DBS surface,
    \item \textbf{$r_{D1avg} $:}  The average reflection efficiency for bands 1-3 from the first surface.
    \item \textbf{$r_{D1min} $:}  The minimum reflection efficiency at a single wavelength for bands 1-3.
    \item \textbf{$<T_{D1}T_{D2}> $:}  The average two surface transmission efficiency for bands 4-6.
    \item \textbf{$T_{D1min}T_{D2min} $:}  The minimum transmission efficiency at a single wavelength in-band.
    \item \textbf{$<r_{D1}r_{D2}>$:}  The average product of the reflection of both surfaces of the DBS in band for bands 4-6.  This quantifies the amplitude of the ghost image seen in transmission.

\end{itemize}

\begin{table*}
    \centering
    \begin{tabular}{cccc}
        \textbf{Property} & \textbf{Requirement} & \textbf{Achieved} & \textbf{Science Driver}\\ \hline \hline
        $\Delta WFE$ & $< 27$~nm RMS & 10~nm RMS & PSF Performance \\ \hline
        $r_{D1avg} ~(0.74~\mu \rm{m} <\lambda < 2.36~\mu \rm{m})$ & $>97\,\%$ &  $98.3\,\%$ & $\eta_{DBS}$ \\ \hline
        $r_{D1min} ~(0.74~\mu \rm{m} <\lambda < 2.36~\mu \rm{m})$& $>90\,\%$ &  $94.4\,\%$ & $\eta_{DBS}$ \\ \hline
        $<T_{D1}T_{D2}> ~(2.46~\mu \rm{m} <\lambda < 5.00~\mu \rm{m})$ & $>96.5\,\%$ &  $97.9\,\%$ & $\eta_{DBS}$ \\ \hline
        $T_{D1min}T_{D2min} ~(2.46~\mu \rm{m} <\lambda < 5.00~\mu \rm{m})$ & $>87.0\,\%$ &  $92.6\,\%$ & $\eta_{DBS}$ \\ \hline
        $<r_{D1}r_{D2}> ~(2.46~\mu \rm{m} <\lambda < 5.00~\mu \rm{m})$ & $<0.015\,\%$ & $0.0062\,\%$ & Ghost Amplitude \\ \hline

    \end{tabular}
    \caption{Performance parameters that had required acceptance levels in DBS component-level screening and their measured values.  All parameters met their specifications.}
    \label{tab:dbs}
\end{table*}

\section{Optical Throughput}
\label{S:opteff}

\begin{figure}
    \centering
    \includegraphics[width=1.00\linewidth]{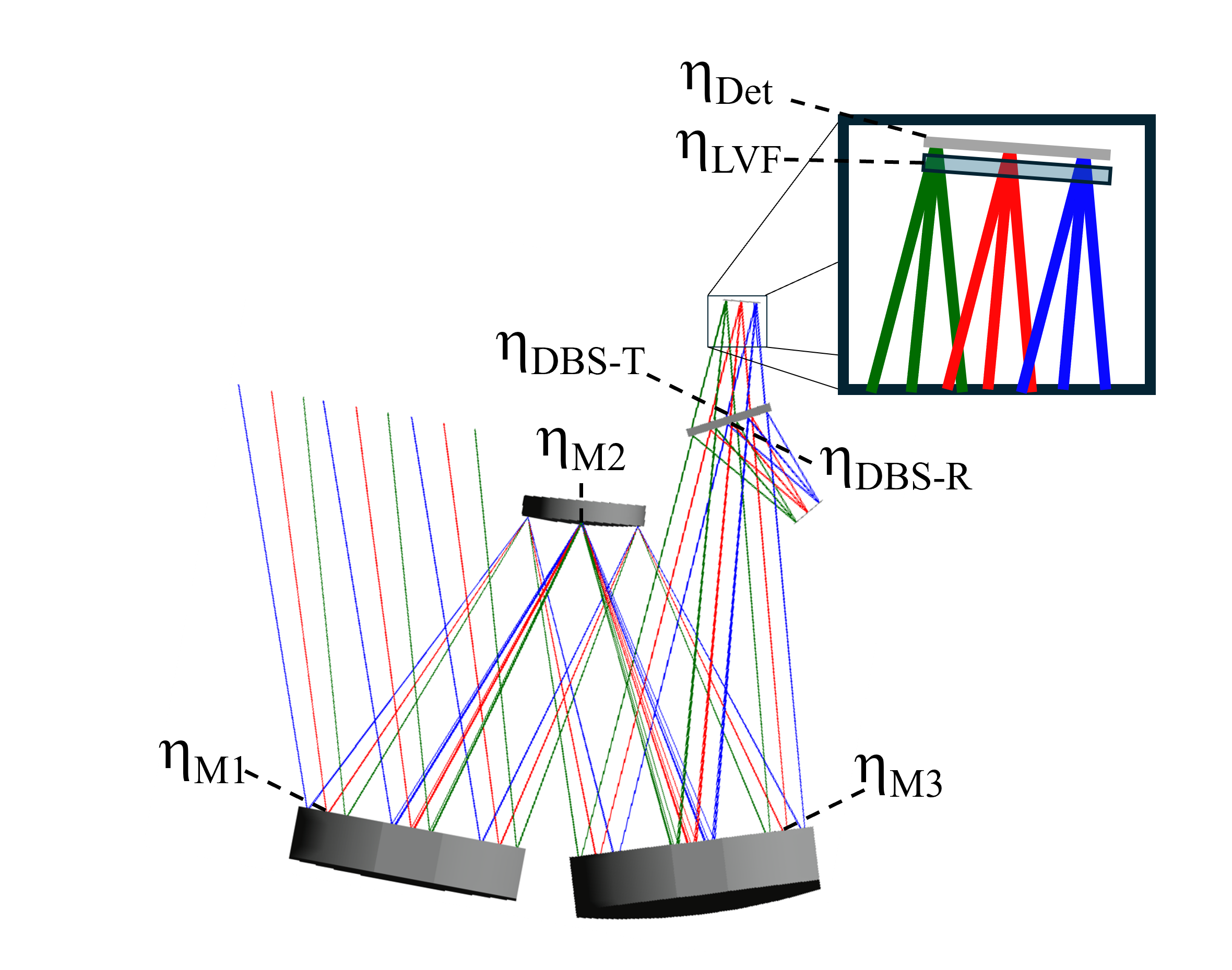}
    \caption{Ray trace through the reflective triplet optics.  The efficiency of each surface that encounters the direct light path is labeled alongside that optic. }
    \label{fig:aeta}
\end{figure}

The efficiency $\eta$, in reflection, transmission or absorption of each surface in the optics contributes directly to the sensitivity and was therefore tracked with requirements at each stage.  During the component-level testing, measurements of each contributor were quantified as a function of wavelength, and we collate their performance here.  The optical ray trace with labeled surface efficiencies is shown in Figure~\ref{fig:aeta}, and we describe the terms below.

    \begin{itemize}
        \item \textbf{$\eta_{M1,M2,M3}$:} The reflection efficiency of the three mirrors.  Measured by BAE Systems at component level after coating.
        \item $\eta_{DBS-R}$ and $\eta_{DBS-T}$ : The efficiency of the DBS.  This is a reflection at wavelengths shorter than 2.42~$\mu$m and a transmission at wavelengths longer than 2.42~$\mu$m.  The spectrophotometry measurements were carried out by VIAVI, as described in section~\ref{ssec:dbs}.
        \item \textbf{$\eta_{LVF}$:} The transmission efficiency of the LVFs across all six bands, measured by VIAVI and described in Section~\ref{ssec:lvf}.
        \item \textbf{$\eta_{Det}$:} The quantum efficiency of the detector system.  Its ability to convert each photon incident on the active surface into an electron measured by the signal chain.  This is measured at discrete wavelengths by Teledyne and described in Section~\ref{ssec:h2rg}.
    \end{itemize}

\begin{figure}
    \centering
    \includegraphics[width=1.00\linewidth]{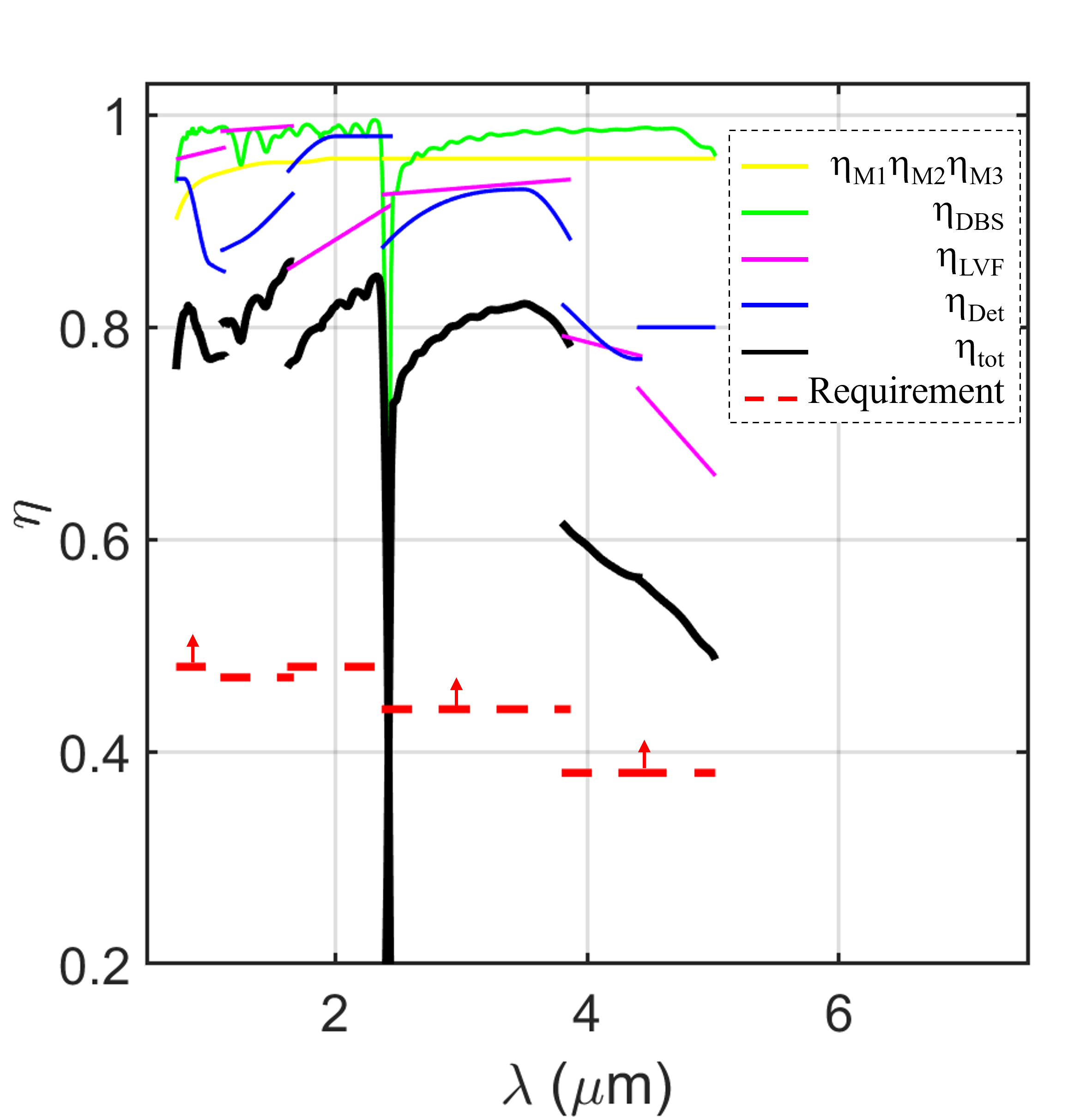}
    \caption{End-to-end efficiency breakdown characterizing each element which interacts with the main beam path corresponding to the surfaces outlined in Figure~\ref{fig:aeta}. $\eta_{DBS}$ as shown here corresponds to a transmission at wavelengths longer than $2.42~\mu$m and a reflection shorter than $2.42~\mu$m.}
    \label{fig:aetameas}
\end{figure}

Figure~\ref{fig:aetameas} compiles the measurements of $\eta$ and plots their cumulative product $\eta_{tot}$ against the requirement levels set by the efficiency allocation to meet top-level mission science requirements.  The requirement is met with substantial margin at all wavelengths. 

\section{Subsystem Level Testing}
\label{sec:subsystem}

\subsection{FPA testing}
\label{ssec:fpas}

The two FPA subsystems house the detectors and LVFs at the focus of the telescope.  To methodically address the thermal, electrical and signal processing concerns, we conducted a tiered testing and assembly campaign consisting of multiple prototypes and models.  These are summarized in Table~\ref{tab:fpaconfig}. A mechanical diagram of the FPA constituents is available in \cite{jamiemission}.  In this section, we walk through the various considerations of FPA quality evaluation, and detail how the measurements were made.

\begin{table*}
    \centering
    \begin{tabular}{cccc}
        \textbf{FPA Unit} & \textbf{H2RGs} & \textbf{LVF} & \textbf{Functionality}\\ \hline \hline
        Ambient Surrogate & - & - & Thermal\\ \hline
        Cold Surrogate & - & - & Thermal\\ \hline
        Brassboard Model& 1 SWIR EM & Band 1 Prototype &  Optical\\ 
                   &           &                  &  Electrical \\ \hline
        Engineering Model & 1 SWIR EM & 1 Band 3 EM & Optical\\
         & 2 Bare ROIC & 2 Uncoated Substrate & Mechanical\\
                            &           &                  &  Electrical \\ \hline
        Flight SWIR & 3 SWIR FM & Bands 1-3 FM & Optical \\
             &           &                  &  Electrical \\ 
             &           &                  &  Thermal \\ 
             &           &                  &  Mechanical \\ \hline
        Flight MWIR & 3 MWIR FM & Bands 4-6 FM & Optical \\
             &           &                  &  Electrical \\ 
             &           &                  &  Thermal \\ 
             &           &                  &  Mechanical \\ \hline
    
    \end{tabular}
    \caption{FPA configurations with their contents and functionalities detailed.}
    \label{tab:fpaconfig}
\end{table*}

\subsubsection{Noise and readout optimization}
\label{sssec:noise}
 
\begin{figure}
    \centering
    \includegraphics[width=1.00\linewidth]{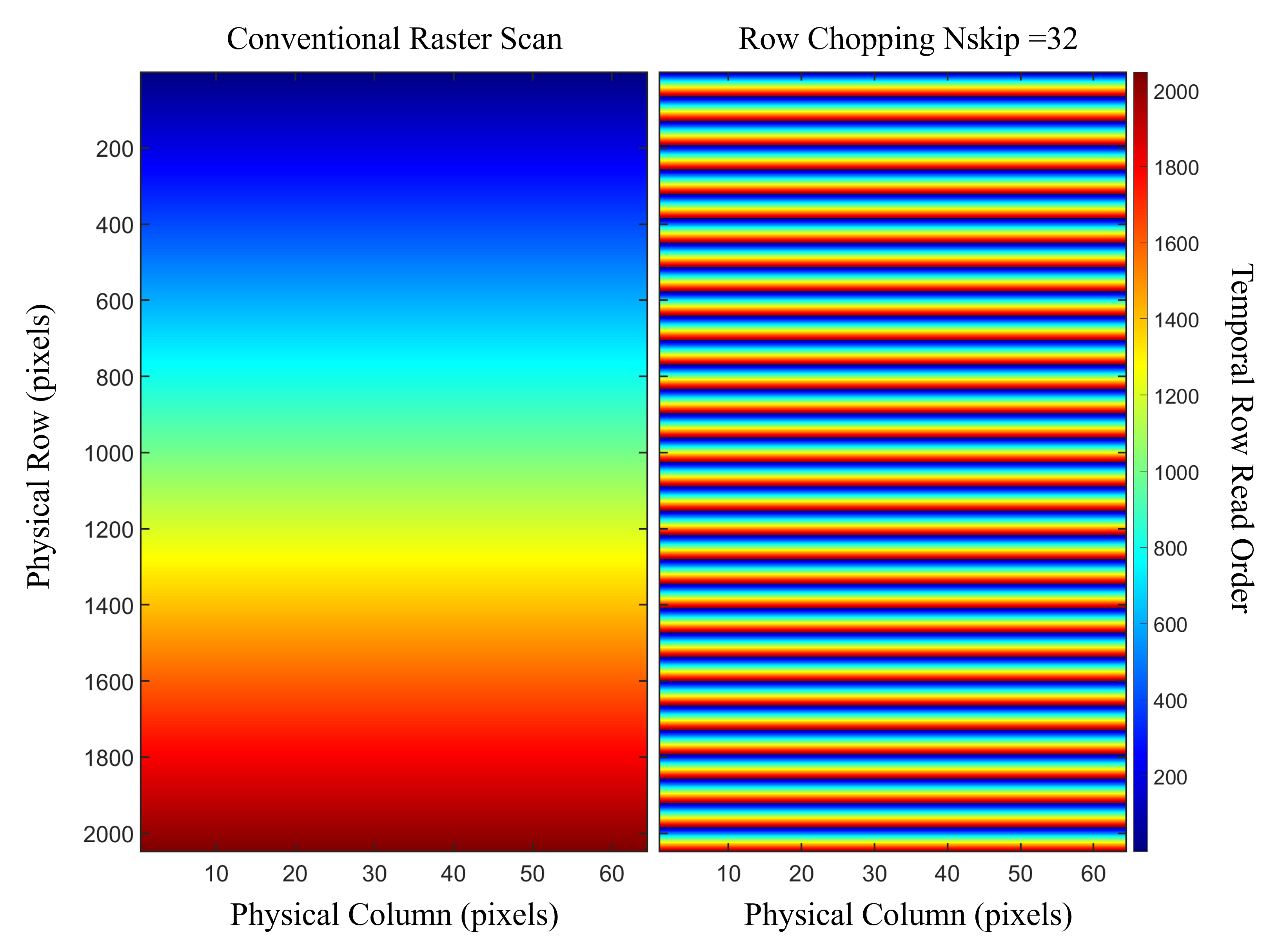}
    \caption{Illustration of the row chopping readout technique designed to alias power from 1/f noise to higher spatial frequencies away from the EBL signal band.}
    \label{fig:rowchop}
\end{figure}

The noise measurement program focused heavily on mitigation of 1/f fluctuations generated in the amplification chain (\cite{h2rgnoise} , \cite{Rauscher2015}).  
The H2RGs are read out in 32 electrical channels that are 64 pixels wide each. Traditional multiplexing schemes for H2RGs read out pixels in a raster pattern that marches across a row consecutively. After the last pixel of a row is read, the first pixel of the next row is read and the pattern continues.  In this scheme, drifts in the amplifier offsets common to all pixels in a 64$\times$2048 channel that are slow compared to the timescale for a single row read will generate correlated noise patterns on the array.  For measurements of the EBL in the case of SPHEREx, we must minimize noise on angular scales between 5\arcmin~and 20\arcmin~where the clustering of galaxies is most readily measured.  To avoid placing the 1/f noise from amplifier drift onto these angular scales, we use a unique row chopping scheme of readout, described in detail in \cite{grig}.  With this method, after a row is read, we skip 32 rows away before the next read and return to fill in the gaps after reaching the bottom of the array.  While the RMS of a resulting noise image is unchanged between conventional and row chopping readouts, the angular scales on which the noise resides in is aliased to higher spatial frequencies, and away from the signal band.  The relationship between physical pixel location and temporal read order in the conventional scheme compared to row chopping is shown in Figure~\ref{fig:rowchop}. 

The laboratory read noise measurement campaign was carried out in several stages.  First, a brassboard prototype FPA containing a single SWIR H2RG was operated in a small test cryostat.  To isolate measurements of the read noise, we reduced photocurrent beneath the dark current level by housing the FPA in a cold, light-tight enclosure.  The brassboard FPA was read out in a single ROB configuration.  This mode archives the raw detector samples directly to a laboratory hard drive, bypassing on-board processing algorithms. This configuration proved ideal to optimize the readout cadence.  To further reduce 1/f noise, the non-photosensitive reference rows of the H2RG, which share an amplifier with the nominal pixels, are read out with additional visits, interspersed in the row chopping scheme.  Additionally, we used ``phantom pixels," which are reads of the Video 8 pre-amplifier drift without input from the H2RG, also interspersed in the multiplexing readout.  To further reduce per-channel noise, we also read the reference pixels at the top and bottom of the arrays redundantly. The resulting scheme reduced the 1/f noise contribution to the EBL signal band by $50\,\%$ (\cite{chinoise}).

\begin{figure}
    \centering
    \includegraphics[width=1.00\linewidth]{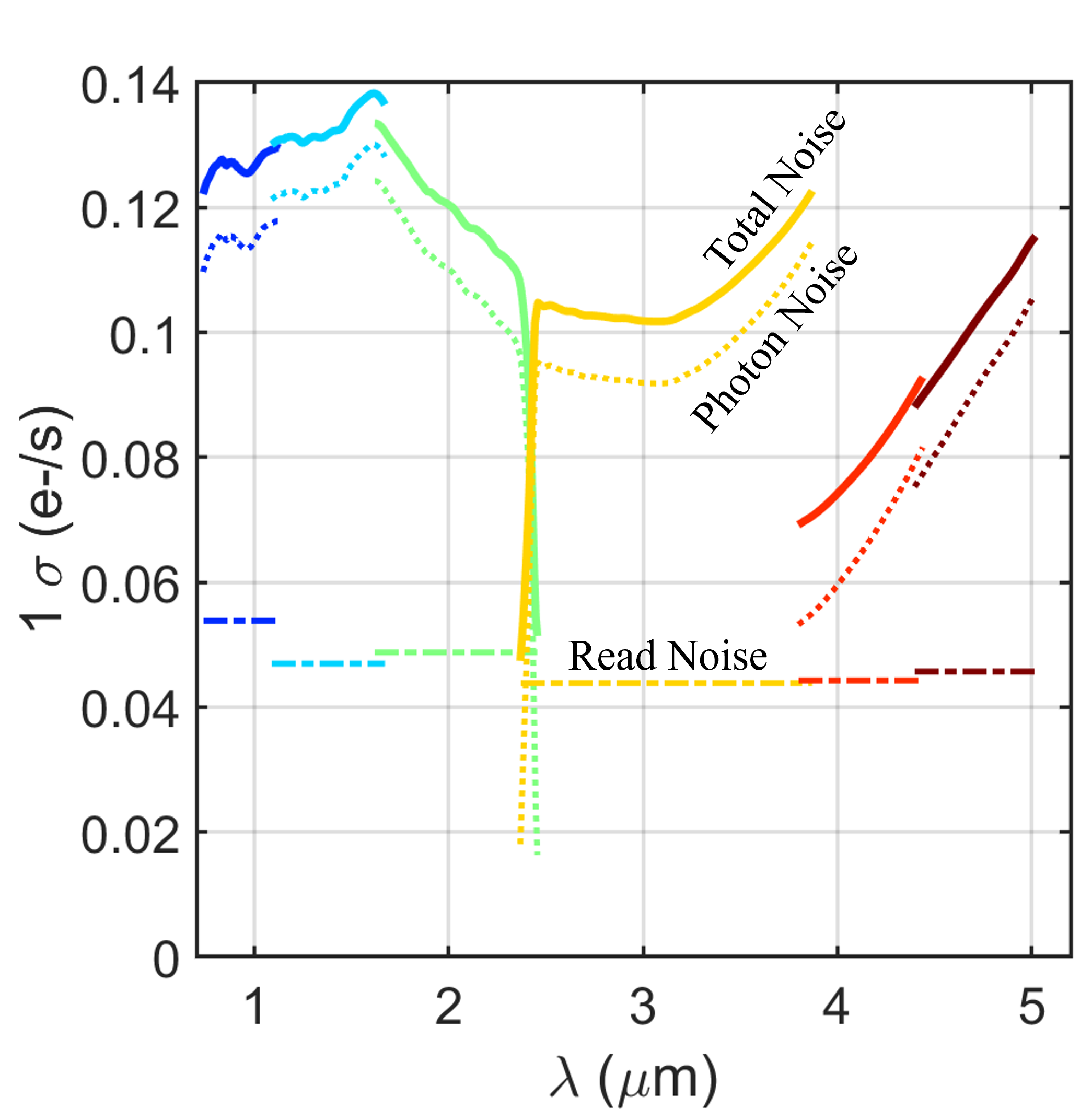}
    \caption{Breakdown of the white noise budget for a single 118~s exposure as a function of wavelength including read noise from the detector chain and the photon noise from ZL.  Each of the six bands is assigned a color.}
    \label{fig:noisebreak}
\end{figure}

The flight SWIR and MWIR FPAs, once fully assembled, were also subjected to dark measurements in the test cryostat.  This allowed for a verification of the noise performance with full SPHEREx readout in the flight-like cadence, to produce a suite of laboratory dark noise exposures.  This clean dark exposure library is invaluable for validating flight noise models during operation, as well as obtaining a statistically significant confirmation of the noise performance at component level.    Combining this dataset with the optical efficiency measurements described in Section~\ref{S:opteff}, we can produce a noise budget across the wavelength range for a single SPHEREx exposure of 118~s.  To estimate the photon noise, we assume a mean ZL intensity using the model generated for a data simulator, as described in \cite{brendansim} and the photon noise calculation using SUR of \cite{garnett}.   The resulting pixel standard deviation $\sigma$, as a function of wavelength is given in Figure~\ref{fig:noisebreak}.  

In addition to noise measurements, FPA-level measurements in the test cryostats enabled quantification of other detector characteristics, such as non-linearity, cross-talk, dark current and image persistence (\cite{candicepersistence}).

\subsubsection{Active thermal stabilization}
\label{sssec:therm}

\begin{table}
    \centering
    \begin{tabular}{ccc}
        \textbf{Timescale} & \textbf{Drift Req} & \textbf{Timescale}\\ \hline \hline
        150~s &  $m < 0.67~\mu \rm{K}\,\rm{s}^{-1}$ & exposure \\ \hline
        8~minutes &  $< 8$~mK &  30$^\circ$ orbital \\ \hline
        6~months &  $< 100$~mK &  survey \\ \hline
        1~year &  $< 200$~mK &  double survey \\ \hline
    \end{tabular} 
    \caption{Timescales for thermal drift stabilization and their respective control requirements. 
    All requirements other than slope $m$ are peak-to-peak.}
    \label{tab:tempreqs}
\end{table}

The FPAs contain an active thermal stabilization mechanism.  This design feature was included to address concerns that thermal drifts could impact two key areas of science. Firstly, photometric gain drifts in time, when modulated by SPHEREx's survey pattern, could produce spurious clustering signal on large scales ($>30^{\circ}$) that generate a systematic error in $f_{NL}$ measurements.  Secondly, the offsets of H2RG channel outputs have a temperature dependence.  Measurements using the brassboard FPA quantified these at the mean level of $\sim150~e^- \rm{K}^{-1}$, where pixel-to-pixel and H2RG output channel variations in this number make a distribution that can place power on the scales of interest for EBL.  To control these systematics, we placed the requirements given in Table~\ref{tab:tempreqs} on an active stabilization system. 

\begin{figure}
    \centering
    \includegraphics[width=1.00\linewidth]{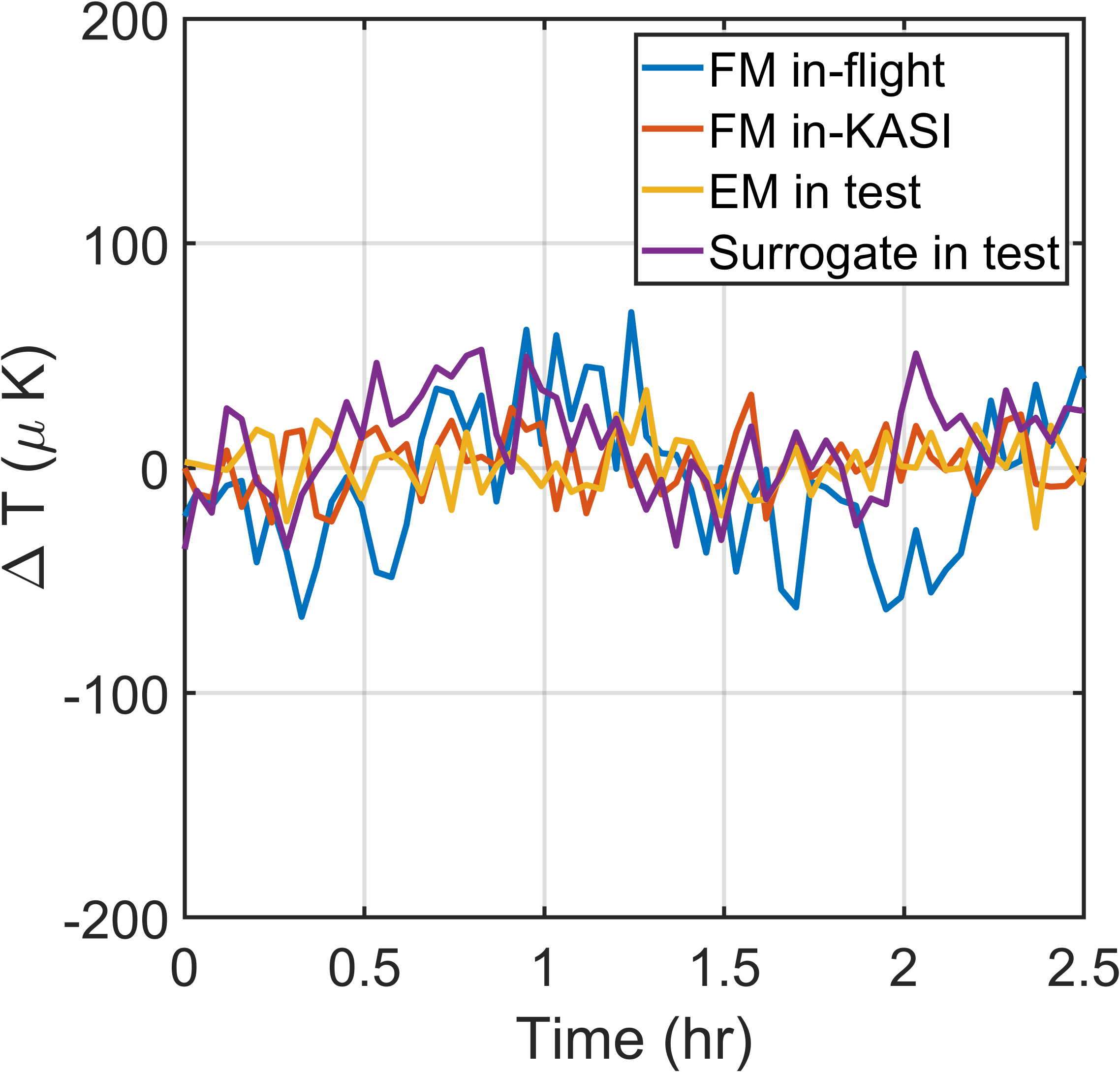}
    \caption{Temperature sampled on flight exposure cadence of the MWIR FPA monitor point collected in multiple configurations during the test campaign. In all cases, the stability is better than required, as specified in Table~\ref{tab:tempreqs}.}
    \label{fig:tempctrl}
\end{figure}

The stabilization system utilizes a proportional, integral and derivative (PID) algorithm controlled by the CEB.  Temperatures are measured by Lakeshore Cernox CX-1080 sensors.  The heater is a Vishay VPR-220Z precision foil resistor. The control point resides on the detector side of a radiator-interface strap to directly regulate the detector and reject incoming thermal disturbances across the strap. Development and verification of the control system followed a tiered path.  First, a room-temperature thermal surrogate equipped with a thermo-electric cooler (TEC) cold sink was used for preliminary controller tuning.  At the next phase of fidelity, a cryogenic surrogate with heat capacity and conductance analogs that replicate the flight materials and couplings to space was set up in the test cryostat.  With the control settings optimized and circuitry demonstrated using these prototype tests, system-level checks were performed at later stages of integration.  These included using an EM FPA in the test cryostat, as well as the full flight system at instrument-level testing during spectral calibration (Section~\ref{ssec:sppeccal}).  
Figure~\ref{fig:tempctrl} shows the variation in temperature on exposure timescales, comparing ground measurements to data collected after launch in flight.  In all cases, the thermal stability of the FPA met its requirements with several orders of magnitude in margin.  For practical considerations, the 6 month and 1 year stability could not be assessed in the laboratory and flight compliance on the survey timescales are discussed in Section~\ref{sec:conc}.

\subsubsection{Mechanical Alignment}
\label{sssec:fpamech}
Accurately setting the alignment of the detectors, filters and coupling hardware within the FPAs was essential to realizing two goals:
\begin{enumerate}
    \item To avoid PSF degradation across the FOV and maximize our PSS.  This in turn drove the necessity to place the active surface of the H2RGs at the location that the telescope comes to its final focus.  This needs to be accomplished within a tolerance of $1\sigma = \pm 50~\mu \rm{m}$ at each location across the FOV. This range is set by the amount of defocus needed to maintain the core PSF within a single 18~$\mu \rm{m}$ pixel at the shortest wavelength. 
     \item To minimize the gap between the detector's active surface and the bottom face of the LVF.  As both surfaces have small but finite reflectances, rays not initially absorbed will reverberate between detector and filter to produce power at large radii in the extended PSF.  This drove us to require the LVF to H2RG gap to be between 50~$\mu \rm{m}$ and 120~$\mu \rm{m}$ where the lower limit was chosen to avoid physical damage during launch vibrations.  
\end{enumerate}

\begin{figure}
    \centering
    \includegraphics[width=1.00\linewidth]{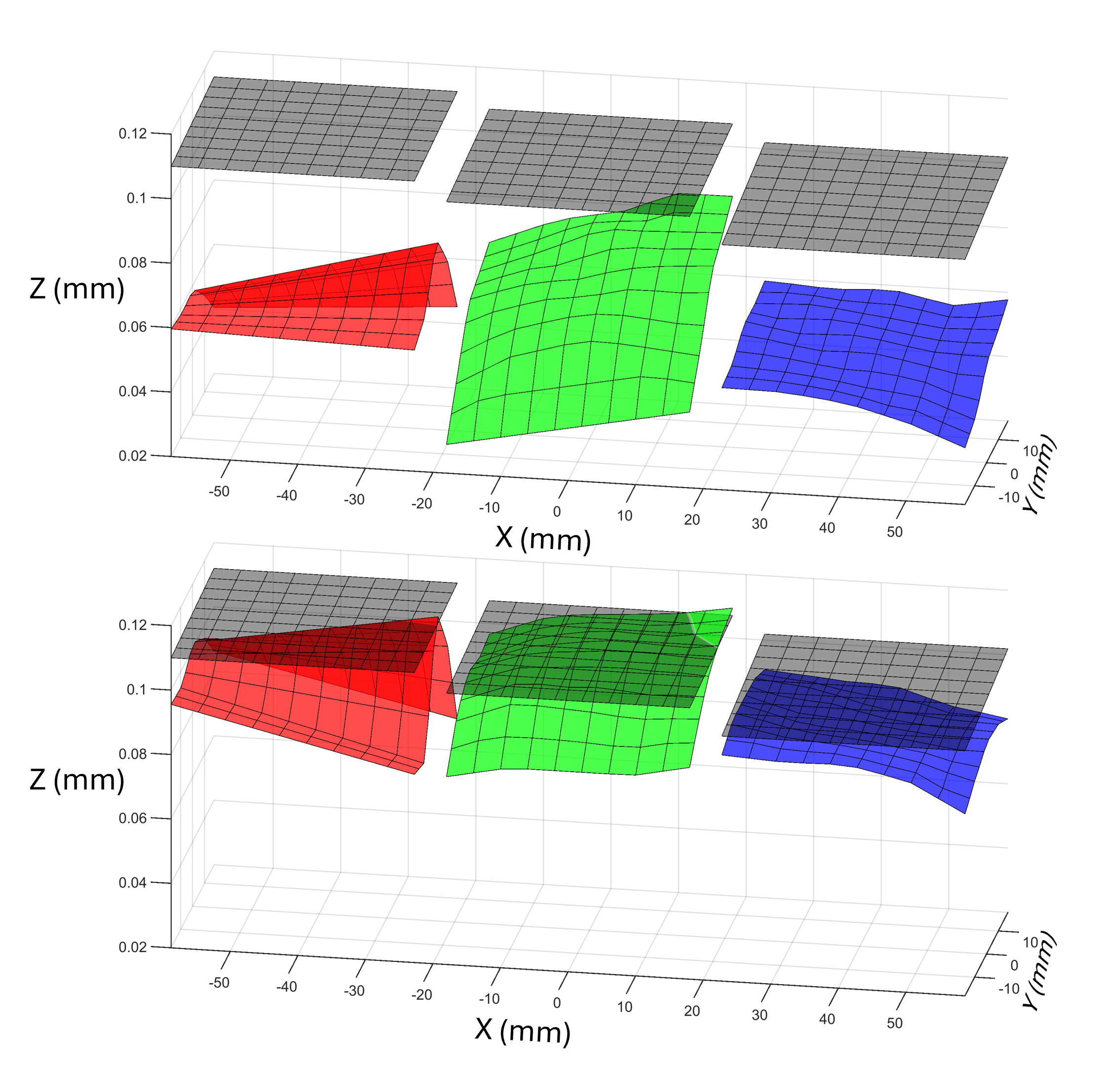}
    \caption{Iterative metrology data of the top surface of the detector from the assembly process of the the MWIR flight FPA.  The grey surfaces represent the target focal surfaces.  The top image was taken after initial assembly and the bottom is the final configuration after applying adjusting shims.  \textit{Blue:} Band 4, \textit{Green:} Band 5 and  \textit{Red:} Band 6. Note that while the $x$ and $y$ axes are matched in scale, the $z$ axis is stretched by a factor of 500.}
    \label{fig:fpashim}
\end{figure}

The FPAs, like the rest of the telescope, were not built with any compensation mechanisms.  Instead, each opto-mechanical interface was kinematically mounted at three locations.  Adjustments to tip, tilt and piston were made with precision shim adjustments at key locations within the FPAs:

\begin{enumerate}
    \item The H2RG arrays are 4-side-buttable packages equipped with three legs that mate with a mosaic plate on the FPA.  Choice of shim beneath each leg allowed us to move the active HgCdTe surface to the desired plane with respect to a datum on the mosaic plate.  To first order, the active surfaces of the three H2RGs need to be coplanar within the required focus tolerance.  However, this approximation is insufficient in the case of SPHEREx since the sapphire substrate and coatings of each LVF have wavelength-dependent indices of refraction.  Because the LVFs are located in a converging beam, a  band-specific correction was accounted for in setting the distance between active surface and mosaic plate.
    \item The LVFs were edge-bonded into a frame with a coplanar surface.  There are no adjustments for individual LVFs relative to this frame. The frame to mosaic plate alignment was globally adjusted to set the displacement between all three LVF- H2RG pairs.
    \item The entire FPA was adjusted relative to the interface to the telescope.  This global focus adjustment was only modified after measurements at instrument-level at operating temperature as discussed in Section~\ref{ssec:focus}.
\end{enumerate}

We used optical metrology in the various stages of alignment to determine key positions of surfaces.  To set the relative heights of the H2RGs, we first installed them into the mosaic plate using uniform thicknesses of shim.  We then placed the assembled mosaic under a precision microscope, and measured the distance between the active surface and a fiducial on the mosaic plate.  A best fit plane per-detector was used to determine the shim sizes required to bring the detector surfaces to the target focal location.  The mosaic was then disassembled, and shims were lapped to the desired thicknesses and reinstalled.  Once the mosaic was reassembled with the new shims in place, we repeated the metrology.  The MWIR mosaic measurements before and after shim adjustment are shown in Figure~\ref{fig:fpashim} compared to the target surfaces.  A single iteration brought the focal surfaces within the desired tolerances across the FOV.  The differences in band-specific target surfaces evident in Figure~\ref{fig:fpashim} illustrate the amplitude of the refractive corrections as a function of wavelength.

\begin{figure}
    \centering
    \includegraphics[width=1.00\linewidth]{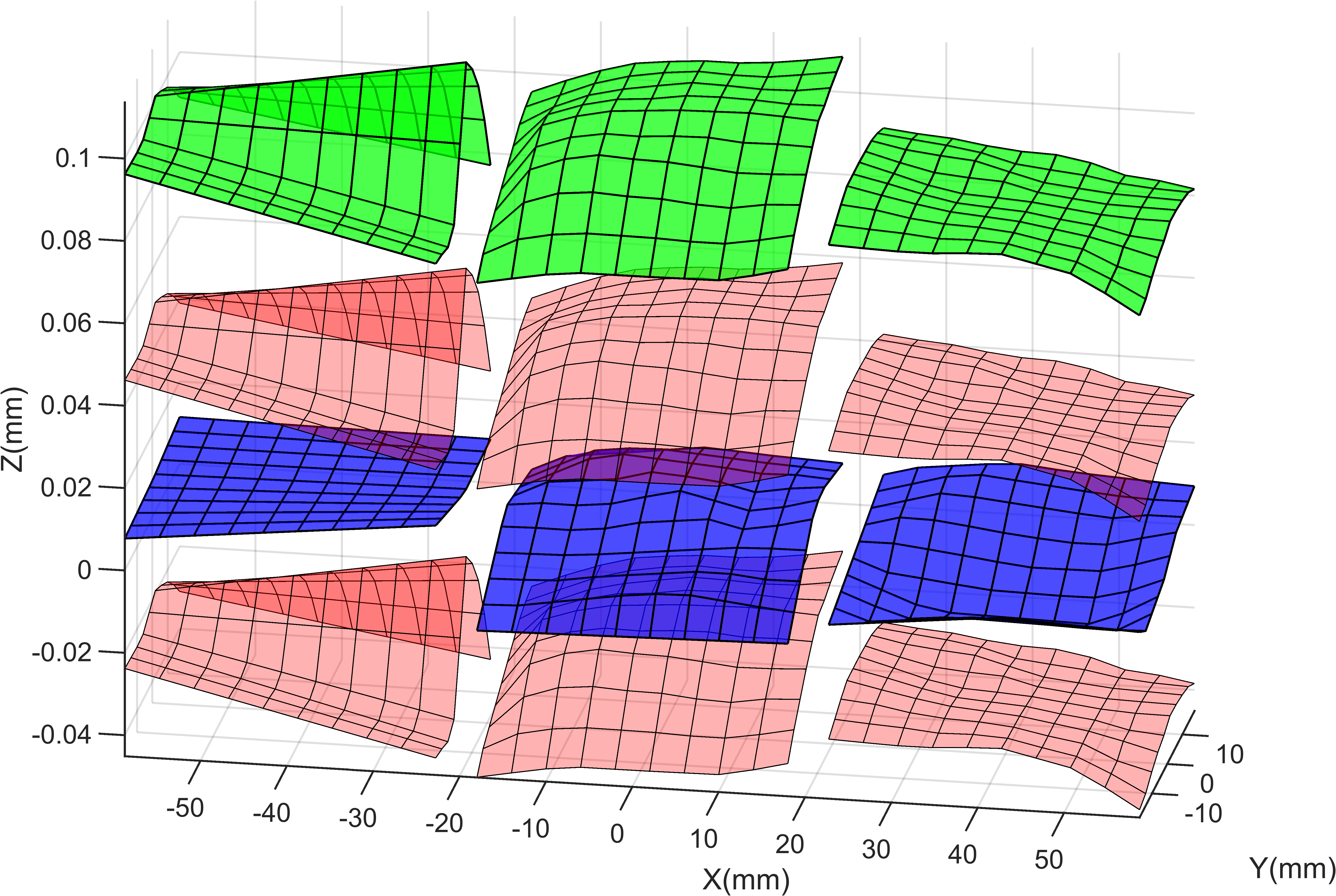}
    \caption{Metrology measurements used to set the gap between the H2RG surfaces (green) and the closest LVF surface (blue).  The red surfaces are shifted copies of the green surface, indicating the required range of gap between $50~\mu$m and $120~\mu$m.  The MWIR FPA is shown here, and the plot positions of bands 4, 5 and 6 are identical to the coordinates given in Figure~\ref{fig:fpashim}. In all cases, the blue surfaces lie within the red boundaries.}
    \label{fig:lvfgap}
\end{figure}

To set the gap optimally between the LVFs and H2RGs, we followed a similar metrology procedure.  After the LVFs were bonded into the frame, the height of the surface of the detector-facing side of the LVFs was measured relative to a fiducial on the LVF frame.  Combining the measurements of the LVF and H2RG allowed for the calculation of optimal shims to set the gap within tolerance.  The combined metrology results are shown in Figure~\ref{fig:lvfgap} compared to the requirements, which are met at all positions.

The metrology and shimming procedures described here were conducted at ambient temperature while the ultimate alignment requirement compliance is needed at the operating temperature of 62~K and 45~K for SWIR and MWIR respectively. The path length between the mosaic plate and LVF holder is spanned by Molybdenum TZM, which experiences a contraction in length, $\frac{\Delta L}{L}=0.09\%$ from 293~K to 45~K.  The same is true for most of the path length of the H2RG surface to the mosaic plate, with the exception of the Copper-Tungsten coplanarity shims. For a worst case $120 \mu$m gap, this contraction would reduce the gap by 0.1$\mu$m, negligible compared to the requirement range.  The same is true for the target H2RG positions within an FPA.  The maximum intra-FPA height difference was between band 1 and band 3, which are separated by $33 \mu$m. The Copper-tungsten contracts by $\frac{\Delta L}{L}=0.15\%$ from 293~K to 62~K, making the differential height impact only 0.05$\mu$m.
The FPA structures were designed to be athermal, maintaining the position of the detectors with respect to the telescope focus position constant with temperature.  This was accomplished using a re-entrant design, combined with bi-metallic scissor jacks (see Figure~13 in \cite{jamiemission}).  The target focal surfaces relative to the FPA/telescope interface were determined empirically at operating temperature from telescope-level cryogenic testing (Section~\ref{ssec:tel}) with calculated corrections for the sapphire refractive index (\cite{sapphire}) and a focus shift due to the coatings, calculated by VIAVI.  The only shims which required CTE corrections for initial sizing were the Titanium shims that couple the FPA to the telescope, which are of order 10~mm, and therefore contract by $15 \mu$m on cooldown.  These shims were the only ones adjusted at instrument-level after cryogenic-optical testing (Section~\ref{ssec:focus}).

\subsection{Telescope}
\label{ssec:tel}
Before integrating with the rest of the instrument system, the telescope underwent a series of alignment and optical performance measurements at BAE Systems in Boulder Colorado (\cite{frater23}). To perform end-to-end measurements of the telescope's optical performance, and to properly align the mirrors, we undertook a 633~nm interferometry campaign.  To accomplish this task, we utilized a surrogate SWIR FPA featuring spherical retro-reflectors placed at nine locations across the FOV. We introduced collimated coherent light into the aperture, and bounced the beam back through the telescope to the interferometer.  We conducted the 633~nm measurements at both ambient and cryogenic operating temperatures. The same measurement was made in DBS transmission as well using a 3.39~$\mu m$ interferometer to measure focal position in MWIR, but the longer wavelength was only done at ambient temperature. The combined measurements quantified deformations and focal shifts induced by thermal stresses.  The system wavefront error (WFE) complied with our imaging requirements of 350~nm rms across the FOV, with the exception of a small corner region corresponding to the short wavelength end of band 3. The resulting WFE map is available in \cite{frater23}.

The results of these tests allowed for sufficiently constrained degrees of freedom to construct an as-built optical model, containing the tolerance of alignments.  It also incorporated mirror-level interferometry that parameterized the figure and roughness of each optic. The measured focal positions relative to the FPA interface were used to select the initial shims in instrument-level assembly.

\subsection{The ICE}
\label{ssec:ice}

The ICE subsystem campaign consisted of tests that ranged from partially populated single ROBs to the full flight unit, assembled in its chassis.  The list of test configurations and their applications is given in Table~\ref{tab:iceconfig}.  In this section we discuss the ICE functionality that needed optimization and verification, and how the tests were carried out.

\subsubsection{On-board processing}
\label{sssec:iceonboard}
The data volume budget is constrained by the available on-board storage in the spacecraft as well as the bandwidth available to telemeter data to the ground.  A simple archiving of the raw pixel reads at 18 bits sampled every 1.5349~s would produce a volume of 34.4~Gb per exposure.  The required maximum volume is 260~Mb per exposure.  Therefore, instead of archiving the raw samples, the rate of charge accumulation per-pixel is determined on-board.  This rate, measured in analog-to-digital converter units per frame (ADU/f) is linearly related to the photocurrent incident on the pixel, and is converted to physical instrument units (e-/s) using knowledge of the detector transimpedance gain, amplification chain and the digitization. We use a sample-up-the-ramp (SUR) algorithm, which calculates a slope by accumulating a series of sums (\cite{zemcovsur}).  To first order, recording slopes provides a factor of 118.1873~s / 1.5349~s  = 77 in data volume reduction over the raw reads.  However, a simple estimate of slope per-pixel is not sufficient, due to the various types of non-ideal effects that may arise during the integration.  At a minimum, the artifacts detected must be flagged so the data user is aware. Several types of flags are archived by the SUR algorithm:  

\begin{itemize}
    \item \textbf{Transient:} Cosmic ray impacts on a pixel impart sudden accumulations of charge.  The on-board SUR algorithm detects sudden jumps, flags the pixel and registers the slope estimated prior to impact.  The amplitude of the jump which triggers this flag is configurable via a commanded parameter.  This flag is triggered by multiple sources of charge jumps, including bright, moving satellites.
    \item \textbf{Overflow:}  The H2RGs have a finite well depth of order $10^5~e^-$.  As the accumulated charge approaches this ceiling, the measured slope deviates from linear and eventually asymptotes. During an integration, when a pixel exceeds a specified charge limit, it is flagged as overflow and the slope prior to overflow is registered. The threshold number of electrons that triggers this flag is controllable by a commandable parameter.
    \item \textbf{Early Transient:}  If a transient is detected before a configurable time threshold, part way through the exposure it is deemed an early transient and flagged as such. This functionality contains an additional parameter called ZEROSLOPEON.  If set to active, pixels flagged have the slope set to zero in the telemetered data. If inactive, the slope prior to impact is returned.  
    \item \textbf{Early Overflow:} If a source is so bright that it reaches the overflow limit before sufficient samples are collected to estimate a slope, then no photocurrent can be reported.  These pixels are flagged and a slope of zero is returned.  In the SPHEREx survey, the pre-exposure reset is done prior to slew settle completion, such that two frames post-reset can be discarded.  This allows for the decay of an electro-thermal transient in the detector which could contaminate the slope estimation.  The sum accumulations for SUR begin in concert with the completion of the slew-settle, and two frames without reaching overflow are minimally required to estimate a slope. 
\end{itemize}

The reliability and parameter sensitivities of the SUR algorithm operating in firmware were tested at several stages, from prototype boards to flight electronics.  The first configuration, utilizing a single ROB to read a band, could be used in conjunction with either the H2RG analog or later, with the cold brassboard FPA. Full functionality of the SUR algorithm could only by tested in conjunction with the full instrument, during focus and spectral calibration testing (Section~\ref{sec:inst}).  In those configurations, the collimator could be used to project scenes onto the detectors to verify correct behavior of the on-board processing. 

\begin{figure}
    \centering
    \includegraphics[width=1.00\linewidth]{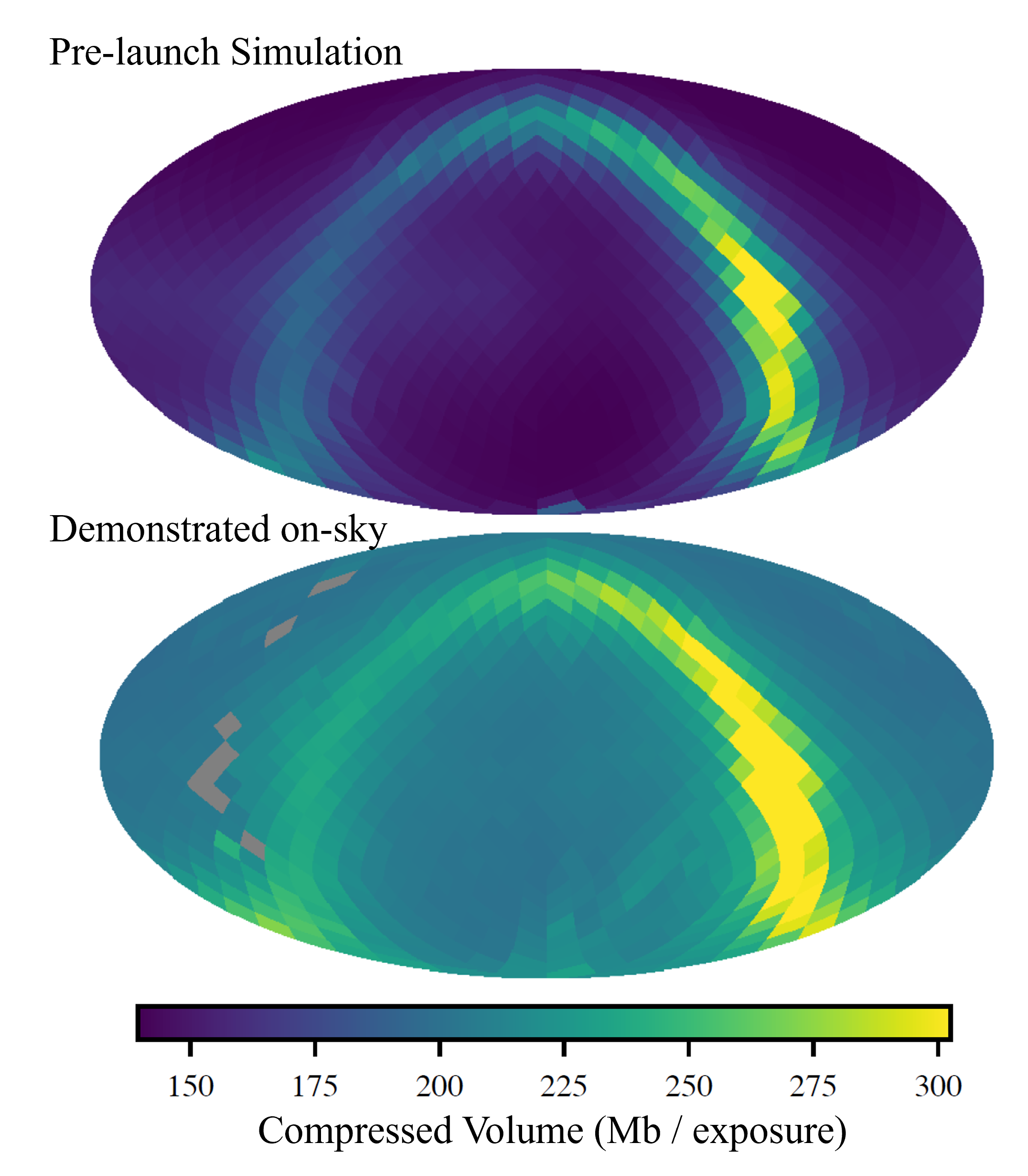}
    \caption{\textit{Top}: Pre-launch map of data volume across the sky generated through simulation.  The mean volume is 166 Mb per exposure but is highly modulated by the entropy of a given exposure.  The sky is shown using a Molleweide projection in ecliptic coordinates.  The Galactic and Ecliptic plane have lower compression factors. \textit{Bottom}: Same as top but generated from on-sky data collected during the first science survey, validating that pre-flight assumptions were sufficient to bound data volume projections.  The mean demonstrated data volume is 221 Mb per exposure, corresponding to a compression factor of 2.86, and meeting the requirement of 260 Mb per exposure.}
    \label{fig:compsky}
\end{figure}

While the SUR algorithm provides a large factor in data volume reduction, an additional compression stage is required to limit the data volume production within the requirement of 260~Mb per exposure.  This requirement is imposed on the average of 1000 consecutive exposures, because the compression factor is expected to vary significantly with entropy in the astrophysical scene, yielding worse performance in crowded fields. The survey plan naturally distributes these regions sparsely in time, such that no period is expected to contain a strong concentration of low-compression Galactic plane fields.  To accomplish this compression, we implemented a lossless Rice algorithm (\cite{rice}) in the CEB.  Testing of the basic functionality was trivial in any lab configuration that included a CEB and cold FPA. However, accurate final data volume estimation in hardware testing was not practical, as we were limited by the inability to project flight-like scenes onto the instrument. Therefore, prior to launch we estimated compression performance using a software-based emulator of the ICE that implemented the compression algorithm on simulated sky-scenes (\cite{brendansim}). This algorithm required a single parameter setting per-band, which determined the number of electrons in the intrinsic maps per integer value in the output.  The resolution was set to ${\sigma}/{2}$, where $\sigma$ reflects the measurement noise, dominated by the ZL foreground (Figure~\ref{fig:noisebreak}).  The resulting compressed data volume per 6-band exposure simulated for a complete sky survey is given in Figure~\ref{fig:compsky}. The simulation predicted a mean volume of 166 Mb per exposure with an average compression ratio of 2.37 for the image data, leaving substantial margin to the 260 Mb requirement. This large margin was essential at the pre-launch stage, as the fidelity of flag statistics and data contaminants included in the simulations were limited.  In Figure~\ref{fig:compsky}, we compare the pre-launch simulation to the actual data volume achieved on-sky during the first survey.  The spatial distribution is consistent with pre-flight expectations. The data have a mean volume of 221 Mb per exposure, for a total compression factor of 2.86.  This volume meets the requirement, and no issues with on-board data storage have arisen in survey operations.

The need for on-board processing driven by data volume constraints is common to multiple missions that fly H2(4)RGs.  The NISP instrument aboard the Euclid mission utilizes a different method from SPHEREx, called the ``multiple accumulated sampling" algorithm, in which the slope is estimated from coadding the the ramp into subgroups of samples seperated in time (\cite{euclidsur}).  The algorithm produces a quality factor to assess deviation from linearity on the ground  to use in flagging (\cite{euclidqf}), but this metric does not distinguish between the types of contaminants as our SUR implementation does, as time resolved information is lost in the averaging.  The Roman space telescope will implement a similar technique to Euclid (\cite{romansur}).

\subsubsection{Housekeeping and thermal control}
\label{sssec:icehk}
The PID feedback loop which stabilizes the FPAs (Section~\ref{sssec:therm}) is controlled with a circuit that resides in a self-contained section of the CEB.  For this reason, we tested the temperature control algorithm and tuned parameters using a single CEB in concert with the various prototype FPAs and thermal surrogates.  Final verification of the thermal control was carried out with the flight ICE in concert with the full instrument during spectral calibration (Section~\ref{ssec:sppeccal}).  

The housekeeping functions of the ICE consist of the monitoring and archiving of performance parameters including temperatures, voltages, current draws and parameter settings.  Each of the six ROBs has a uniquely valued but identically formatted set of parameters to monitor that correspond to the board and array operations. The CEB returns its own set of parameters, including the 32 auxiliary thermometers that are distributed throughout the instrument and payload.  

\subsubsection{Enabling other calibration measurements}
\label{sssec:icecal}
A working readout system was needed to facilitate the calibration tasks throughout the laboratory campaign, which had different considerations from flight operation.  In the high signal-to-noise regime of focus and spectral calibration measurements, exposure times could be much shorter (tens of seconds). If operating with the full ICE in flight conditions, the time required for the compression computation ($\sim100~$s per exposure) would be burdensome for the thousands of exposures needed. Additionally, on the ground, data storage was ample, making on-board SUR and compression unnecessary.  Therefore, for these measurements, we used a single ROB, reading one detector at a time archiving individual samples.  For the final suite of noise measurements and operational testing, where replication of flight configuration was mandatory, we used the full ICE.

\subsubsection{Noise and clocking}
\label{sssec:icenoise}
\begin{figure}
    \centering
    \includegraphics[width=1.00\linewidth]{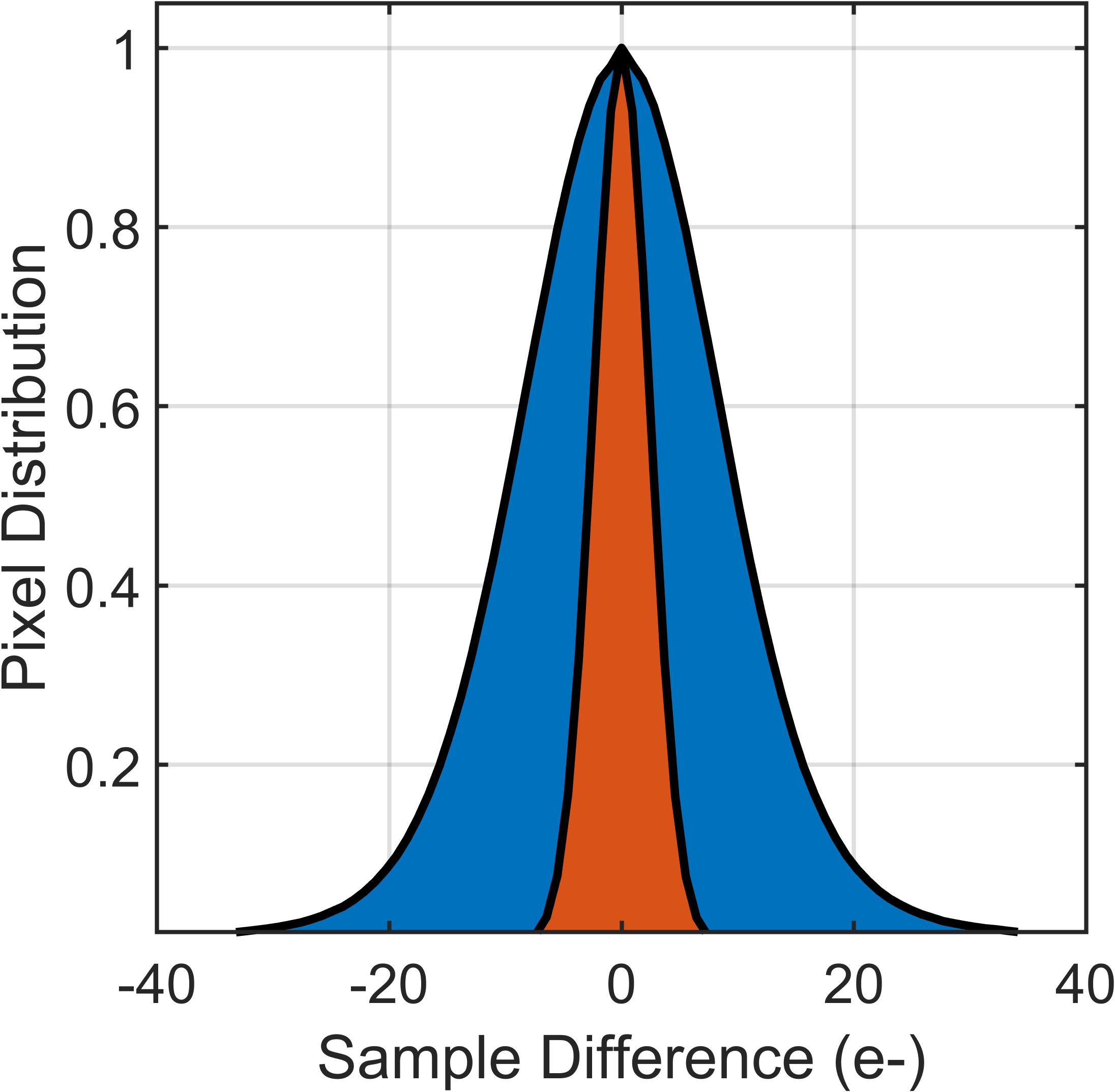}
    \caption{Double sample noise distribution for nominal pixels (blue) and phantom pixels (orange) which contain only the read noise contributions from the ICE.  The noise level of the readout is a factor of $\sim 3$ smaller than the contributions from the H2RG. The readout contribution to the total is minimal when added in quadrature with the nominal noise.  Distributions are peak normalized.}
    \label{fig:phannoise}
\end{figure}

The most basic functionality required of the ICE is the ability to read, bias and clock the H2RGs.  
We require the random and correlated noise contributions from the amplifier chain to be far sub-dominant to the read noise in the ROIC.  The detector bias for each band is generated in a thermally stabilized region of the ROB and is conditioned for minimal 1/f noise.  In addition to the row-chopping and referencing techniques discussed in Section~\ref{sssec:noise}, we utilized a scheme that included the reading of four phantom pixels during each row visit in the multiplexing cycle.   These phantom pixels are not physical on the array, but are merely allocations of the 10~$\mu $s reads without an input from the H2RG coupled into the Video 8s.  They therefore provide a measure of the ICE 1/f contributions that can be subtracted out in data reduction.  The phantom pixels are passed through the SUR and compression algorithms identically to the regular H2RG pixels, although the compression bin widths used are much smaller to reflect the lower read noise.  The equivalent contribution of correlated double sample noise, $\delta Q_{CDS}$ in the phantoms was below 3~$e^-$, compared to the $\sim 10~e^-$ generated by the ROIC that adds in quadrature with photon noise.  Distributions of double sample differences of phantom and nominal pixels are shown in Figure~\ref{fig:phannoise}. These were acquired using the full flight ICE and the band 4 flight H2RG.  The correlated noise performance from the ICE is negligible, as any residual structure generated by 1/f drifts in the Video 8 preamplifiers are removed by subtraction of the phantom photocurrents from the normal pixels in data analysis.  Presentations of power spectra in the lab and on the sky are given in \cite{chinoise} and \cite{jamiemission}.

\begin{table*}
    \centering
    \begin{tabular}{cccc}
        \textbf{ICE Configuration} & \textbf{Boards} & \textbf{Processing} & \textbf{Functionality}\\ \hline \hline
        Prototype  Read& 1 ROB (single Video 8) & - & Early noise and readout\\ \hline
        Single Read & 1 ROB & - & Noise and readout for 1 array\\ \hline
        Single Central & 1 CEB & - & HK and FPA thermal control\\ \hline
        Single Band & 1 CEB & SUR and Compression & Processing tests\\ 
         & 1 ROB &  & \\ \hline
        Software Model & - & - & Simulates on-board processing\\ \hline
        Engineering Model & 1 CEB & SUR & HK and FPA thermal control\\ 
             & 6 ROB & Compression & 2 FPA readout and processing \\ 
             & 1 LVPS &  & Mechanical housing $+$ radiator\\ 
            & &  & Spacecraft communication\\ \hline
       Flight Model & 1 CEB & SUR & HK and FPA thermal control\\ 
             & 6 ROB & Compression & 2 FPA readout and processing \\ 
             & 1 LVPS &  & Mechanical housing $+$ radiator \\ 
             & &  & Spacecraft communication\\ \hline
     \end{tabular}
    \caption{ICE configurations utilized throughout the testing campaign .}
    \label{tab:iceconfig}
\end{table*}

\section{Instrument-level testing}
\label{sec:inst}

The test campaign at instrument level provided the highest fidelity demonstration of system performance prior to launch.  The telescope, focal planes, mechanical structure, V-groove radiators and flight harnessing were all assembled in flight configuration, and no mechanical or electrical joints were de-mated between the end of instrument testing and launch.   The program was designed to ensure that the optics and FPAs were aligned within tolerance at operating temperature, and to measure the spectral response functions for each of the 24 million pixels.

To support instrument-level testing, a custom cryogenic chamber was provided by the Korean Astronomy and Space science Institute (KASI) and delivered to the test venue at Caltech in Pasadena, California.  This chamber could be configured to either accept collimated light from the laboratory through a large aperture sapphire window for focus testing (optical mode), or with a cryogenic Winston cone (\cite{Winston1970}) uniformly illuminating the telescope with tunable monochromatic light for spectral characterization (dark mode).  This test strategy is built on the successful implementation from the CIBER sounding rocket campaigns (\cite{ciber1}, \cite{ciber2}).  An overview of the full KASI-chamber test campaign is given in \cite{korngut24}.   We use the moniker ``dark mode" to describe the spectral calibration mode because there is very little path length for ambient 300~K radiation to reach the telescope, and light levels are minimal when no monochromatic light is input to the cold Winston cone.

\subsection{Focus testing}
\label{ssec:focus}
The focus testing campaign consisted of a sequence of three cooldowns of the KASI chamber in optical mode with the instrument installed.  In all cases, the deviation from collimation at infinity was quantified by scanning the position of a pinhole through focus of a laboratory collimator that had a large magnification with respect to the telescope. Images of the projected pinhole were collected at each focus position and the beam volume calculated.  For a properly aligned system, the minimum beam volume is observed when the pinhole is located at the focus of the collimator.  The offset of the pinhole position from collimator when the best beam was observed informed us of the shim size needed to bring the FPA to focus.  In each cooldown, the deviation from focus at infinity was measured at nine positions per band.   For a comprehensive description of the design and execution of the cryogenic focus testing campaign, including the control of systematic errors in the measurement, see \cite{condon24}.  The cooldown sequence was constructed as follows:
\begin{enumerate}
    \item \textbf{TVAC-1:} Initial measurements of the alignment after the instrument was assembled.  This included the metrology from FPA and telescope level testing combined with calculated corrections for the temperature dependence of several indices of refraction and residual athermality in the FPA and telescope.  The optimal shim corrections were calculated by fitting a plane to the measured deviation across each FPA.  The initial alignment required average piston corrections of $-90~\mu$m for SWIR and $+150~\mu$m for MWIR, with significant tip and tilt components as well.
    \item \textbf{TVAC-2:}  After the shim corrections were determined in TVAC-1, the instrument was cooled down again and focus remeasured.  At this juncture, the focus of both SWIR and MWIR were sufficiently close to optimal, with all points within the required $\pm 50~\mu$m set by the effective depth of focus at the SPHEREx pixel scale.  After the completion of TVAC-2, the instrument was sent to the Jet Propulsion Laboratory (JPL) for vibration testing, simulating the launch loads it would experience on the rocket.
     \item \textbf{TVAC-3:} No adjustments were made to the focus after TVAC-2, and TVAC-3 was intended to asses the vulnerability of a focus shift on launch.  No systematic shifts were observed in either FPA, and the system was determined to be sufficiently focused to proceed towards launch without further focus testing.
\end{enumerate}

While the focus testing campaign demonstrated the telescope and FPA system were optimally aligned, it did not provide an unambiguous measurement of the effective PSF.  This is because the collimator itself produced a beam with significant stray light sources, and the resulting projected images incident to the SPHEREx aperture were not completely unresolved.  Since
a direct measurement of PSF volume from these data would be overestimated, we chose to use a calculation method for PSF estimation prior to launch.

\begin{figure}
    \centering
    \includegraphics[width=1.00\linewidth]{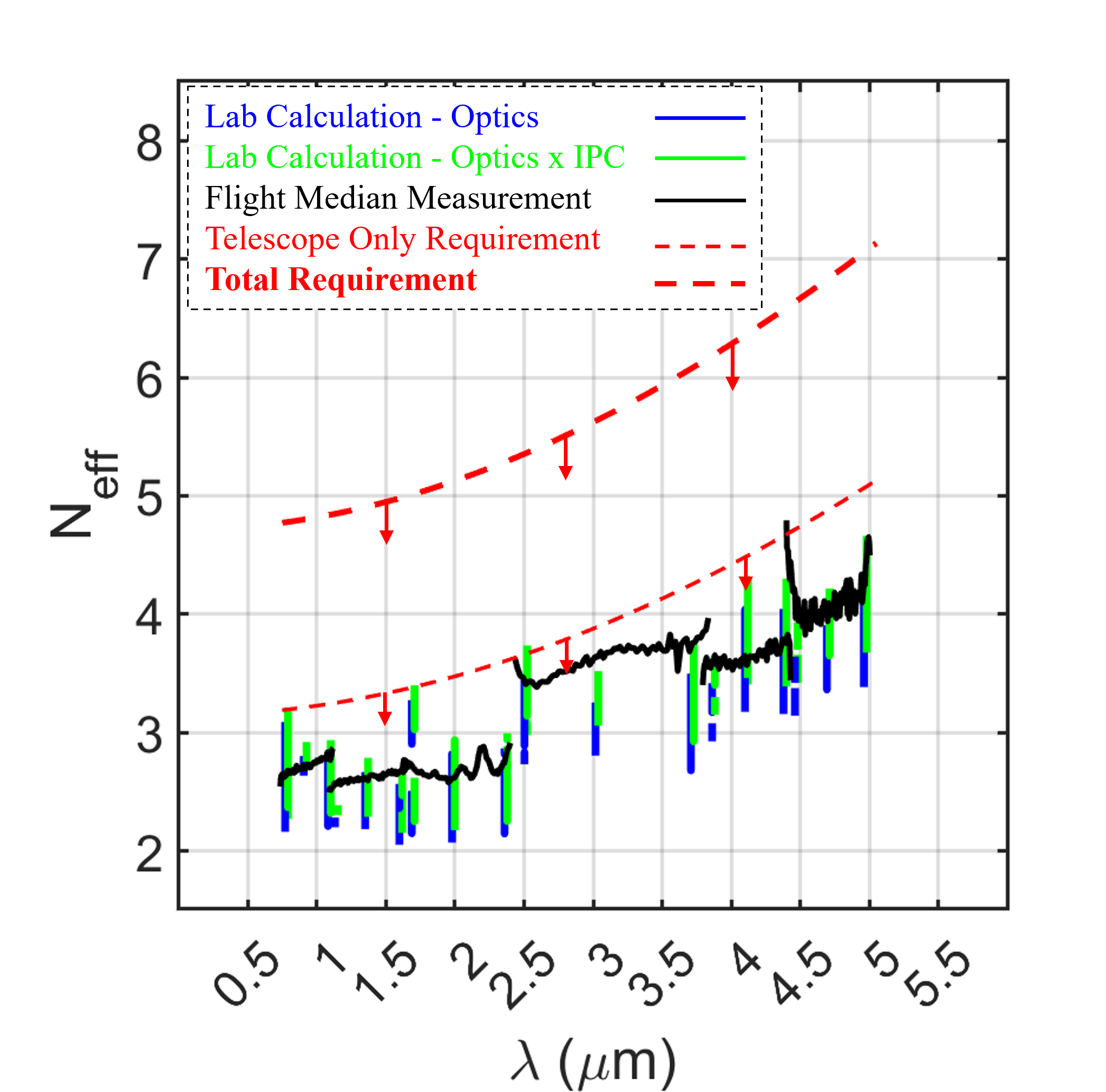}
    \caption{Calculated $N_{eff}$ based on laboratory measurements and the optical model with (green) and without (blue) the effects of IPC. Measurements from flight data are shown in black, estimated from direct integration on millions of isolated bright stars in the linear regime.  Also shown for comparison are the requirement levels with an allocation solely for optical effects, as well as the combined requirements including the random pointing jitter of the spacecraft.  The flight measurements, which benefited from excellent pointing stability (\cite{jamiemission}), should be compared with the total requirement.  In all cases, the pre-flight estimates are consistent with the in-flight measurements, and the requirements are met with margin across the wavelength range.}
    \label{fig:neff}
\end{figure}

SPHEREx has large pixels that under-sample the PSF.  This was a deliberate design choice taken to optimize the signal-to-noise on point sources, a common tradeoff in spectroscopic measurements.  Therefore, traditional metrics like full width at half maximum (FWHM) or encircled energy are difficult to use, as they require information below the pixel size and are consequently poorly defined at our spatial resolution.  Therefore, to quantify the beam volume encompassed in the effective number of pixels that a sampled PSF occupies, we define the quantity
    \begin{equation}
         N_{eff} =  \frac{1}{\sum p_i^2},
    \end{equation}
    where $p_i$ is the fraction of total flux in each pixel.  This quantity is a convenient metric to track, because in the limit of being photon noise dominated by a uniform background, the PSS scales as
    \begin{equation}
         \Delta F_{\nu} \propto \sqrt{N_{eff}}.
    \end{equation}
   
To estimate the expected flight $N_{eff}$ as a function of wavelength and position in lieu of a clean laboratory measurement, we carried out an analysis using the optical model of the as-built telescope described in Section~\ref{ssec:tel}.  Using ray tracing software including diffraction, we produced simulated optics-only PSFs at each of the nine locations sampled in focus testing.  Because the focus testing had a finite combined statistical and systematic uncertainty of $\sigma P = \pm 20~\mu$m, we produced three PSFs per location to span the allowed range.  The simulated optical PSFs were generated on a $1~\mu$m pixel scale, a factor of 18 oversampled from the size of an H2RG pixel.  This high resolution was necessary to account for the significant variation in $N_{eff}$ arising from the random alignment of a source and the pixel grid.  Due to sub-Nyquist sampling, a source aligned with the center of a pixel will produce a smaller $N_{eff}$ than a source aligned with the corner intersection of four pixels.

Figure~\ref{fig:neff} shows the projected flight $N_{eff}$ based on the optical modeling combined with focus position data.  Because the PSFs are so undersampled, the contribution of inter-pixel capacitance (IPC) in the H2RGs, which introduces correlation between neighboring pixels (\cite{ipc}), must be included.  The bars in this figure represent the averages of 100 random alignments of source and pixel grid, and show the estimate with (green) and without (blue) IPC.  Except for the short wavelength end of band 3, the telescope-only requirement is met.  The violation is due to an increase in WFE at the extreme corner, and was expected based on the telescope measurements at subsystem level (\cite{frater23}).  Also shown in this figure is the comparison to actual measurements of the $N_{eff}$ in-flight as measured in IOC (Section~\ref{ssec:firstlight}). The projected pre-flight calculations based on the telescope model and actual flight measurements are consistent.  The top level requirement of effective PSF is met with margin, because the pointing stability of the spacecraft was better than required (\cite{jamiemission}), and the PSF is dominated by the optics.

\subsection{Spectral calibration}
\label{ssec:sppeccal}

TVAC-4 consisted of the full instrument installed in the KASI chamber configured to dark mode.  In this test, we measured the spectral response function for each pixel in the FOV by coupling the aperture of the telescope into the output of the Winston cone illuminator.  A tunable monochromatic signal, generated outside the chamber, was collimated and steered through a 2 inch diameter cryostat window into the pinhole entrance to the illuminator.  We then scanned the wavelength of the monochromatic source across the entire SPHEREx band, while measuring the instrument response.  The details of this measurement are featured in \cite{howard} and are summarized in \cite{korngut24}.

\begin{figure}
    \centering
    \includegraphics[width=1.00\linewidth]{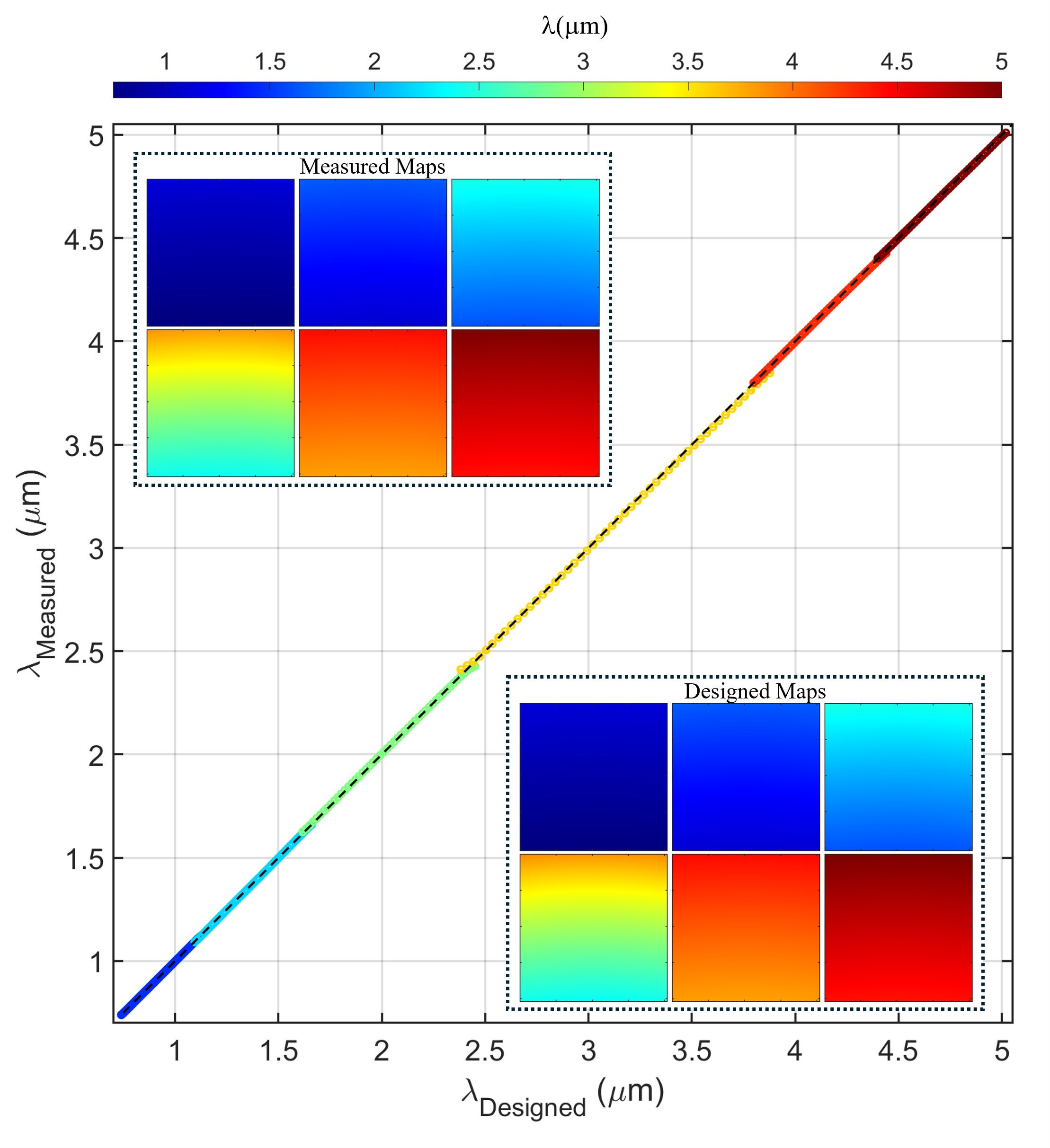}
    \caption{Pixel correspondence of the designed and measured central wavelengths across the six spectrometer bands.  The inset in the upper left shows maps of laboratory measured wavelength progressions with bands 1-3 on the top and 4-6 on the bottom.  The bottom right inset shows all six bands with the designed wavelength progression maps.  Both insets are matched in color scale following the single color bar at the top of the figure.}
    \label{fig:lamcompare}
\end{figure}

\begin{figure}
    \centering
    \includegraphics[width=1.00\linewidth]{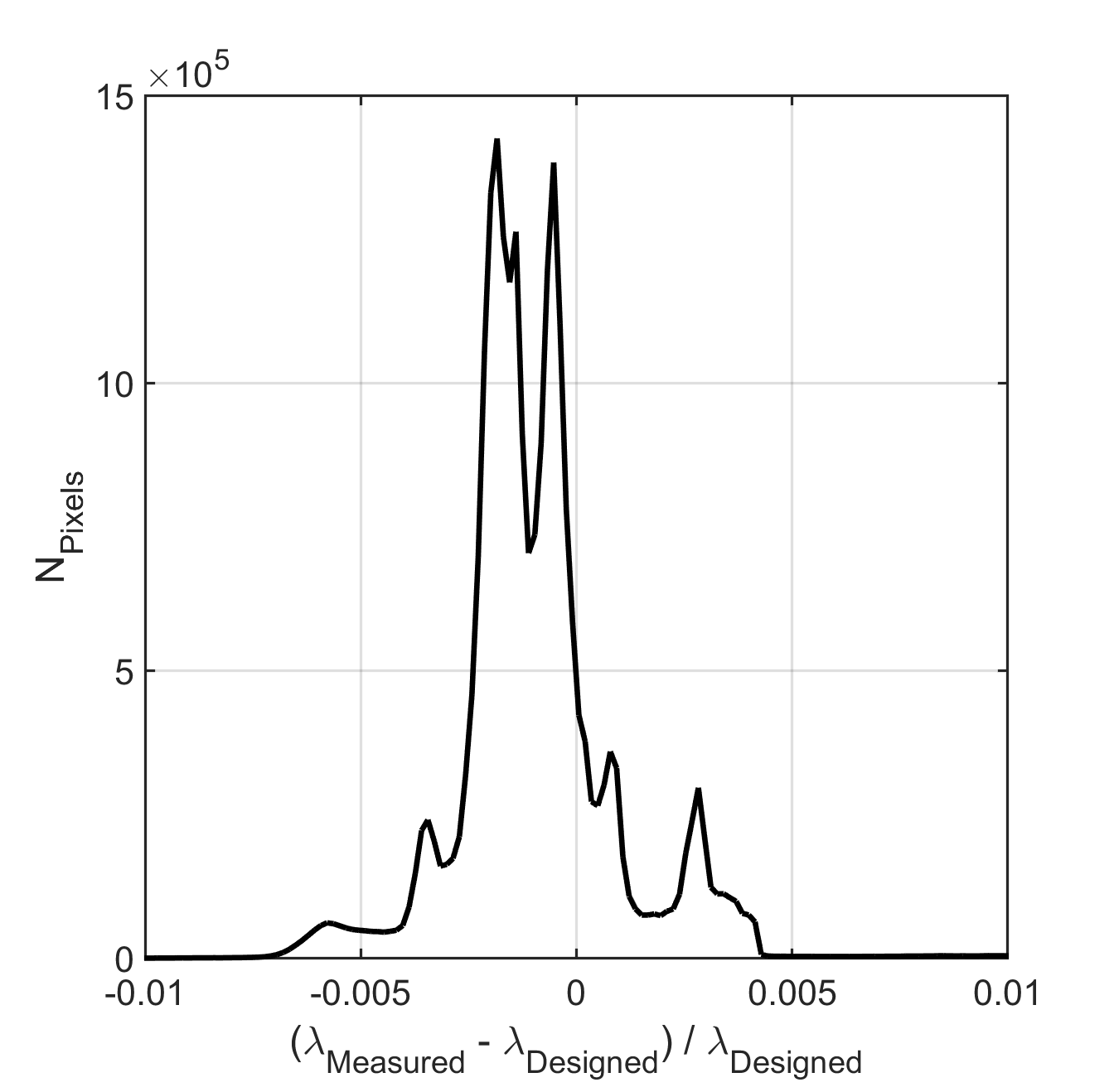}
    \caption{Pixel histogram across all six bands of the residual error in central wavelength between target design and the measured values output of instrument-level spectral calibration.  The residual error is $1\sigma = \pm 0.26\%$ }
    \label{fig:lamresid}
\end{figure}

At the conclusion of the spectral calibration testing, a response function with a dynamic range greater than $10^4$ was produced for each of the $24$ million pixels in the SPHEREx FOV.  By fitting a central wavelength to each bandpass function, we produced a wavelength progression map for each of the six bands.  Figure~\ref{fig:lamcompare} shows these maps compared to the ideal wavelength progressions that follow Equation~\ref{eqn:lamprog} with design target values of $\lambda_{min}$, $\lambda_{max}$ and smile.  The pixel distribution of residual errors from the design progressions is given in Figure~\ref{fig:lamresid}, which has $1\sigma = \pm 0.26\,\%$. The regions that dominate the error distribution are located adjacent to the DBS transition wavelength at the long end of band 3 and the short end of band 4.  This is because the DBS contribution to the bandpass is falling or rising rapidly at these wavelengths.  This effectively squeezes the response functions of the LVFs, making a higher resolving power effective function with a central wavelength shifted to the blue in band 3 and to the red in band 4. This effect was anticipated, and is accounted for in the estimated sensitivity.

While the earlier measurements taken on the LVF at component-level (Figure~\ref{fig:lamprogmeas}) indicated the filters themselves were in excellent agreement with the design, measuring the response function at instrument level quantifies the contributions from additional as-built characteristics such as mechanical alignment shifts between the LVFs and H2RGs, illumination geometry of the filters, and any spectral contributions from the DBS and upstream optics.  

\section{Observatory Level Testing}
\label{sec:obs}
After instrument-level cryogenic testing, the fully assembled instrument was packed and shipped to the spacecraft provider, BAE Systems, in Boulder Colorado.  There, it was mechanically integrated with the photon-shields to complete the payload, and mated to the spacecraft bus to complete the SPHEREx observatory. 

\begin{figure*}
    \centering
    \includegraphics[width=1.00\linewidth]{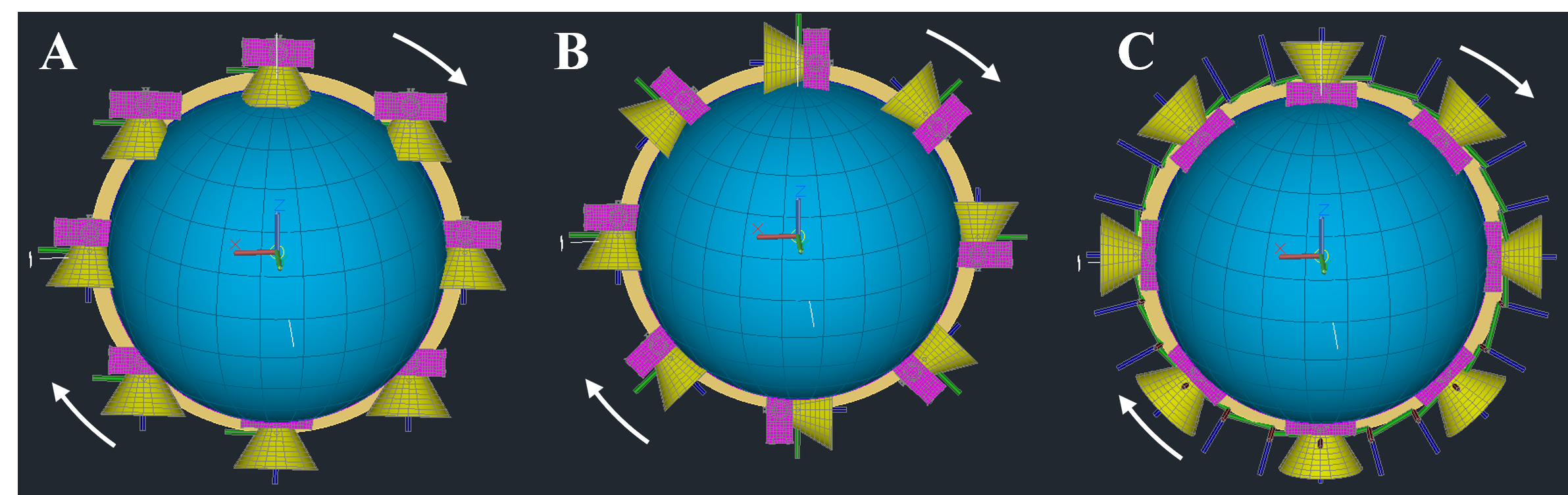}
    \caption{Attitude configurations utilized during IOC.\textit{A:} Inertial orbit maintains a direct view of Earth from the inside of the photon shields for half of the orbit to modulate the instrument temperature for decontamination. \textit{B:} Anti-RAM orbit is used for the ejection of the dust cover to make sure the orbital debris does not re-encounter the flight hardware. \textit{C:} Zenith pointed orbit maintains a view of cold space to allow the radiators to cool the flight hardware to operating conditions.}
    \label{fig:orbits}
\end{figure*}

Ambient temperature instrument testing at observatory level demonstrated end-to-end signal chain functionality and proper communications with the spacecraft. Thermal vacuum testing was also carried out, but the fidelity of instrument performance measurements was limited.  This is due to the higher than flight operational minimum temperature achieved of 100~K.  Unlike testing in the KASI chamber, the primary motivation for the test was to demonstrate functionality of the spacecraft under its representative thermal environment.  The radiative environment inside the test chamber was set by the liquid nitrogen-cooled black shroud, much warmer than the $2.7~$K heatsink provided by deep space.  While not cold enough for low noise and background-free measurements, this environment was sufficient to get below the maximum operating temperatures of the H2RGs, and to demonstrate end-to-end signal path integrity. We were able to operate the ICE near it's flight operating temperature of 0$^\circ$~C, and noise performance in the phantom pixels verified performance in flight configuration.

\section{Launch and In Orbit Commissioning}
\label{sec:ioc}

On March 12th, 2025 UTC, SPHEREx was launched aboard a Falcon-9 rocket from Vandenberg Space Force Base in Lompoc, California.  SPHEREx shared the fairing with PUNCH, a NASA small explorer heliophysics mission that operates from a similar low earth orbit (\cite{punch}).  All events on ascent proceeded nominally and the separation and orbit insertion proceeded without anomaly.   

The period spanning March 12th to May 1st 2025 UTC was reserved for execution of in-orbit commissioning (IOC).  This initial period was carefully scheduled to carry out a staged activation and optimization sequence for all systems on-orbit.   IOC, as well as all mission operations were conducted out of the Earth Orbiting Missions Operation Control center (EOMOC) located at JPL in Pasadena, California. 

From an instrument perspective, the IOC plan was constructed to achieve two goals:
\begin{enumerate}
    \item To optimally configure all on-board and survey parameters, such that they should not need any alteration for the duration of the two-year baseline mission. 
    \item To conduct measurements of instrument characteristics that are otherwise inaccessible from flight survey data, such as the far off-axis response of the telescope.  To address this, we included targeted experiments that utilized the Moon and Earth to systematically quantify these effects.
\end{enumerate}

\subsection{Initial Activations}
\label{ssec:initial}

After separation from the upper stage of the launch vehicle, the spacecraft successfully completed its autonomous initialization (auto-init) sequence.  At the conclusion of the sequence, the spacecraft was maintaining a safe attitude, and the S-band transmitter and receiver were communicating with the ground systems.  

\begin{figure*}
    \centering
    \includegraphics[width=1.00\linewidth]{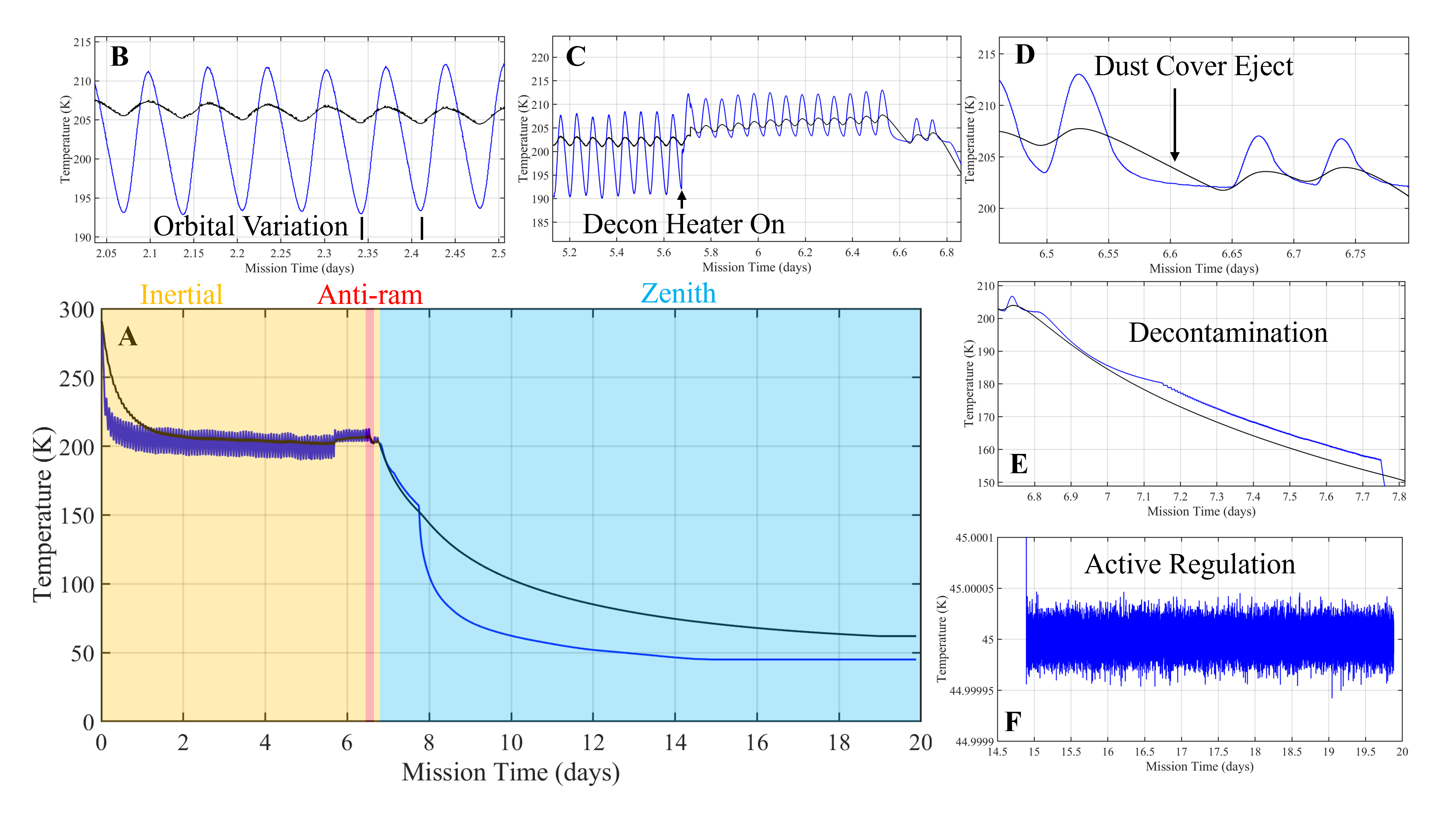}
    \caption{Cooldown curve for the SWIR (black) and MWIR (blue) FPAs during IOC. \textit{A:} Full scale curve going from launch conditions to final operation.  The colored regions correspond to changes in the spacecraft attitude as described in Figure~\ref{fig:orbits}. \textit{B:} Zoomed in temperature plot while the instrument was at equilibrium in inertial pointing. \textit{C:} Zoom in on the moment of activation of the MWIR decontamination heater that prevents the MWIR radiator from cooling below 200~K on the cooler side of the orbit. \textit{D:} Zoom in on the 1.5~orbits of anti-ram attitude, during which the dust cover was ejected. \textit{E:}  Zoom in on the decontamination period, where the heater was controlled to maintain all optical surfaces above the telescope housing temperature. \textit{F:} Zoom in on the MWIR temperature when active PID control is stabilizing to a setpoint of 45~K.}
    \label{fig:cooldown}
\end{figure*}

In IOC, we utilized three predetermined spacecraft attitudes with respect to the Earth and Sun, as illustrated in Figure~\ref{fig:orbits}.  First, the auto-init sequence placed the spacecraft in an inertial configuration (Figure~\ref{fig:orbits}A), where thermal radiation from the Earth periodically warms the instrument, driving off adsorbed water from the optics and keeping the cover deployment mechanism above 200~K.
The inertial attitude also maintains solar illumination of the photovoltaic panels to keep power-positive.  Figure~\ref{fig:cooldown} shows the temperature curves for the SWIR and MWIR focal planes throughout IOC.  Because the MWIR FPA has its own radiator with a relatively low thermal mass compared to the telescope, the amplitude of the MWIR sinusoid from orbital variation is much higher. 

The inertial attitude was maintained for the first several days of IOC.  The attitude determination and control system (ADCS) was commissioned by activating and calibrating the star trackers and the global positioning system receiver which allow for celestial navigation and pointing.  With these key components incorporated into the ADCS solutions and fed back into control loops with the reaction wheel speeds, a fixed target on the Earth or sky could be tracked.

Once attitude commanding and control were robustly established, SPHEREx was poised to eject the telescope dust cover mounted on top of the optical baffle.  To accomplish this, SPHEREx was commanded into an anti-RAM attitude, where the RAM direction is defined along a vector pointing in the direction of motion (Figure~\ref{fig:orbits}B). This attitude is adopted such that that the ejected cover would not recontact the spacecraft.   The deployment mechanism has spring loaded latches on either side of the baffle that are commanded separately.  To allow for the damping of oscillations after the first latch was released, the two commands were separated in time by 90~s.  The activity was conducted while line of sight to the spacecraft was maintained over a ground station, so real-time telemetry could be monitored.  Confirmation of the deployment was obtained by monitoring the attitude error signal. The response of the system to the impulse received by the ejection was observed at very high signal to noise, with the time constant expected set by the control loop.  With the cover successfully deployed, the mission was cleared to proceed to the Zenith pointed attitude and to begin cooling down the cryogenic components of the instrument, as shown in Figure~\ref{fig:orbits}C.

\subsection{Decontamination and cooldown}
\label{ssec:cooldown}

The freezing of water ice onto optical surfaces in spacecraft has been a source of efficiency loss and optical degradation in several previous optical and infrared space missions (e.g. \cite{euclidice}).  SPHEREx was optimized for the detection of water ice absorption in interstellar space using the 3.0~$\mu$m absorption band, and is therefore particularly sensitive to water contamination on the optics.  Using results from laboratory measurements of the infrared absorption of water ice (\cite{lynch05}), we set the requirement for total accumulation of ice on the optical path to be less than 100 $\mathring A$, limiting absorption to $<2.5\,\%$.  Using detailed modeling of the molecular transport of water throughout the spacecraft, we developed a cooldown strategy (\cite{alred24}) similar to what was done for the James Webb Space Telescope (JWST) (\cite{jwstice}).  The basic principle is to carefully control temperature gradients within the instrument while it cools from 200~K to 150~K.  At temperatures above this, water will sublimate into space and at temperatures below, it will freeze in place.  When the optical surfaces cool, we apply current to decontamination heaters to maintain their temperatures above the surrounding telescope housing.  This ensures that ice settles out of the optical path and does not imprint its absorption feature.  Once at operating temperatures, the water is stably frozen in place on the housing and will remain there unless the telescope warms back up beyond 150~K.  

\begin{figure*}
    \centering
    \includegraphics[width=1.00\linewidth]{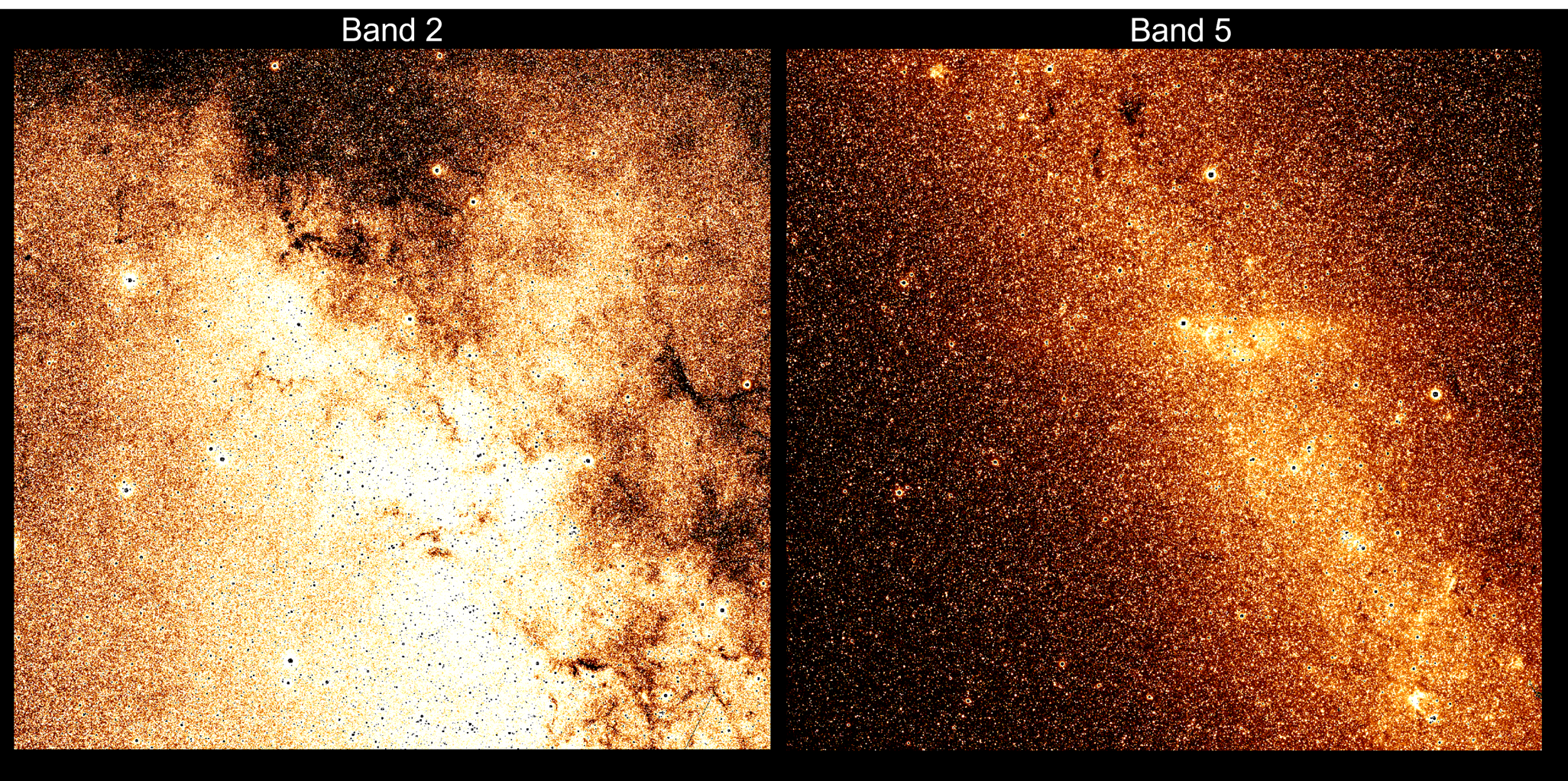}
    \caption{Simultaneously collected raw spectral images in array coordinates from band 2 (left) and band 5 (right) collected during the first practice survey and the first crossing of the Galactic plane. Bands 2 and 5 view the same sky across the DBS.  Each image spans the full 3.5$^\circ$ FOV of the array with wavelength increasing from bottom to top.  The horizontal banding structure observed observed in band 5 corresponds to diffuse Bracket Alpha emission in the Galaxy at 4.05~$\mu$m.}
    \label{fig:firstlight}
\end{figure*}

As depicted in Figure~\ref{fig:cooldown}C, the decontamination heaters were activated 13 orbits prior to the transition from inertial to anti-RAM attitude for cover eject.  This kept the optical surfaces above 200~K during the cold side of the orbit, preventing water ice from forming before cooldown begins.  After cover ejection, the spacecraft returned to inertial pointing autonomously to allow for confirmation of successful deployment by ground operators.  As the ejection was unambiguous, the spacecraft was commanded to zenith pointing and the instrument temperatures cooled towards their operating temperatures. %

The final stage of the cooldown procedure was to activate the FPA thermal regulation.  We chose setpoints of 62~K and 45~K for the SWIR and MWIR FPAs respectively.  These temperatures were selected prior to launch as the coldest temperatures we expected to achieve while maintaining sufficient control margin to regulate at the end-of-life radiator performance.  We intend to maintain these setpoints for the duration of the mission for gain stability considerations.  While the demonstrated orbital-performance of the passive cooling system indicates that lower FPA temperatures are achievable, we opted to remain at the nominal setpoints such that flight temperatures match those used in ground calibration activities.

\subsection{First light}
\label{ssec:firstlight}

Once the FPAs cooled passively below their target setpoints, the H2RGs in both the SWIR and MWIR FPAs were powered on.  Real-time monitoring of housekeeping measurements that contained the current draws, voltages and control heater voltages all read nominal values and the basic health of the entire system was verified.  

After a brief collection of diagnostic photocurrent images indicated the detectors and ICE were performing nominally, we proceeded to command the collection of the first survey-like exposures.  This initial set of observations consisted of 120 exposures collected with survey and instrument parameters configured to default. Visual inspection of the resulting images made it clear that the end-to-end mission functionality was nominal.  The images validated the timing of internally sent on-board commands, the compression of on-board data, the transfer of data from the ICE through the spacecraft, transmitter, ground station and processing at the SPHEREx science center.  An image pair, collected in this survey towards the Galactic plane is given in Figure~\ref{fig:firstlight}.  At the shorter wavelengths in band 2, much of the FOV is lit up with bright diffuse galactic light (DGL), originating from interstellar photons that scatter off dust.   The densest regions have high optical depth at these wavelengths, and extinction is observed in filamentary structures in the image.  At the longer wavelengths measured by band 5, the optical depth is reduced, and the stellar density peak is revealed along the Galactic plane which runs diagonally from top left to bottom right.  The horizontal stripe of bright signal in band 5 corresponds to Bracket alpha emission.  Also visible in band 5 are dark absorption regions at 4.3~$\mu$m, indicative of carbon dioxide ice absorption.

In addition to confirming expected on-orbit behavior, we observed emission associated with the orbital environment.  Line emission from metastable Helium in the Earth's thermosphere at 1.083~$\mu$m was visible in each exposure with an amplitude larger than the ZL background \citep{hui26Airglow}. Additionally, a spectrally resolved diffuse emission, with peaks at 3.0~$\mu$m and 4.5~$\mu$m was observed.  This emission was brightest in the first exposure of a small slew group and diminished in amplitude through the next three.  Because the survey strategy looks ahead in the orbit earliest in the slew group (\cite{seansurvey}), the RAM angle is lowest in the first exposure.  At low RAM angle, the telescope's baffle comes in contact with more atmosphere containing atomic oxygen. At higher RAM angles, the telescope is obscured to the atmosphere by the photon shields. We therefore suspected the emission to be caused by a shuttle glow phenomenon (\cite{shuttleglow}), originating from the fluorescent interaction of paint on the baffle excited by the atomic oxygen.  This RAM angle dependence indicated that including a new avoidance criterion in the survey plan could reduce susceptibility to this emission. We decided to quantify the effect further in a dedicated IOC test (Section~\ref{ssec:earthshine}).  Template spectra of the helium emission and shuttle glow are available in \cite{jamiemission}.

\subsection{Survey diagnostics}
\label{ssec:surveydiag}

\begin{figure}
    \centering
    \includegraphics[width=1.00\linewidth]{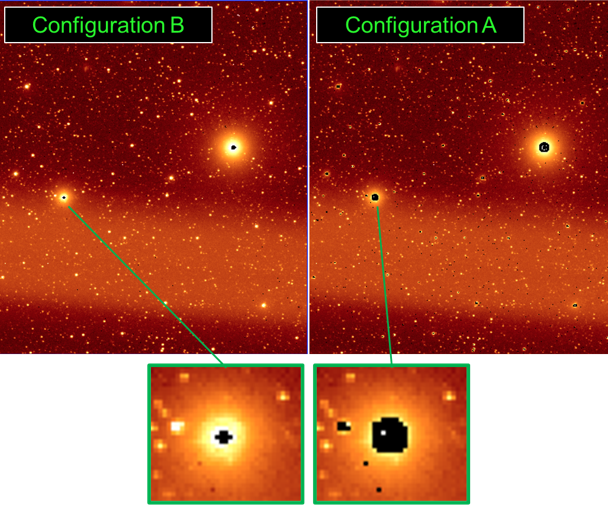}
    \caption{Diagnostic mode data processed on the ground with varied values of the ZEROSLOPEON parameter. Black corresponds to zero value. In Configuration A (right), the ZEROSLOPEON parameter is enabled, and pixels that trigger the early transient flag have a value of zero.  In Configuration B (left), the pixels with early transient flag triggered have a slope estimate from before the trigger.  Configuration B was enabled during IOC and is maintained for the survey operations.}
    \label{fig:zeroslope}
\end{figure}

With basic functionality established, the focus of IOC shifted towards optimizing performance.  To better inform parameter choices, a series of exposures were taken in diagnostic mode, where SUR processing is bypassed, and individual detector samples are archived.  Data generated in this mode are much larger than survey data, and require a period of twenty minutes in-between exposures to transfer the data from memory in the ROBs to the spacecraft.  We commanded the spacecraft ADCS to execute a one-day survey with diagnostic instrument data collected during a subsample of 24 pointings.  For these targets, detector reads were archived beginning 12 seconds before arrival on-target.  This allowed us to track bright stars as seen by the FPAs to monitor the settling directly.  Data from the ADCS already indicated that the stabilization performance was better than specifications (See Figure~16 of \cite{jamiemission}), but this test confirmed that any additional motion between the telescope and spacecraft body was negligible.  We therefore chose to shorten the survey design's slew settle time, and lengthen exposure times from the pre-IOC value of 112~s, to 118~s for increased sensitivity.

\begin{figure}
    \centering
    \includegraphics[width=1.00\linewidth]{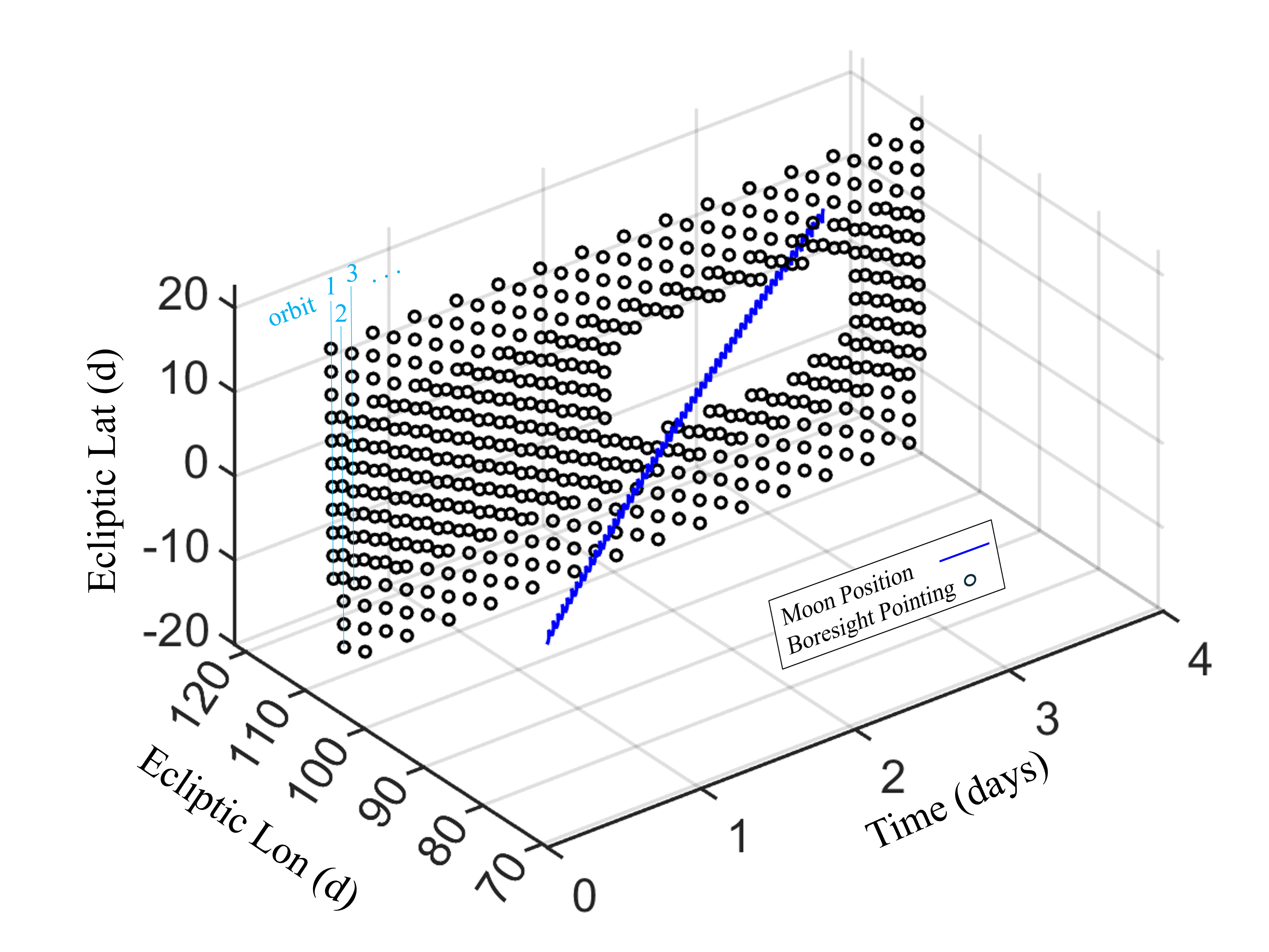}
    \caption{Boresight pointings designed for the moonshine activity compared to lunar motion during the period of observations. This strategy allows for repeated observation of celestially fixed pointings while their proximity to the Moon varies with time.  The background can therefore be differenced out, leaving only the stray light from the Moon measured at high signal to noise. Each point represents an exposure collected.  Cyan labels denote exposures taken during a single orbit, viewed in this plot as a vertical line.}
    \label{fig:moontgt}
\end{figure}

\begin{figure*}
    \centering
    \includegraphics[width=1.00\linewidth]{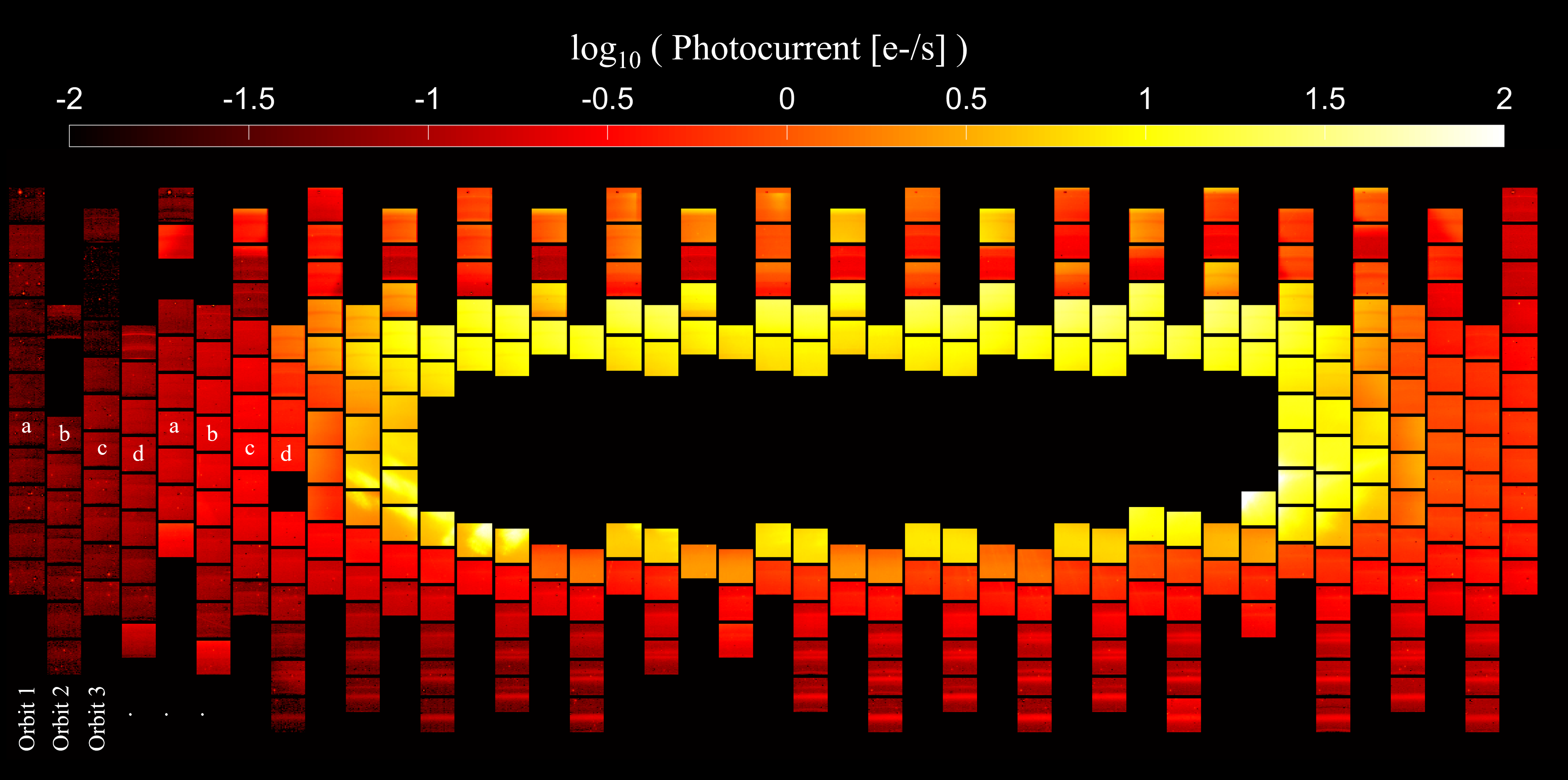}
    \caption{Measured photocurrent in band 2 of exposures taken during the moonshine measurement.  Each square represents the 3.5$^\circ \times$ 3.5$^\circ$ FOV of the single SPHEREx band.  Columns show array-wide photocurrent of exposures taken in a single orbit along a constant line of ecliptic longitude.  Because the FOV is larger than the spacing between targets and we wish to avoid overlapping measurements, we have added space between squares for clarity. The dimensions of this figure do not linearly correspond to angular separation.  Each square corresponds to a target exposure location in Figure~\ref{fig:moontgt}.  Lower case letters \textit{a} through \textit{d} denote the unique field in celestial coordinates imaged at multiple epochs, with the Moon at varied relative distance.}
    \label{fig:moonmeas}
\end{figure*}

In addition to optimizing the settle time, these diagnostic observations provided a set of fully sampled flight data to test parameter choices in the SUR algorithm, including ZEROSLOPEON.  
Running the SUR algorithm in software on the flight diagnostic data revealed that early transient flags were being activated around bright sources, caused by charge blooming effects in the H2RGs and information was being lost. Disabling ZEROSLOPEON returned this information, and the flight data showed that pre-flight data-volume concerns from frequent early transients were unwarranted. Figure~\ref{fig:zeroslope} shows the effects of ZEROSLOPEON around bright sources in photocurrent maps generated from diagnostic data.  The remaining black pixels with ZEROSLOPEON disabled are lost to rapid overflow.

\subsection{Extended sidelobe quantification through moonshine}
\label{ssec:moonshine}

Stray light from bright sources outside the FOV can be a source of spurious diffuse light fluctuations that inhibit EBL measurements. A controlled laboratory off-axis measurement campaign was prohibitive in scope, and our pre-flight analyses relied on numerical simulations. To quantify performance in the as-built system, we measured the off-axis response during IOC, with a targeted experiment utilizing the Moon.  The lunar stray-light observations quantify the response at mid-range angles, where bright stars could leave a systematic signal in the EBL measurements.  The goal  was to either show the level is negligible, or to provide sufficient data for its removal.

We planned the observations by designing a grid of celestially-fixed pointings in ecliptic coordinates. 
The grid contained four groups, each with a series of 11 points, evenly spaced in latitude along a line of constant longitude.  The four groups were offset from each other in latitude, by amounts chosen to provide a sampling when combined that was fine at the center while maintaining a large dynamic range. The observation concept was to take exposures towards every pointing in one group per orbit.  The groups were observed in consecutive orbits requiring four orbits to provide one sample towards each of the celestial pointings.  After all four were observed, the first group was returned to and the cycle repeated. This process was carried out continuously for a period of four days while the moon moved with respect to the celestial background, as shown in Figure~\ref{fig:moontgt}.  The base set of eleven targets per line was the maximum achievable number of exposures per orbit given Earth and Sun avoidance criteria.  To avoid saturation resulting in persistent photocurrent contaminating measurements, no exposures were collected while the moon center was closer than 5$^\circ$ from any location on the FOV.  

The moonshine activity was successfully executed during the first quarter window beginning on April 2nd, 2025.  Figure~\ref{fig:moonmeas} shows the photocurrent response for band 2 during the observations.  
Each tile in this figure represents the 3.5$^\circ \times$ 3.5$^\circ$ FOV of band 2, and each column an orbit's worth of data.
Because the targets were repeated in celestial coordinates, the background could be subtracted by using the first exposure per target as a reference.  The asymmetry observed in the response pattern, with higher signal towards the top of the figure was expected from simulations, and is attributed to ray paths that reach the tertiary mirror in the off-axis design directly.  The largest peaks have structures smaller than the FOV, and are most notable at the left and right edges of Figure~\ref{fig:moonmeas}. As described in the SPHEREx data explanatory supplement\footnote{\url{https://irsa.ipac.caltech.edu/data/SPHEREx/docs/overview_qr.html}}, this can cause rare spurious artifacts in the survey data when a bright star of order $M_{AB} \sim< 3$ is nearby.  The numerical analyses of the stray light performance, informed by these moonshine observations will appear in \cite{darren}.

\subsection{Earthshine and shuttle glow quantification}
\label{ssec:earthshine}

\begin{figure}
    \centering
    \includegraphics[width=0.8\linewidth]{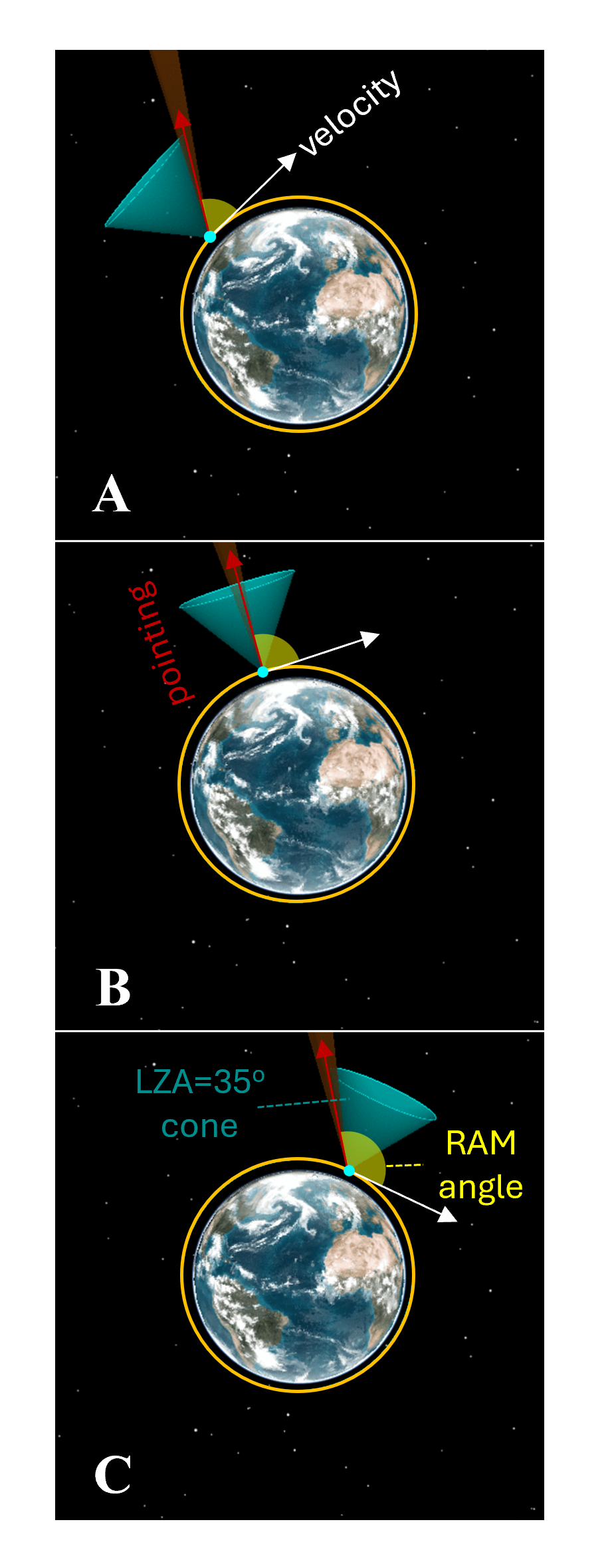}
    \caption{Illustration featuring the configuration for three exposures during a single orbit of the Earthshine and shuttle glow experiment.  In all three panels, the red arrow points to a location fixed in celestial coordinates.  The white arrow points along the velocity direction, the cyan cone represents the LZA = 35$^\circ$ surface and the yellow angle represents RAM. \textbf{A:} A configuration with RAM = 35$^\circ$ and LZA = 35$^\circ$, subject to the most Earthshine and the most shuttle glow. \textbf{B:} A configuration with RAM = 90$^\circ$ and LZA = 0$^\circ$, subject to the smallest Earthshine and moderate shuttle glow. \textbf{C:} A configuration with RAM = 125$^\circ$ and LZA = 35$^\circ$, subject to the most Earthshine and smallest shuttle glow. }
    \label{fig:earthdesign}
\end{figure}

\begin{figure}
    \centering
    \includegraphics[width=1.00\linewidth]{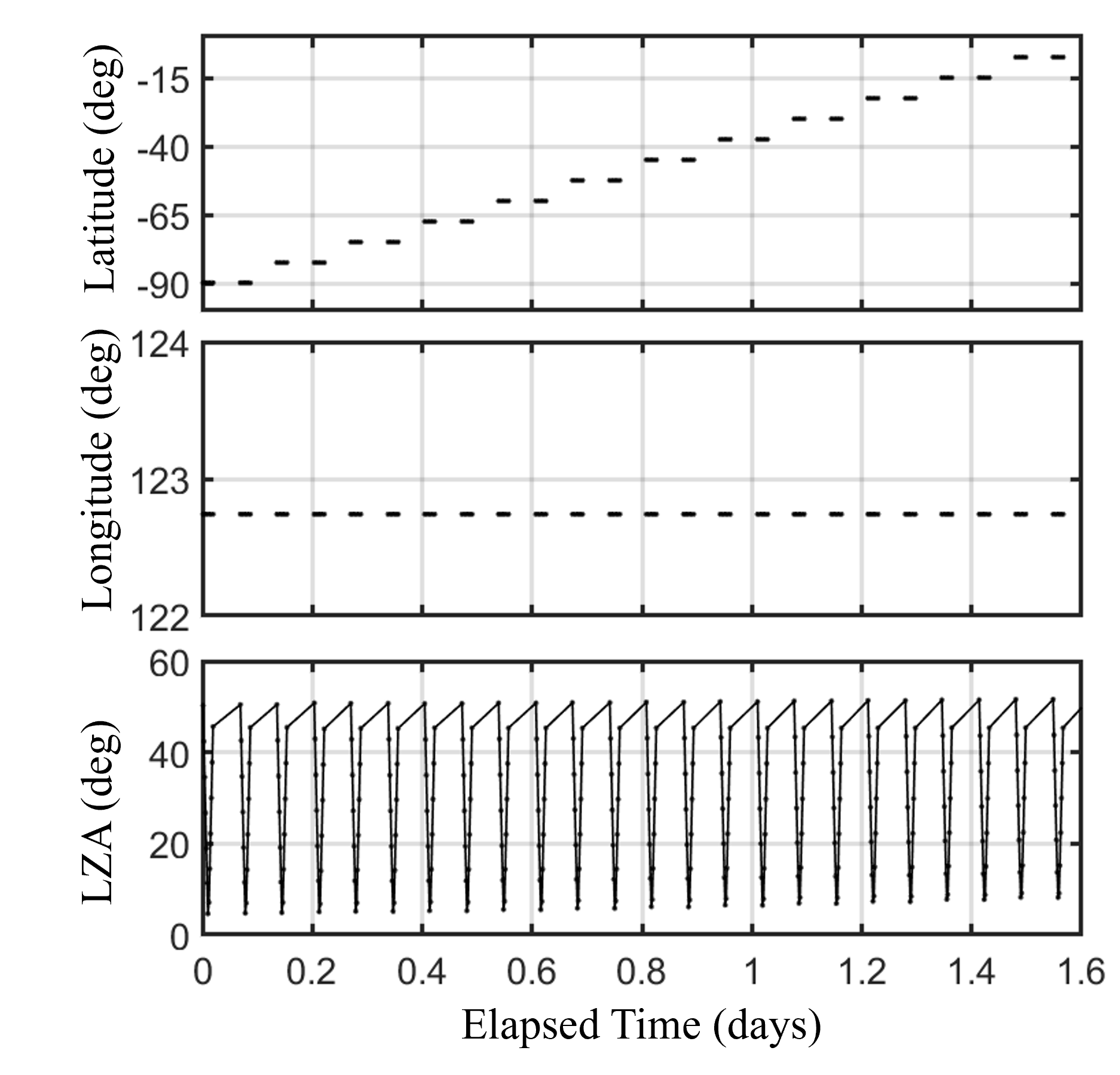}
    \caption{Pointing characteristics of the Earthshine measurement during IOC.  This measurement maintains the boresight fixed on a celestial target while collecting exposures.  The exposures are repeated as the local zenith angle is modulated.  }
    \label{fig:earthtgt}
\end{figure}

The pre-IOC survey strategy included an avoidance criterion such that the local zenith angle (LZA), defined as the angle between the vector normal to the payload and zenith at spacecraft location, be less than $35^{\circ}$ during all exposures.  At LZAs larger than 24$^\circ$, a section of the inner photon shield is directly illuminated by the Earth's limb.  The conical shield geometry is designed such that the specular Earth rays are reflected out to space, but scattered rays can still reach the telescope. Numerical simulation suggested that the level of scattered Earthshine reaching the FPAs at 24$^\circ <$ LZA $< 35^\circ$ would be small, and tolerable in EBL studies.  To quantify both the photocurrent introduced by Earthshine as well as by the aforementioned shuttle glow effect, we constructed a specialized IOC observation. This allowed us to optimize the science survey to maintain survey completeness while using updated avoidance criteria that minimize the contributions of these contaminants. The survey completeness as a function of avoidance criteria was estimated using simulations from the survey planning software of \cite{seansurvey}.  Figure~\ref{fig:earthdesign} illustrates the technique and geometry involved.  In each orbit of the experiment, a fixed celestial target which transited near zenith was tracked with repeated exposures, as the LZA of the spacecraft was modulated up to 55$^\circ$. This strategy, is roughly symmetric about zenith, so naturally modulates the RAM angle as well. Because the celestial target is fixed, photocurrent modulations can be isolated to the contaminants and their angular dependencies.   The executed pointing plan is shown in Figure~\ref{fig:earthtgt}. 

\begin{figure}
    \centering
    \includegraphics[width=1.00\linewidth]{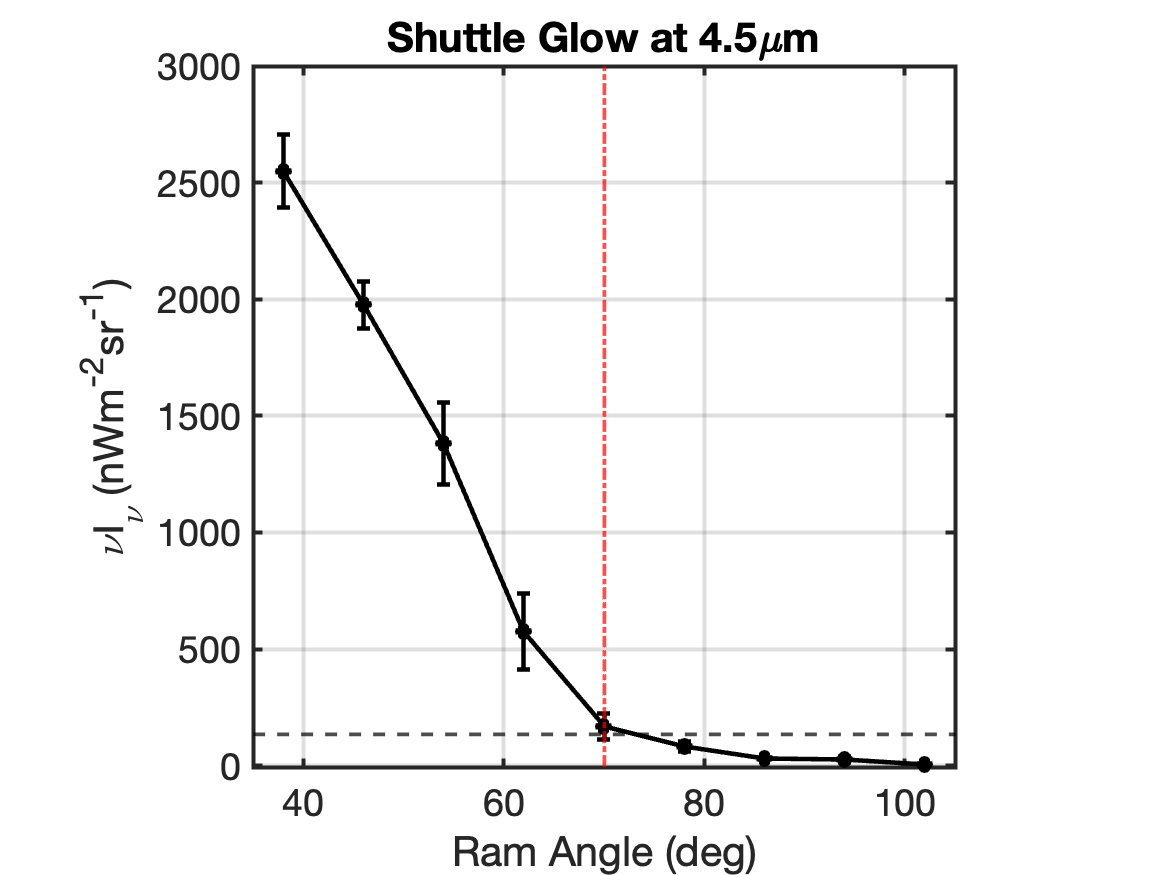}
    \caption{Intensity of the shuttle glow emission at $4.5~\mu$m as a function of ram angle as measured in the Earthshine experiment.  The red line indicates the avoidance criterion of 70$^\circ$ imposed on the survey as a result of this test.}
    \label{fig:shuttle}
\end{figure}

The measured dependence of shuttle glow amplitude on RAM angle is shown in Figure~\ref{fig:shuttle}.  Based on this, we chose to impose a new avoidance criterion of RAM $<70^{\circ}$ on the survey.  Simulations showed that adding this restriction while maintaining the LZA$<35^{\circ}$ limited the available sky too harshly, and survey completeness would be impacted.  When pointed at LZA$=42^{\circ}$ in the anti-RAM direction, the diffuse earthshine signal was measured to have a modest amplitude of $0.05~e^-/s$, peaking in the short wavelength bands. At $\sim 40\times$ fainter than the ZL, this level was deemed acceptable. We therefore opted to loosen the LZA requirement from $35^{\circ}$ to  $42^{\circ}$ and maintain 100$\%$ survey completeness.
While the 70$^\circ$ avoidance eliminates the bulk of shuttle glow contamination, a small amount remains, and is present in the survey data. This type of contaminant may be more serious for EBL studies, however the redundancy of the deep field observations allows us to discard exposures that contain significant shuttle glow or Earthshine.

\subsection{Parameter optimization and practice surveys}
\label{ssec:practice}
We optimized parameter settings and survey performance through a series of practice surveys, distributed throughout the IOC period. They began with the default configuration in first light, and steadily increased in length and complexity, incorporating information gleaned from previous activities.  Each time a characteristic was changed, we conducted a practice survey to verify that the consequences were as intended. The resulting seven survey sequence is summarized in Table~\ref{tab:practice} with key milestones noted. The sixth and seventh practice survey parameters were identical to those in the science survey, and the collected data are included in the released data set.

\begin{table*}
    \centering
    \begin{tabular}{c|c|c}
        \textbf{Practice Survey} & \textbf{Duration (days)} & \textbf{Features and demonstrations} \\ \hline \hline
        1& 0.21 & First light and ``out-of-the-box" performance observed\\ \hline
        2& 1.7 & Full day of survey with downlinks\\ 
        &  & Initial optical measurements from stars\\ 
        &  & validation of on-board SUR and compression\\ 
        &  & quantification of shuttle glow\\ \hline
        3& 1.0 & Disabled ZEROSLOPEON in SUR \\ \hline
        4& 3.0 & Final uplink and survey command cadence \\ \hline
        5& 2.5 & Transfer of coverage to date in survey plan \\ \hline
        6& 3.3 & Final compression bin widths commanded\\ 
        &  & Final LZA and ram angle avoidance\\         
        &  & Demonstration of final configuration\\ \hline
        7& 3.0 & Demonstration of final configuration\\ \hline

     \end{tabular}
    \caption{Practice surveys conducted during IOC to progressively increase the fidelity of the survey.}
    \label{tab:practice}
\end{table*}

The practice survey data allowed for the first statistical quantification of instrument optical performance.  By excising $7\times7$ pixel images centered on 1.1 million isolated stars with photocurrents brighter than  $200~e^-/s$ in the absence of SUR flags, we produced the $N_{eff}$ curves of Figure~\ref{fig:neff} that indicate the imaging performance of the telescope meets requirements. 

Measures of the spectral dependence of median photocurrent distribution across the sky contained no detectable feature at $3.0~\mu m$, indicating the decontamination procedure successfully kept ice from forming on optics.  Data from practice surveys 6 and 7 displayed strongly reduced shuttle glow contamination, as expected from the avoidance criteria changes. 

Several instrument parameters were changed in between practice surveys, and the expected consequences were verified without exception.  These include the disabling of ZEROSLOPEON (Section~\ref{ssec:surveydiag}), matching the bit resolution of photocurrent used in compression to the observed flight noise, and changes made to setpoints in thermal regulation of boards in the ICE.

\section{Conclusions}
\label{sec:conc}

On May 1st, 2025, upon the completion of in-orbit commissioning activities, the SPHEREx mission began its all-sky spectroscopic survey. It has continued without interruption, completing its first map of the celestial sphere in December of 2025.  

\begin{figure}
    \centering
    \includegraphics[width=1.00\linewidth]{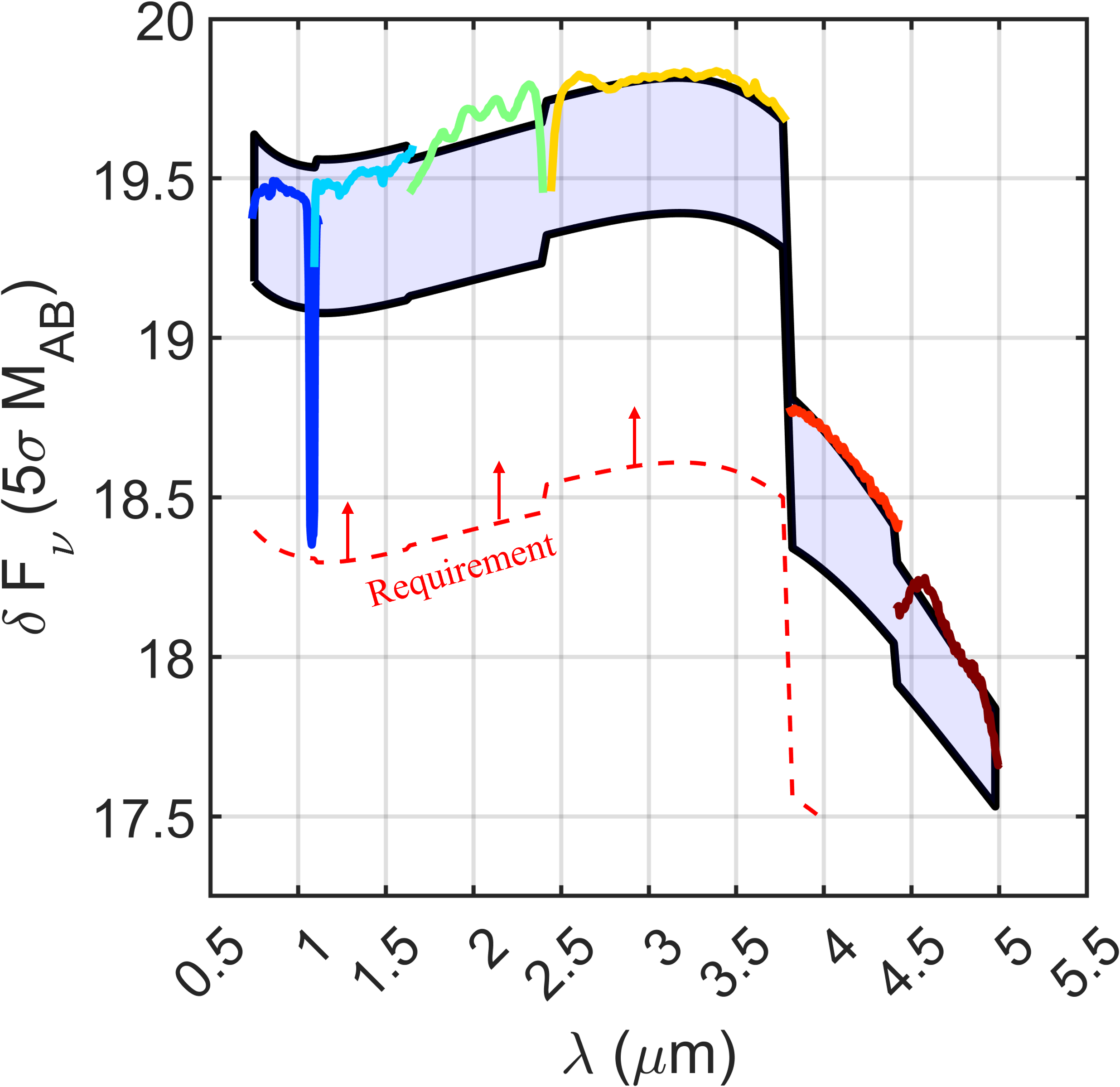}
    \caption{Projected all sky sensitivity based on flight performance assuming 4 observations per wavelength and the median measured photocurrent across the sky.  The colored lines indicate the wavelength range covered in each of the six bands. The gray area encompasses the projected mission sensitivity and uncertainty at the time of submission of the SPHEREx concept study report in 2018.  In all cases other than the helium contaminated 1.083~$\mu$m channel, the actual performance is as-good-as or better than the original projected performance estimates and meets requirements (red dashed line) with margin.}
    \label{fig:sens}
\end{figure}

The survey data allows for additional evaluation of on-sky performance through the statistically significant sample of exposures and repeated observations of celestial sources. 
Figure~\ref{fig:sens} shows the projected PSS at the completion of the four survey baseline mission estimated from flight derived products (\cite{jamiemission}).  Embedded in this calculation are the empirical conversion from flux units to photocurrent (\cite{matt}), the median photocurrent across the sky (\cite{jamiemission}), the effective PSF volume (Figure~\ref{fig:neff}), instrument noise (\cite{jamiemission}, \cite{chinoise}) and the spectral response functions (\cite{howard}).  Across all six SPHEREx bands, in every channel save the one contaminated by 1.083~$\mu$m helium emission, the performance is as good as the original projections at the submission of the proposal to NASA in 2018, meeting the requirements with margin. 
The original estimates assumed longer integration times (134 to 146~s) than realized in flight (118~s), but the higher efficiencies of the LVFs and detectors, along with smaller effective PSFs, compensated to give a similar final sensitivity.

\begin{figure*}
    \centering
    \includegraphics[width=1.00\linewidth]{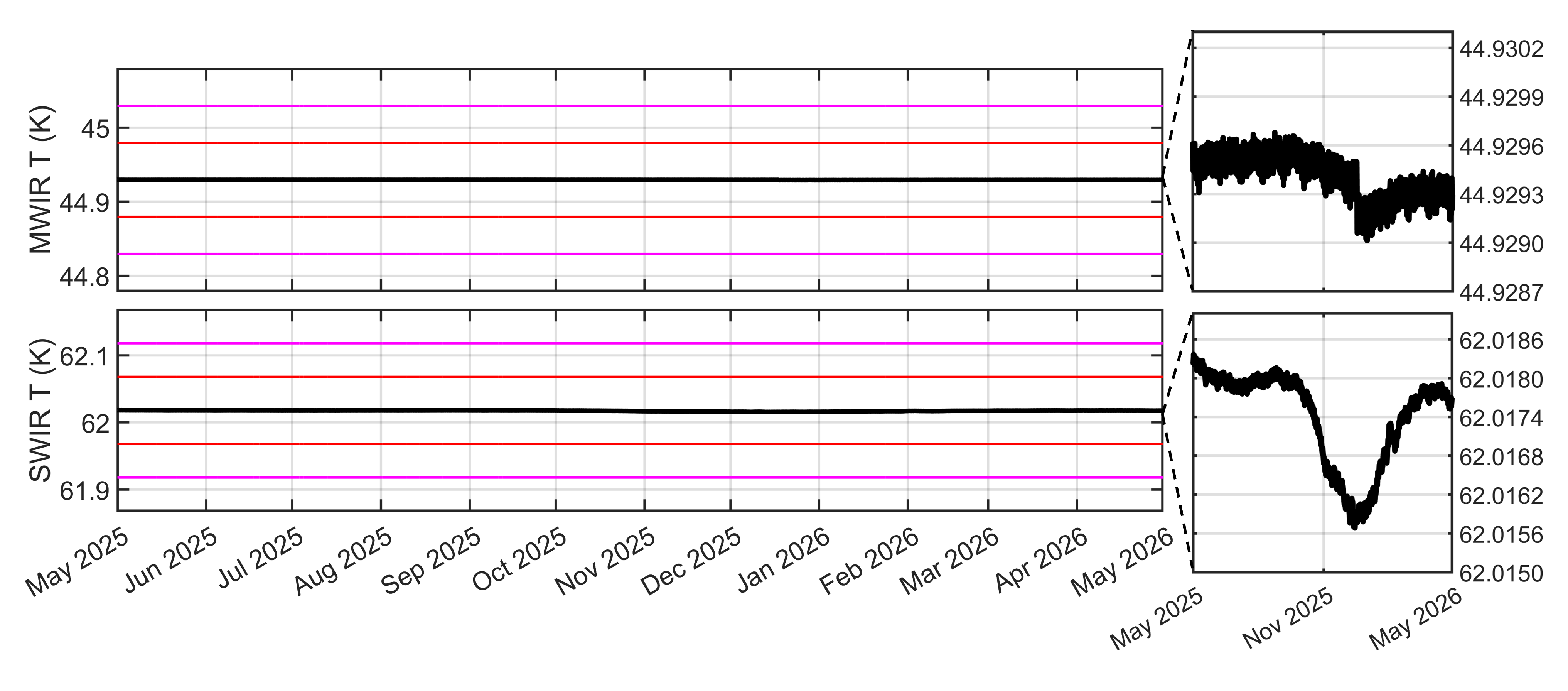}
    \caption{The FPAs exhibit thermally stable performance on survey-long timescales in both the MWIR (top) and SWIR (bottom). Black lines show the temperature data, while the red envelope indicates the 100~mK peak-to-peak requirement for 6 month temperature drift, and the magenta is the 200~mK one year limit.  Both requirements are met with nearly two orders of magnitude in margin.}
    \label{fig:6month}
\end{figure*}

The thermal stability of the instrument on all ranges of timescale meets requirements.  The most stringent of which are the 100~mK over 6~months, and 200~mK over 1 year peak-to-peak limits on FPA temperature drift.  Compliance with these long timescale requirements can only be verified through extended survey data.  Figure~\ref{fig:6month} shows FPA temperatures across the entire first and second surveys. The measurements here come from a monitor thermistor, located at the interface of the FPA and the heatstrap through which thermal disturbances enter.  The worst case peak-to-peak from the SWIR FPA was $<2.5$~mK, nearly two orders of magnitude better than the requirement.  A detailed description of the broader thermal performance of the instrument on-orbit is available in \cite{bradthesis}.

Throughout the first survey and into the second, over $1.1\times 10^6$ exposures have been successfully collected without interruption. The optical performance of the telescope is stable, and the point spread functions easily meet requirements.  The noise performance in operation is indistinguishable from measurements in the laboratory and the thermal stability of the FPAs on-orbit is such that detector gain fluctuations are negligible. As the SPHEREx survey progresses, the data are processed, archived and continuously made available to the broader astronomical community (\cite{racheldata}). Key inputs to the analysis pipeline such as the spectral calibration, response functions and noise properties were all carefully crafted deliverables from the integration and test campaign described here.  There have been no indications of performance degradation, and the instrument is well poised to continue observations near peak performance throughout the baseline mission and years into any extension.

\begin{acknowledgements}
We acknowledge support from the \spherex\ project under a contract from the NASA/Goddard Space Flight Center to the California Institute of Technology.  Part of the research described in this paper was carried out at the Jet Propulsion Laboratory, California Institute of Technology, under a contract with the National Aeronautics and Space Administration (80NM0018D0004).  This work would not have been possible without the support of our industrial partners at BAE Systems, Teledyne Technologies Inc., VIAVI Solutions Inc., Photon Engineering, and SpaceX.  We'd also like to thank the United States Space Force guardians that supported our launch at Vandenberg SFB.  The Korean contribution to SPHEREx was supported by the Korea Astronomy and Space Science Institute (KASI) under the Korea AeroSpace Administration (KASA), Republic of Korea.
\end{acknowledgements}

\bibliography{main}{}
\bibliographystyle{aasjournalv7}

\end{document}